\documentclass[twocolumn,twocolappendix]{openjournal}
\usepackage{graphics}
\usepackage{amssymb}
\usepackage{bbm}
\usepackage{amsmath,textcomp,array}
\usepackage{url}
\usepackage{hyperref}
\usepackage[ruled]{algorithm}
\usepackage{color}
\usepackage[normalem]{ulem}
\usepackage[dvipsnames]{xcolor}
\usepackage{multirow}
\usepackage[flushleft]{threeparttable}

\newcommand\minfall{$M_{\rm{infall}}$}
\newcommand\rdisrupt{$R_{\rm{disrupt}}$}
\newcommand\rmerge{$R_{\rm{merge}}$}
\newcommand\mass[1]{$M_{#1}$}
\newcommand\radius[1]{$R_{#1}$}
\newcommand\redmapper{redMaPPer }
\newcommand\hmpc{$h^{-1}$Mpc}
\newcommand\hkpc{$h^{-1}$kpc}
\newcommand\hmdot{$h^{-1}$M$_\odot$}
\newcommand\lstar{$L_\star$}
\newcommand\mstar{$M_\star$}
\shorttitle{Modeling the Galaxy Distribution in Clusters using Halo Cores}
\shortauthors{Korytov et al.}

\begin{document}

\title{Modeling the Galaxy Distribution in Clusters using Halo Cores\vspace{-1.25cm}}
\author{Danila Korytov\altaffilmark{1}, Esteban Rangel\altaffilmark{2}, Lindsey Bleem\altaffilmark{1}, Nicholas Frontiere\altaffilmark{2}, Salman Habib\altaffilmark{1,2}, Katrin Heitmann\altaffilmark{1}, Joseph Hollowed\altaffilmark{1}, Adrian Pope\altaffilmark{2}}

\affil{$^1$HEP Division, Argonne National Laboratory, 9700 S. Cass Ave.,
 Lemont, IL 60439, USA}

\affil{$^2$CPS Division, Argonne National Laboratory, 9700 S. Cass Ave.,
 Lemont, IL 60439, USA}

\begin{abstract}
The galaxy distribution in dark matter-dominated halos is expected to approximately trace the details of the underlying dark matter substructure. In this paper we introduce halo `core-tracking' as a way to efficiently follow the small-scale substructure in cosmological simulations and apply the technique to model the galaxy distribution in observed clusters. The method relies on explicitly tracking the set of particles identified as belonging to a halo's central density core, once a halo has attained a certain threshold mass. The halo cores are then followed throughout the entire evolution of the simulation. The aim of core-tracking is to simplify substructure analysis tasks by avoiding the use of subhalos and, at the same time, to more easily account for the so-called `orphan' galaxies, which have lost substantial dark mass due to tidal stripping. We show that simple models based on halo cores can reproduce the number and spatial distribution of galaxies found in optically-selected clusters in the Sloan Digital Sky Survey. We also discuss future applications of the core-tracking methodology in studying the galaxy-halo connection.
\end{abstract}



\section{Introduction}
\label{sec:intro}
In the fiducial $\Lambda$CDM model, small initial density fluctuations in the matter distribution collapse gravitationally into localized clumps, called halos. Halos are dominated by dark matter and are the nurseries within which galaxies form~\citep{Rees1977, White1978, Fall1980}. The formation of halos is a hierarchical process: larger halos are the result of mergers of smaller halos. The history of the formation of a halo, its `merger tree', and the local environment in which it lives, both play important roles in the complex set of physical processes which determine a halo's galaxy distribution (for reviews, see \citealt{Mo_2011,Somerville_2015,Wechsler_2018}). Individual galaxies form at the center of halos where the gas density is the highest for star formation~\citep{White1978, Blumenthal1986}; the most massive galaxies are the result of multiple galaxies merging into one object. As halos merge, the remnants of previously accumulated halos can survive as localized overdensities called subhalos. Subhalos are subjected to tidal forces, dynamical friction, and ram-pressure stripping, which strip them of bound mass and may lead to their eventual disruption. These effects are particularly pronounced inside massive galaxy cluster-sized halos. Many of the galaxies within the cluster halo did not form inside it, but were gathered through merger events with other halos. 

The intertwined dynamics of halos, and the formation of galaxies within, is a highly complex multi-scale nonlinear process, accessible in detail only via numerical simulations that incorporate gravity, gas dynamics, and a number of astrophysical processes (\citealt{vogelsberger2019cosmological} present a recent review). The spherically-averaged halo matter density profile can be described reasonably well by a broken power law with a dense center $(\rho\sim1/r)$, but with much less dense outskirts $(\rho\sim1/r^3)$, the so-called NFW profile (after Navarro, Frenk, and White,~\citealt{Navarro1996}). The NFW profile is determined by two parameters, which can be taken to be the halo mass and the halo concentration, a density scale parameter. The observed distribution of galaxies tends to follow the NFW profile~\citep{Lin2004, Budzynski2012,Shin2021}, although this profile does not have to trace that of the underlying dark matter. In contrast, the distribution of subhalos does not follow an NFW profile, as the subhalo population density, relative to the overall matter density, is suppressed at smaller halocentric radii~\citep{Gao2004, Guo2010}, the origin of `orphan' galaxies invoked in semi-analytic models of galaxy formation.  As the above indicates, directly associating galaxies with subhalos can be problematic for several reasons. First, the current subhalo mass does not directly reflect the mass of the original halo when it was accreted by the main halo (the `infall' mass). More than 90\% of the infall mass can be lost from tidal stripping, dynamical friction, or other interactions~\citep{Wetzel2010}. Second, finding subhalos and determining their properties robustly in simulations is not straightforward~\citep{2011MNRAS.410.2617M,Onions2012, Knebe2013,2018MNRAS.474..604H}. Third, subhalos can be disrupted (or not found) due to the inherent force and mass resolution limitations of the simulation~\citep{2014MNRAS.437.3228G,2018MNRAS.475.4066V}.

The aim here is to establish a simple description of galaxy populations in halos that avoids direct reliance on subhalos. The idea is to follow a group of simulation particles in gravity-only simulations that can be explicitly tracked and collectively serve as a proxy for a galaxy position. The basic notion can be traced back to \cite{kaiser1984} and \cite{white1987}, focusing on following substructure density peaks as potential locations for galaxies. Several additional assumptions underlie the procedure followed here. As in subhalo abundance matching (SHAM;~\citealt{Kravtsov2004, Val2004, Conroy2006}), we first assume that all galaxies form as central galaxies in halos and become satellites only through subsequent merging with other halos. 
Second, as the stellar component is dense enough that hydrodynamic forces are negligible and the stars are sparse enough that they are effectively collisionless, we assume that once a galaxy is formed the main force that dictates the
bulk motion of the galaxy is gravity.  With these two assumptions, the trajectory of the central or `core' set of simulation particles of a halo should have a very similar trajectory to that of a galaxy that originated at the center of the same halo. We call this central set of simulation particles the `halo core'. Following an individual halo core not only yields the putative galaxy position and velocity inside the halo, but also informs us if the galaxy experienced strong tidal forces that may have led to its disruption --- a combination of dynamical friction and tidal forces will determine the eventual fate of a satellite galaxy via merging into a central galaxy and/or contributing to the intra-cluster light (ICL) as it is disrupted. Finally, core mergers can be studied as proxies for galaxy mergers. 

Tracking of halo cores has been implemented within the Hardware/Hybrid Accelerated Cosmology Code (HACC) framework~\citep{HACC, Habib:2014uxa}, as described in \cite{Rangel2018} and \cite{LJ1}. It has been extended in \cite{LJ2} to include a mass loss model for substructures, called SMACC (Subhalo Mass-loss Analysis using Core Catalogs). 

In this paper, we focus on using halo cores to model the spatial distribution of galaxies in cluster-scale halos. We describe our method for identifying and following the halo cores using two cosmological simulations; a smaller one for subhalo comparisons, development, and testing, and a very large run that has sufficient volume for carrying out cluster-scale analyses. The simulations have sufficient mass resolution to adequately follow the evolution of cluster-scale halo substructure, an essential requirement for core-tracking. Halo core modeling, combined with a simplified approach to populating galaxies in halos, is used to reconstruct the projected radial profiles of galaxies in clusters found in the Sloan Digital Sky Survey (SDSS), with very good results. The use of halo cores for modeling the galaxy content at lower halo masses -- with more complex models than the approach used here for clusters -- will be described elsewhere.

The procedure for tracking halo cores is integrated within our halo merger tree construction process. It allows for the implementation and testing of physically motivated models for associating galaxies with halos. The work reported here is a first investigation beginning with a simple prescription controlled by a single free parameter (the infall halo mass associated with a core, the value of which serves as a proxy for a threshold in the associated galaxy luminosity), to two-parameter models (infall mass plus core disruption and infall mass plus core mergers), and a three-parameter model (infall mass plus core disruption plus core mergers). We find that the first type of model is inadequate to describe the galaxy distribution within SDSS clusters as expected, but the other models provide very good fits, with core disruption (corresponding to satellite galaxy disruption and/or merger with a central galaxy) proving to be an important ingredient.

The SDSS clusters were obtained from an optically identified
redMaPPer cluster catalog \citep{Rykoff2014}. redMapPPer is a red sequence-based cluster finder~\citep{Gladders_2000} designed for photometric surveys; its galaxy richness estimates correlate well with other observational halo mass proxies. In our work, the observed halo mass is a modeling input; for this purpose we utilize several mass-richness scaling relations \citep{Rykoff2014, Farahi2016, 2017MNRAS.466.3103S, 2019MNRAS.482.1352M} to study the potential systematic effects induced by different methods for estimating cluster masses.

As can be easily imagined, halo core-tracking may be straightforwardly extended in a number of directions -- as an underlying infrastructure for semi-analytic models of galaxy formation, for incorporating galaxy positions and velocities in structure formation probes of cosmology, and in the construction of empirical models that describe galaxy properties in a halo as a function of the halo history and environment.

The organization of the paper is as follows. The N-body simulations, halo core tracking algorithm, merger tree construction, and comparison of cores to subhalos are discussed in Section~\ref{sec:methods}. We present details of several simple core-based galaxy models in Section~\ref{sec:fitting}. Section~\ref{sec:sdss_data} covers how the observed data sets from SDSS are converted to galaxy cluster surface density profiles.  In Section~\ref{sec:results}, we then use the observational data to derive fitting results for different core-based galaxy models, and discuss various tests of robustness that were performed in Section~\ref{sec:robustness_convergence_tests}. Concluding remarks and an outlook regarding future work, including cores in the context of hydrodynamic cosmological simulations, are to be found in Section~\ref{sec:conclusion}.

\section{Simulations and Core-Tracking Approach}
\label{sec:methods}

In this section we describe the cosmological simulations used for the halo core analysis, and present the procedures employed for core identification and tracking, along with a list of measured core properties. A comparison of the core distribution against that of subhalos is investigated and discussed.

\subsection{N-body Simulations}

We employ two N-body gravity-only simulations, carried out using HACC. The overall parameters of the simulations (box sizes, number of particles and the resulting mass resolution) are listed in Table~\ref{tab:simulations}. The simulations share the same spatially flat $\Lambda$CDM cosmology with parameters consistent with WMAP7 results~\citep{2011ApJS..192...18K}:
$\Omega_{\rm tot}=0.265$, $\Omega_b = 0.045$, $h = 0.71$, $\sigma_8 = 0.8$, and $n_s = 0.963$. All calculations in this paper are carried out with these cosmological parameters.

The smaller-volume AlphaQ simulation was used for development of analysis methods and codes, as well as for the subhalo-core comparisons. The much larger Outer Rim simulation~\citep{OuterRim} provided the primary data
source to perform fits to cluster profiles. Both simulations have similar mass resolution to ensure that the tests with AlphaQ are directly relevant to the Outer Rim analysis. We note that, amongst other applications, the Outer Rim run has been used in support of eBOSS analyses~(see, e.g., \citealt{rossi}) and is the basis for the LSST DESC DC2 synthetic galaxy catalog~\citep{dc2}.

\begin{table}[b]
  \begin{center}
  \caption{N-body simulation parameters}
    \begin{tabular}{lrrr}\label{tab:simulations}
      Simulation & Box Length & Particles & Mass Resolution\\
      \hline
      \hline
      AlphaQ & 256~\hmpc & $1024^{3}$ & $\sim1.6\times10^9$ \hmdot \\
      Outer Rim & 3000~\hmpc & $10240^{3}$ & $\sim1.7\times10^9$ \hmdot \\
    \end{tabular}
  \end{center}
\end{table}

\begin{table*}{
  \begin{center}
  \caption{Number of cluster-sized halos in the simulations and the
    \redmapper catalog}
    \begin{tabular}{crrrr}\label{tab:halo_counts}

      Mass & AlphaQ & Outer Rim & redMaPPer \\
      \hline
      \hline
      
      14.00 $<$ log($M_{200c}$) $<$ 14.25 &     92 &   175735 &  2746  \\
      14.25 $<$ log($M_{200c}$) $<$ 14.50 &     26 &    60719 &   888  \\
      14.50 $<$ log($M_{200c}$) $<$ 14.75 &      7 &    16203 &   230  \\
      14.75 $<$ log($M_{200c}$) $<$ 15.00 &      1 &     3018 &    49  \\
      15.00 $<$ log($M_{200c}$) $<$ 15.25 &      1 &      318 &     5  
    \end{tabular}
  \end{center}
  \begin{tablenotes}
          \item 
          \centering\small The richness-mass scaling uses the relationship from~\cite{2019MNRAS.482.1352M}. The original redMaPPer masses are measured with respect to the mean density (\mass{200m}); the conversion to \mass{200c} assumes an NFW profile and the mass-concentration
relationship specified in~\cite{Child2018}. 
        \end{tablenotes}}
\end{table*}

Both simulations were run and processed similarly: the simulations start at redshift, z$_{in}\sim$200, and evolve to $z=0$. About 100 snapshots of the simulations are used for the core-related analyses carried out between $z\sim10$ to $z=0$. Halos in each snapshot are identified using the Friends-of-Friends (FoF) halo finder algorithm~\citep{Davis1985} with a link length $b=0.168$. Spherical over-density (SOD) halos~\citep{1994MNRAS.271..676L} with density 200 times the critical density (\mass{200c}) are identified using the FoF halo potential minimum as the SOD halo center. We use \mass{200c} as the mass definition throughout the paper for consistency. When converting to an SOD mass measured with respect to the mean density, as is the case for the redMaPPer clusters, we assume that the halo follows an NFW profile and obeys the concentration-mass relationship specified in~\cite{Child2018}. The number of cluster-sized halos in each simulation is given in Table~\ref{tab:halo_counts}. The statistics from the SDSS redMaPPer clusters (Section~\ref{sec:sdss_data}) are also listed to provide context; the Outer Rim run is comfortably large enough to make statistically robust predictions for the SDSS clusters.

Merger trees are constructed by comparing common particles between FoF halos in adjacent time steps to establish progenitor and descendant relationships. In most cases, halos merge into a single halo but some individual halos split into multiple halos, which is a physical process in the simulation~\citep{Stuart2005, Sales2007, Ludlow2009,2017MNRAS.468..885V}. In our merger trees, any halo that splits into two or more descendant halos at a later time step is separated into a corresponding number of `fragment' halos equal to the number of descendant halos. The `fragmentation' of halos effectively unravels the halo merger event until the halos permanently merge. The resulting halo merger trees are untangled and each tree may be considered fully independently of any other tree. Further details on our merger tree construction process can be found in~\cite{Rangel2018}.

\subsection{Identifying and Tracking Halo Cores}
\label{sec:core_tracking}
Halo core tracking is performed in two stages. In the first stage, particles are identified that represent cores of individual halos at each simulation snapshot. These core particles are tracked for the remainder of the simulation. During the second stage, halo merger trees and the tracked particles are combined to follow the evolution of cores as halos evolve and merge. The first stage is carried out on the fly (as the simulation is running) while the second stage is carried out in post-processing. As described further below, halo core-tracking adds very little computational and memory overhead to the simulation, since it involves only neighbor-finding for a small set of particles and keeping a running record of particle labels. The core-augmented merger-tree construction in post-processing is also straightforward and adds only a small overhead to our usual procedure of halo merger tree construction. 

\begin{figure}[b]
  \begin{center}
    \includegraphics[width=0.9\linewidth]{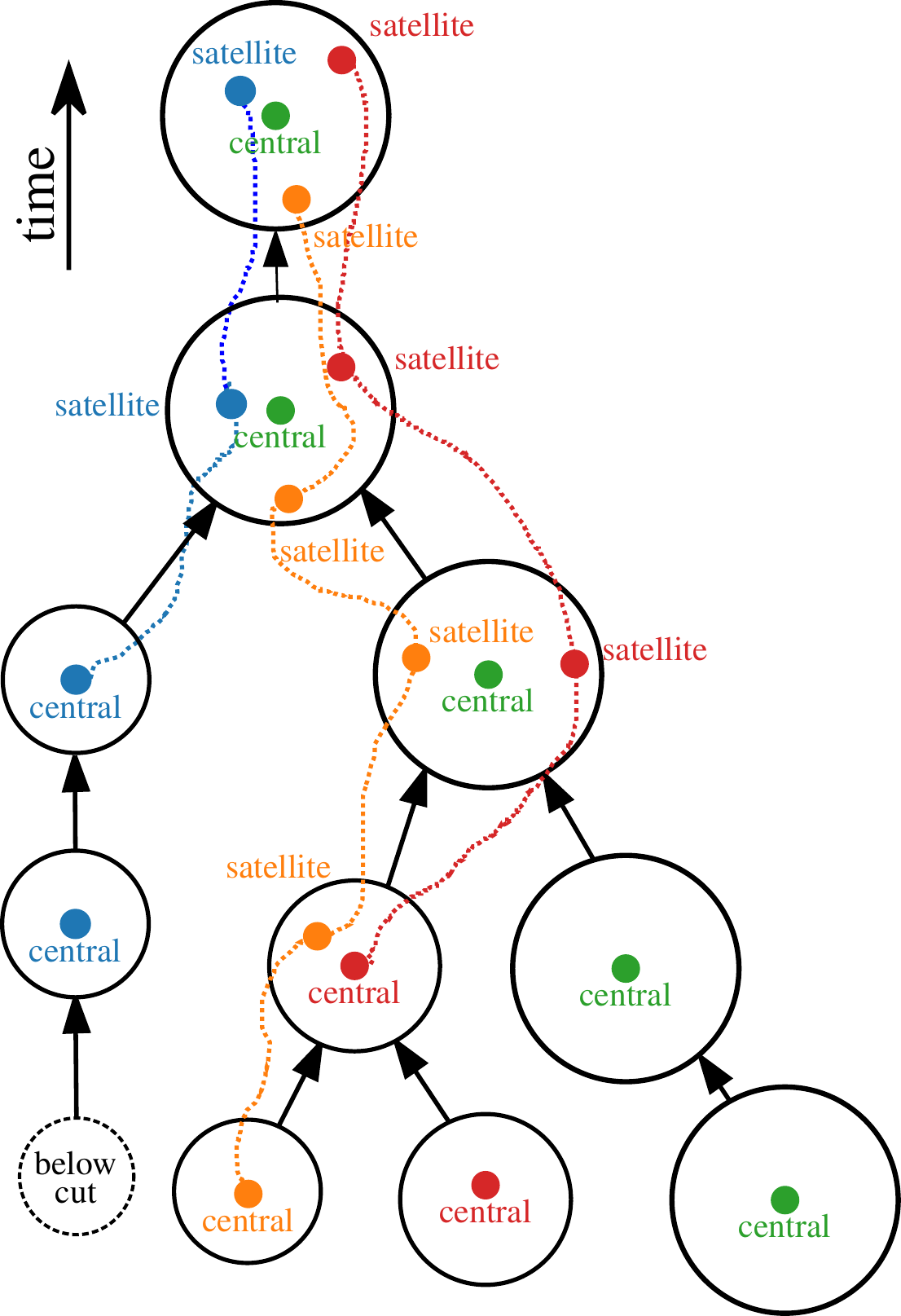}
  \end{center}
  \caption{Cartoon showing how cores are assigned to halos
    using merger trees. Circles represent halos and arrows represent halo descent across adjacent simulation snapshots. Colored dots represent halo cores, each color marking a separate core. Cores linked by dashed lines are represented by the same set of simulation particles. Each halo hosts up to one central core and any number of satellite cores. All cores start out as isolated centrals in halos above the chosen mass threshold (solid circles). Halos below the mass threshold (dashed circles) do not have their halo core tracked. Central cores are updated
    with a new set of particles at each simulation snapshot. Central cores become satellite cores after a merger with a more massive halo. We note that core disruption and merger processes are not shown in this figure for sake of clarity.}
  \label{fig:cartoon_merger_tree}
\end{figure}

For the two simulations used here, AlphaQ and Outer Rim, we apply a halo selection threshold of $M_{\rm{FoF}}\sim 2\times10^{11}$\hmdot, corresponding to 100 simulation particles. For every FoF halo above this mass threshold, we identify its halo core as the spatially closest 20 simulation particles to the halo center (we have investigated using fewer and greater numbers of particles, the results are stable with this choice). This process is repeated for each FoF halo at each simulation snapshot. We use the criteria of neighbor-finding to determine cores because it is computationally very efficient and does not require a detailed computation of the local potential field to determine bound particles. 

We note that a passively evolving halo with no mergers will have a new set of core particles identified for every simulation snapshot. Particle migration across cores is possible as the same simulation particle can be identified to belong to multiple cores at different time steps. 

Starting from an early simulation snapshot and including all subsequent snapshots, every particle that has been flagged as belonging to a halo core is added to a running list of particles. As particles in this set can belong to multiple halo cores, the membership to halo cores is not stored. At each snapshot, the position and velocity of these accumulated particles is updated. The accumulation of particles in this manner allows for the process to be carried out on the fly during the first analysis stage.

The second stage of core tracking links cores across snapshots to follow the evolution of halos and substructure within halos. Cores that were identified in the first stage are linked across simulation snapshots to create a trajectory using the halo merger trees. A cartoon of the process is shown in Figure~\ref{fig:cartoon_merger_tree}. We will first describe the general case of core-augmented merger tree construction, followed by a discussion of the complications related to fragment halos and other effects. 

We define three basic types of halos: halos below the chosen mass threshold for core tracking mentioned above, isolated halos above the mass threshold, and composite halos above the mass threshold. Once a halo crosses the mass threshold from below, we start tracking its halo core particles. At an early stage, such a halo would typically be isolated, with only a core at its center and no satellites. Isolated halo cores are assigned the set of particles that were most recently identified as being the closest to the halo center at that time, except in the rare cases of halo fragmentation described below. When two or more halos containing central cores merge into one composite halo at a later time step, one core -- the one belonging to the most massive merging FoF halo -- is designated as the central core while the others become satellite cores. Satellite cores
are permanently locked to the last set of particles assigned just prior to the merger (we note that there is no assumption that this set of core particles will remain compact during subsequent evolution); whereas the central core is updated with new particles from the halo in the same way as if it was an isolated halo.

\begin{figure}[b]
  \begin{center}
    \includegraphics[width=3.5in]{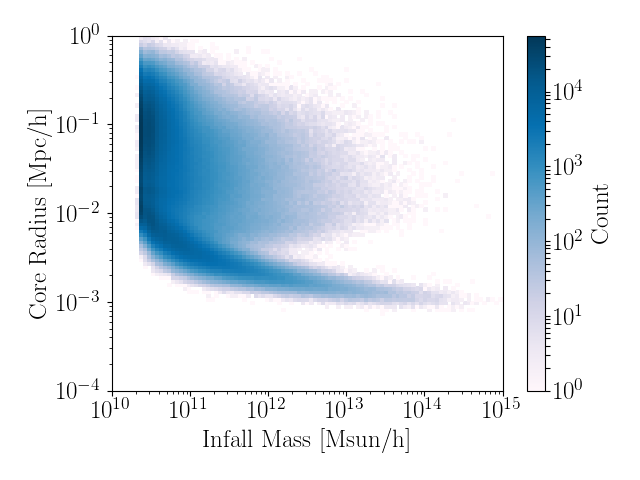}
    \includegraphics[width=3.5in]{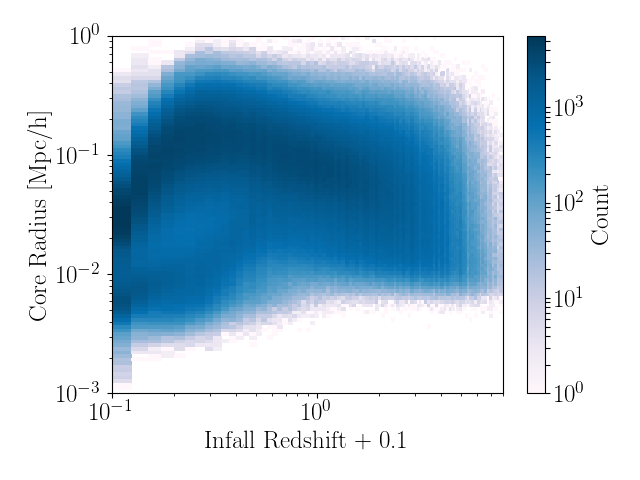}
  \end{center}
  \vspace{-0.6cm}
  \caption{{\it Top panel:} Population density as a function of halo core radius and core infall mass in the AlphaQ simulation at $z=0$. Several distinct features are visible. The bottom group with the smallest radii are central cores. The particles representing the core have been refreshed to be the spatially closest to the potential minimum of the current host halo. The population above the centrals are (dispersed) satellite cores which have had time for the core particles to spread out. {\it Bottom panel:} Core radius as a function of core infall redshift for all cores in AlphaQ at $z=0$. Recently merged satellite cores have a smaller radius as they have not experienced strong disruptive forces in the host halo. }
  \label{fig:core_radii_pop}
\end{figure}

There are several edge cases that need to be handled. Firstly, halos may have two or more local potential minima that are comparably deep and the halo potential center may therefore flip-flop between them across snapshots, adding an essential ambiguity in the assignment of the central core. To avoid this type of jump, we only update the central core particle assignment if its position has not moved past a distance threshold. 

Another edge case occurs when halos that are the result of temporary mergers are separated into a number of fragment halos.
Most fragment halos do not have well defined positions -- with the exception of the largest fragment, which is assumed to share the same potential center as the halo it was fragmented from. Fragment halos are not present in the FoF
halo catalogs nor do they have well-defined centers, so no core particles are associated with them for that time step (except for the largest fragment halo). 
Instead, these halos adopt core particles from the first non-fragment progenitor in their assembly history. In very rare cases, this adoption method does not produce any particles as all of the progenitors are fragments. (For further details, see \citealt{Rangel2018}.)

Lastly, halos that have once passed the mass threshold may fluctuate below it at some future time step. For these halos, we follow a similar procedure as above of adopting previous core particles and including them in our analysis. As a result some cores will have a recorded infall mass (described below) that is less than the mass threshold for tracking cores.

Once all the edge cases are resolved and all halo cores have their particles assigned, we compute the properties of each core. The full list of these properties as computed here is given in
Table~\ref{tab:core_catalog}. (Other quantities can be added and current ones modified, for example, the core velocity -- a variable not used in this work -- can be replaced by an average over neighbors to produce a less noisy estimate.) Satellite cores have additional
information from when they were last a central core in a host halo, just before their host fell into a more massive halo. The infall redshift, the identity of the infall halo, and the mass of the infall halo are recorded. For central cores, the associated infall mass is simply the current host halo mass. As will become apparent soon, these infall masses play an important role in our modeling of the cluster galaxy population at different luminosity thresholds using cores.

\begin{table}[t]
  \begin{center}
  \caption{\label{tab:core_catalog}Properties identified or computed for each core.}
    \begin{tabular}{ll}
      Property & Description\\
      \hline
      \hline
      Core tag & unique identifying number (integer) \\
      Host halo tag & identifier for the current host halo  \\
      Core position & position of center-most particle\\
      Core velocity & velocity of central core particle \\
      Core radius & compactness measure\\
      Peak core radius & maximum radius during time evolution \\
      Infall halo tag & identifier of an ``infall'' halo \\
      Infall mass & mass of the halo at infall\\
      Infall time & time step at halo infall\\
      Infall peak mass & maximum mass attained by an infalling halo\\
      Central flag & identifies a core as currently a central core
    \end{tabular}
  \end{center}
\end{table}

\begin{figure*}[t]
  \begin{center}
    \includegraphics[width=0.333\textwidth]{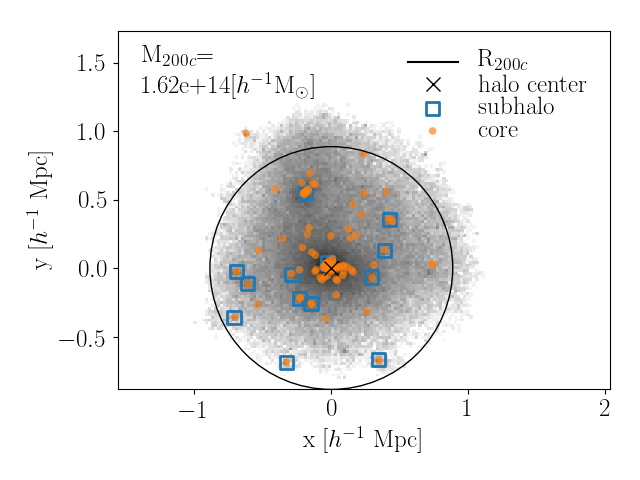}\includegraphics[width=0.33\textwidth]{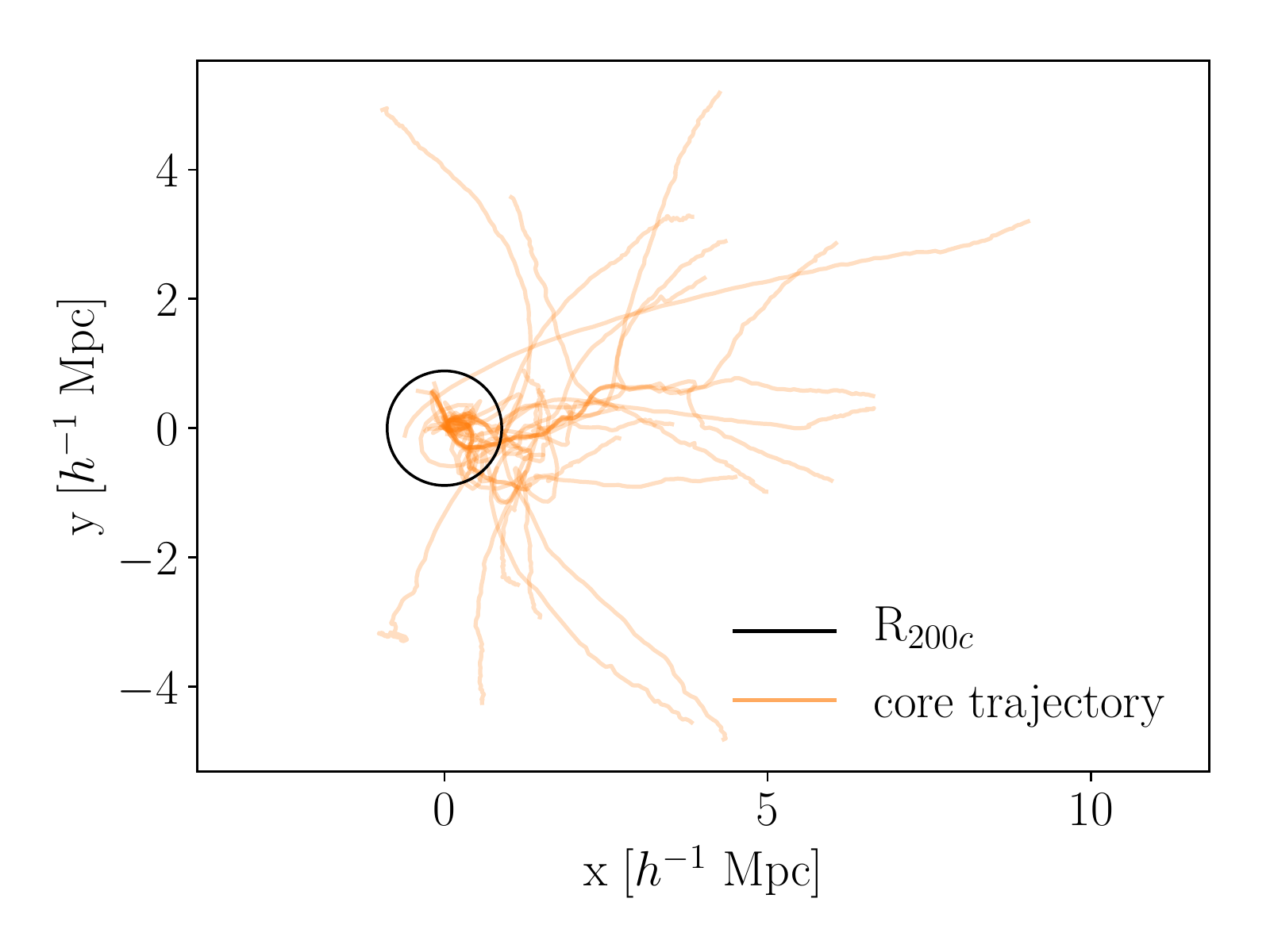}\includegraphics[width=0.34\textwidth]{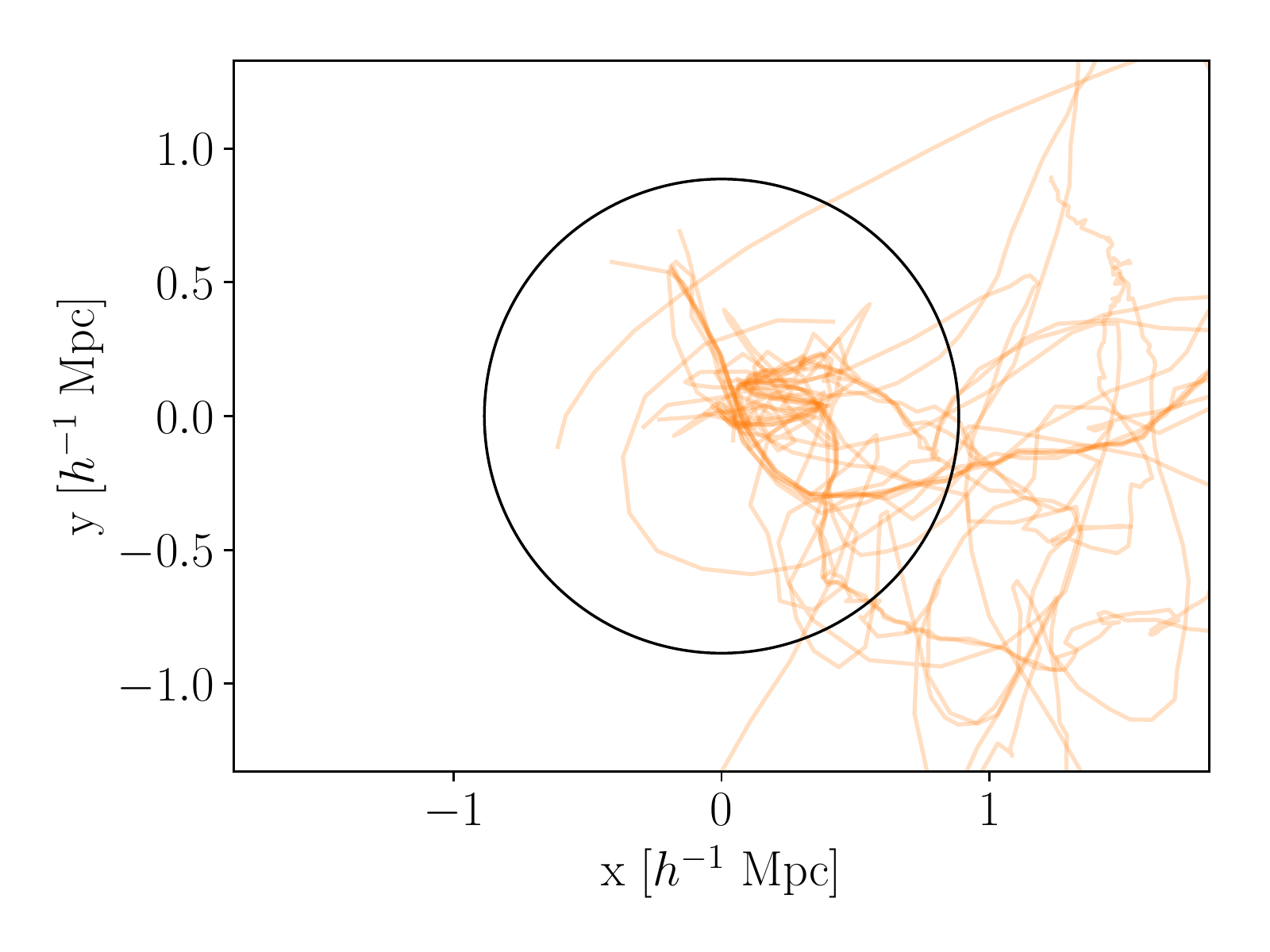}
    \includegraphics[width=0.333\textwidth]{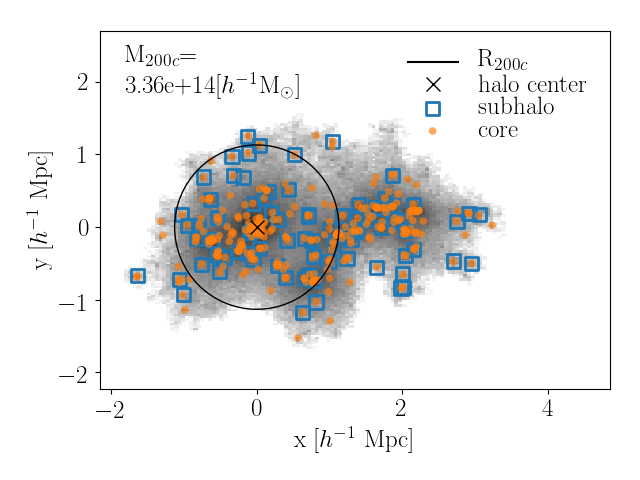}\includegraphics[width=0.33\textwidth]{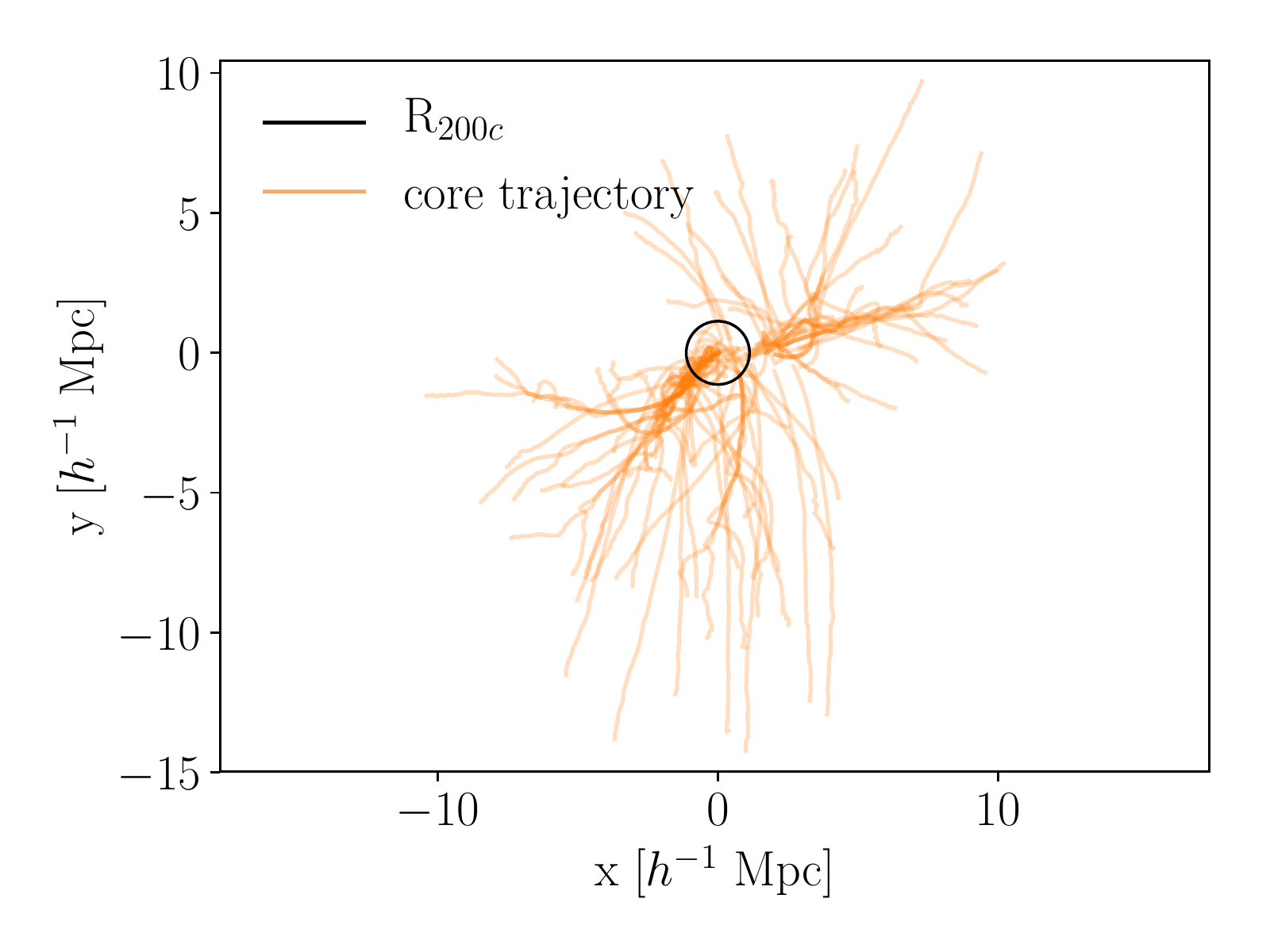}\includegraphics[width=0.34\textwidth]{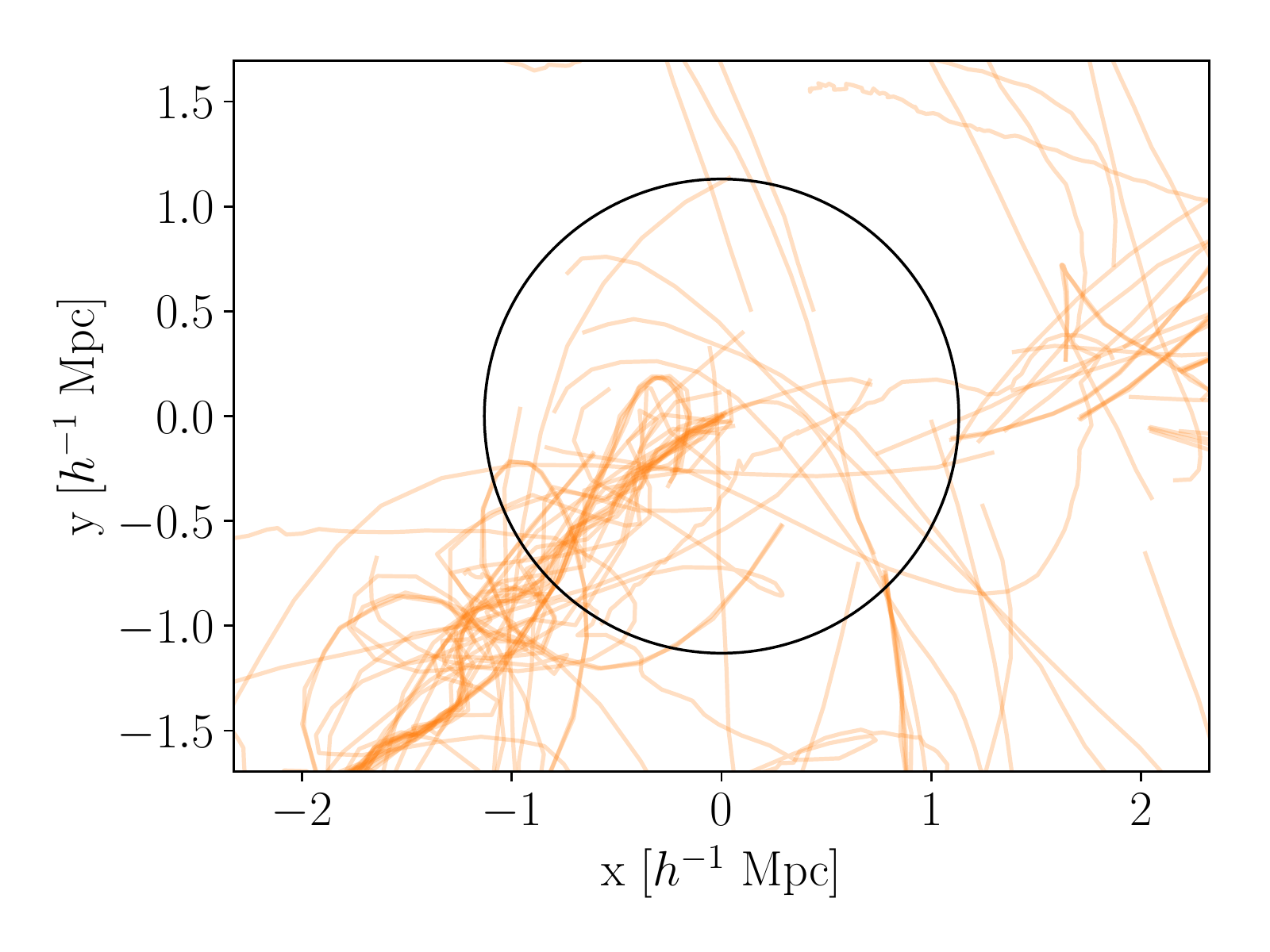}
  \end{center}
  \caption{Halo core distribution and assembly for two cluster-scale mass halos at $z=0$. Each row presents information for the same halo, with its \radius{200c}\ at
    $z=0$ marked by the black circle. The left panels show the distributions of cores, subhalos  (see Section~\ref{sec:subhalo_compare}) and (FoF) dark matter particles (density in grayscale). The subhalos and cores that are displayed have a current mass and infall mass, respectively, of at least $\sim1.6\times 10^{11}~h^{-1}$M$_\odot$. The middle and right panels display the
    trajectories of cores throughout the assembly history of the two halos. For clarity, only cores with an infall mass above $10^{12}$ \hmdot are shown. The right panels show a zoom-in of the middle panels into the $z=0$ position of the halo, showing the complex nature of the core trajectories as viewed in projection.}
    \label{fig:core_halos}
\end{figure*}

For all cores, the position and velocity of the core is defined to be the position and velocity of the core particle that has the minimum sum of linear distances to all the other core particles. The effective radius, a measure of compactness, is taken to be the {\em rms} of the standard deviation of the core particle positions in each dimension:

\begin{equation}
  R_{\rm{eff}}  = \frac{1}{3}\sqrt[]{ \frac{1}{N}\sum_{i=0}^{N}
    (\bar{x}-x_i)^2+(\bar{y}-y_i)^2+(\bar{z}-z_i)^2},
  \label{eq:core_radius}
\end{equation}
where $N$ is the number of particles in the core, $(x_i,y_i,z_i)$ is the
position of the $i^{th}$ particle, and $(\bar{x}, \bar{y}, \bar{z})$ is the average position of all the particles in the core, respectively. The distribution of core radii and infall masses for
AlphaQ at $z=0$ are shown in Figure~\ref{fig:core_radii_pop}. 

We explored several definitions of effective radius, including choices of how many particles to include in the core definition. By systematically increasing the number of particles in a core we found that by 50 particles the results were well-converged, and that results from 20 particle cores were equally stable, within statistics. The different core size definitions performed similarly and correlated well $(r>0.9)$ with cores sampled with 50 particles. These definitions were the average
particle distance from the central particle, and the 50th, 60th and 80th percentile of particle distances from the center.  Applying Chauvenet's criterion for outlier detection~\citep{Taylor_1997}, we found that outlier rejection for all of the methods did not improve the convergence properties of the radius definition. We used 20 particles for the core.

Once the core properties have been computed, the core catalog is assembled. The core-halo membership is determined from this catalog and the assembly history of the cores may now be followed. The $z=0$ distribution of cores and the assembly of core
trajectories for two example halos are shown in
Figure~\ref{fig:core_halos}. Halos typically have a tight cluster of cores that have fallen into the potential center of the halo. In observations, the centers of galaxy clusters are often populated by the brightest cluster galaxy (BCG)~\citep{Jones1984,Ho2009,Mehrtens2012}. These BCGs are massive elliptical galaxies and are the end-result of a complex merger process, including the effects of dynamical friction and tidal stripping~\citep{Dubinski1998,Gao2004b,DeLucia2007, Wetzel2009}. Cores follow a similar pattern where many cores resulting from mergers concentrate at multiple dominant potential minima. The modeling of core disruption and mergers to yield a single satellite galaxy proxy object is described in Section~\ref{sec:core_merger}; we do not treat BCG formation in this paper -- or more generally, very massive elliptical galaxies, which involves a separate modeling procedure.

\subsection{Comparison of Cores to Subhalos}
\label{sec:subhalo_compare}
In this section, we compare cores to subhalos as found by a dedicated subhalo finder; a related study was carried out in~\cite{LJ2}. We focus on the AlphaQ simulation for this comparison. The expectation is that at larger radii and at higher subhalo masses, the matching of cores to subhalos would be essentially perfect, while near deep local potential minima, there would be a concentration of core particles and remnants of disrupted subhalos, where the correspondence between subhalos and cores would break down. We have also carried out a detailed study of mass modeling for the cores based on the post-infall evolution of subhalo masses. These results, however, are not directly relevant for the present purpose and have been presented elsewhere~\citep{LJ2}.

\begin{figure}[t]
  \centering
  \includegraphics[width=0.45\textwidth]{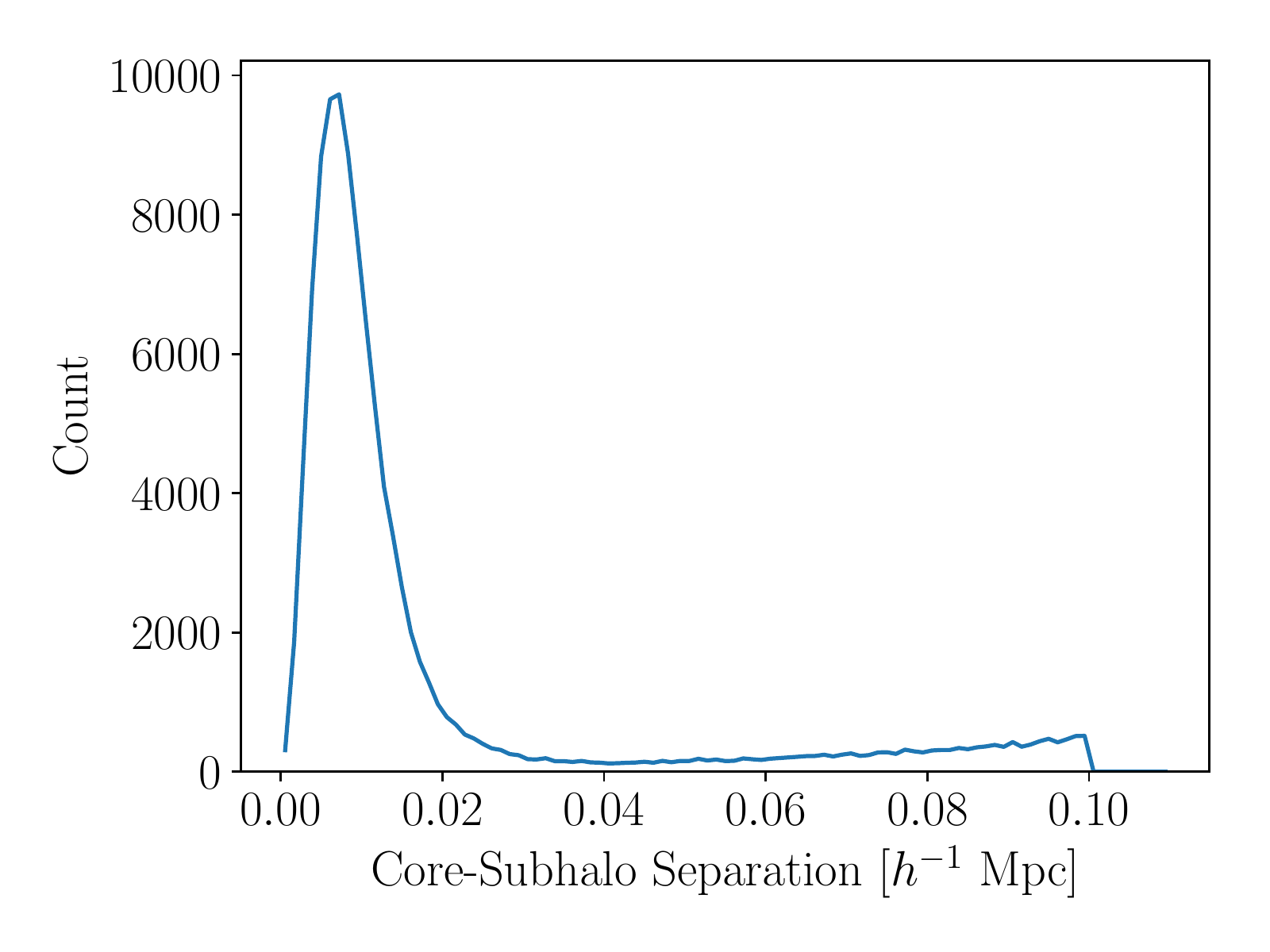}
  \caption{The distance distribution between matched subhalo centers and associated cores (closest single core to a given subhalo), showing a tight confinement within $20~h^{-1}$kpc. The tail arises from subhalos that are not well-localized.}
  \label{fig:subhalo_core_distance}
\end{figure}

Subhalos are found using a method that combines spatial density and phase space information, as described in~\cite{LJ2}. 
In order to compare the subhalo distribution to that of cores without
resorting to subhalo merger trees, we match the two spatially. The distance distribution between matched cores and subhalo centers (closest core to a subhalo center) is shown in Figure~\ref{fig:subhalo_core_distance}, and is tightly confined to less than $20~h^{-1}$kpc. Outliers at larger distances are due to the complex structure of subhalos. A further discussion of completeness in the core vs. subhalo comparison (as a function of subhalo mass) is provided in~\cite{LJ2}.

\begin{figure}[t]
  \centering
  \includegraphics[width=0.45\textwidth]{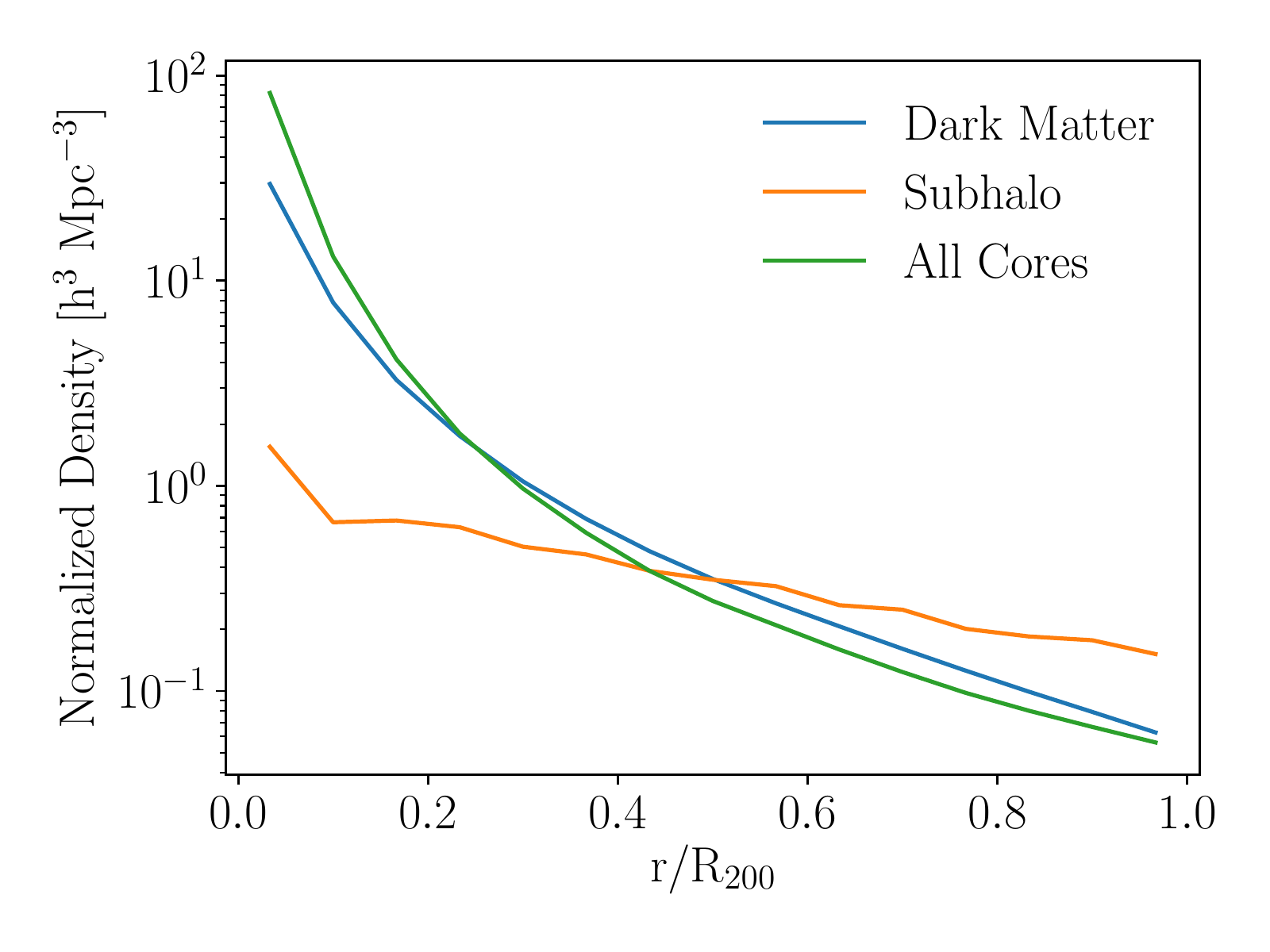}
  \caption{Radial density profile of matter, subhalos and cores. The
    profiles are normalized to integrate to unity at $r=R_{200}$. The
    subhalo profile is much flatter than the matter or the core profile,
    lacking a strong peak at the center. The cores are more centrally
    concentrated than the matter distribution. (We note that core mergers and disruption are not accounted for in this figure.)}
  \label{fig:dm_subhalo_core_profile}
\end{figure}

We compare the radial density profile of matter, subhalos and cores in Figure~\ref{fig:dm_subhalo_core_profile}. As previously discussed, the subhalo distribution is significantly flatter than the core profile, which roughly tracks the matter distribution in its shape. The (naive) core distribution is more centrally concentrated than the matter distribution at smaller distances; this accounting does not include the physically important process of core mergers, which will be further discussed below.

\section{Modeling the Galaxy Population in Halos}
\label{sec:fitting}

There are a number of empirical models for populating halos with
galaxies; two popular classes are halo occupation distribution (HOD)
approaches~\citep{Peacock2000, Berlind2002, Zheng2005} and subhalo
abundance matching (SHAM) methods~\citep{Kravtsov2004, Conroy2006,
  Behroozi2010, Moster2010}. The simplest HOD model describes
the number of galaxies (satisfying some selection criterion) expected
in a halo as a function of halo mass alone. The functional forms of HOD models are parametrically tuned to match the density and two-point correlation
function of the specific galaxy populations of interest, separated into central and satellite galaxies. The satellites
are distributed spatially in a halo following a prescription related to the
halo's density profile or to an NFW profile that uses a different
value of the concentration for the galaxy distribution.

SHAMs rely on abundance matching to populate halos with central and satellite galaxies. In the simplest SHAM model, a monotonic one-to-one relationship between
subhalo masses and the galaxy luminosity (or some other property) is assumed. Typically, the
subhalo peak mass or peak maximum circular velocity is used instead of the (post-merger) subhalo mass and maximum
circular velocity since subhalos may lose more than 90\% of their initial
bound mass after merging with a larger halo \citep{Wetzel2010,
  2018MNRAS.475.4066V}; this mass loss is not directly connected to the luminosity evolution of the modeled galaxy. Due to the association with individual subhalos, galaxies in SHAM can be tracked across simulation snapshots. 
  
In comparison with HODs,
a SHAM model requires much higher mass resolution as the
subhalos need to be resolved and tracked (see, e.g., \citealt{Wetzel2010}). While SHAM methods have fewer parameters
and include more dynamical information, they do have a number of shortcomings. Aside from the main abundance matching assumption, subhalo identification and construction of subhalo mergers have additional complexities as mentioned earlier \citep{2011MNRAS.410.2617M,Onions2012, Knebe2013}, require high simulation resolution \citep{2014MNRAS.437.3228G,2017MNRAS.468..885V,2018MNRAS.475.4066V}, and are computationally expensive to run on large simulations. Additionally, subhalo-based models do not recover the correct small-scale clustering unless orphan galaxies are incorporated in the modeling process~\citep{2018MNRAS.477..359C}. Orphan galaxies (galaxies that have lost most of their dark matter halo) must be modeled using an alternative method.

In this paper, the modeling of the galaxy distribution in cluster-scale halos relies on using halo cores as the essential element, rather than subhalos. We follow the spirit of the SHAM  approach in terms of tracking substructure, but not in terms of abundance matching. Instead of rank ordering core infall masses, we simply use a threshold in infall mass as a proxy for a threshold in galaxy luminosity (Section~\ref{sec:core_m_infall}). 

The main simulation used here simultaneously possesses sufficient mass
resolution to model bright galaxies in clusters, as well as large
enough volumes to provide excellent statistics. Additionally, the
demonstrated robustness of cores relative to subhalos and the
increased ability to model galaxies within the halo's virial radius
provide a set of natural parameters (e.g., describing core disruption
and central/satellite core merging) that can be used to extend a mass threshold parameter, as employed in SHAM approaches. 

\subsection{Modeling Cores as Galaxies}
\label{sec:core-galaxies-models}

Once the simulation output has been reduced
to the core catalog (Section~\ref{sec:methods}), we aim to match the
halo cores to galaxies above a given luminosity threshold. The matching to observations will use the measured halo mass as well as the projected galaxy distribution, dominated by satellite galaxies. By construction, the work here does not attempt a detailed description of BCG formation (or of other very bright galaxies in clusters).

We construct a sequence of models to fit the cluster galaxy profiles as described further below.  All of
these models share a common aspect, viz., the core needs to originate from an
infall halo above an associated mass threshold, \minfall\, to be considered as a ``marker'' for a single central galaxy, which after infall becomes a satellite in the larger halo. In the simplest model, we will assume nothing more, i.e., make no allowance for the physical processes of core disruption and mergers. This model will then be extended by allowing for two separate mechanisms, core disruption (expected to be significant in the high-density central region of the halo) and core mergers, which can be turned on separately, or considered simultaneously.

In total, we investigate four models, each of which associates cores with galaxies, labeled as:
\begin{itemize}
\item \textbf{Mi}: Cores with infall mass greater than \minfall\ are
  labeled as galaxies, with no allowance for disruption or mergers.
\item \textbf{MiRd}: Cores with infall mass greater than \minfall\ and
  radius less than \rdisrupt\ are labeled as galaxies, therefore allowing for disruption but without treatment of potential mergers. 
\item \textbf{MiRm}: Cores with infall mass greater than \minfall\ that
  can merge with other cores above \minfall\ if they are within a
  distance \rmerge~of each other (the actual merger scenario is certainly more complex, this criterion may be viewed as a crude first approximation). The unmerged and merged cores are
  labeled as galaxies. There is no treatment of disruption.
\item \textbf{MiRdRm}: Cores with infall mass greater than
  \minfall\ and radius less than \rdisrupt\ are kept. The cores that
  pass both thresholds can merge into a single object if they are
  within \rmerge\ of each other. The final model allows for both mergers and disruption, combining all three previous assumptions.
\end{itemize}

\subsubsection{Modeling Galaxy Profiles from Cores}
In the following subsections, we investigate the different options for associating galaxies with cores based on using core profiles for cluster-scale halos; surface density radial profiles of cores are constructed for each
halo in the simulation with mass above
$10^{14}~h^{-1}$M$_{\odot}$. 
To include projection effects, radial profiles are built using all cores
within a $10~h^{-1}$Mpc sphere around the halo cluster center and then additional merging or disruption criteria are applied, depending on the model under investigation. We have found that varying the inclusion radius does not change the best-fit parameters when fitting to observational data, as long as the radius is chosen to be greater than $4~h^{-1}$Mpc (see Section~\ref{sec:cluster volume} for more details). Cores are modeled as galaxy proxies using the different model choices described above.

The radial surface density profile is calculated as follows. For each modeled
galaxy and radial bin, a weight proportional to the probability of the
galaxy being projected into that radial bin is determined. The
probability is proportional to the surface area of a sphere that would
be projected onto the two-dimensional radial bins:
\begin{equation}
  \rm{Rad}_i = \sum_{j=1}^{n}
  \begin{cases}
    \cos\theta_{ij} - \cos \theta_{(i+1)j} & r_i < R_j,\\
    \cos\theta_i & r_i \geq R_j, \\
  \end{cases}
\end{equation}
where $\rm{Rad}_i$ is the $i^{{\rm th}}$ cluster radial density bin with a lower and
upper bin edge of $r_i$ and $r_{i+1}$, respectively, $R_j$ is the real (3D) 
distance from the cluster center to the $j^{{\rm th}}$ core of $n$ total, and $\theta_{ij} =
\arcsin r_i/R_j$.

\subsubsection{Infall Mass Threshold}
\label{sec:core_m_infall}
In this section, we investigate the Mi-model and with it the effect of varying \minfall\ on the cluster profile. Following our analog of the SHAM assumption, the infall mass of the core serves as a proxy for the luminosity of the corresponding galaxy. Since we are modeling profiles of galaxies above a luminosity threshold, cores need to be above a mass threshold, labeled as \minfall, to be considered as a galaxy above the luminosity threshold. 

There is considerable freedom in choosing the particular quantity that
represents the best infall mass proxy for the purpose of galaxy
modeling. Aside from the direct halo mass at infall, one may use $V_{\rm{max}}$, the
maximum circular velocity (e.g., \citealt{Zehavi19}), or the maximum mass attained by the
infalling halo at any time in its past history (the `peak' mass) or, correspondingly, the peak value of $V_{\rm{max}}$. For our current purposes, we
use the FoF halo mass as defined in our merger trees as the simplest infall
mass definition that provides a good fit to the data. We found that using the peak mass along the merger tree gave similar results; the two options are covered in more detail in
Section~\ref{sec:peak_infall_mass}. In future work, especially when going beyond matching to galaxy profiles in cluster-scale halos, and when extending our work to two-point galaxy statistics, we will investigate the use of $V_{\rm{max}}$ analogous to the approach of \cite{Zehavi19} for HOD modeling.

\begin{figure}[t]
  \centering
  \includegraphics[width=0.45\textwidth]{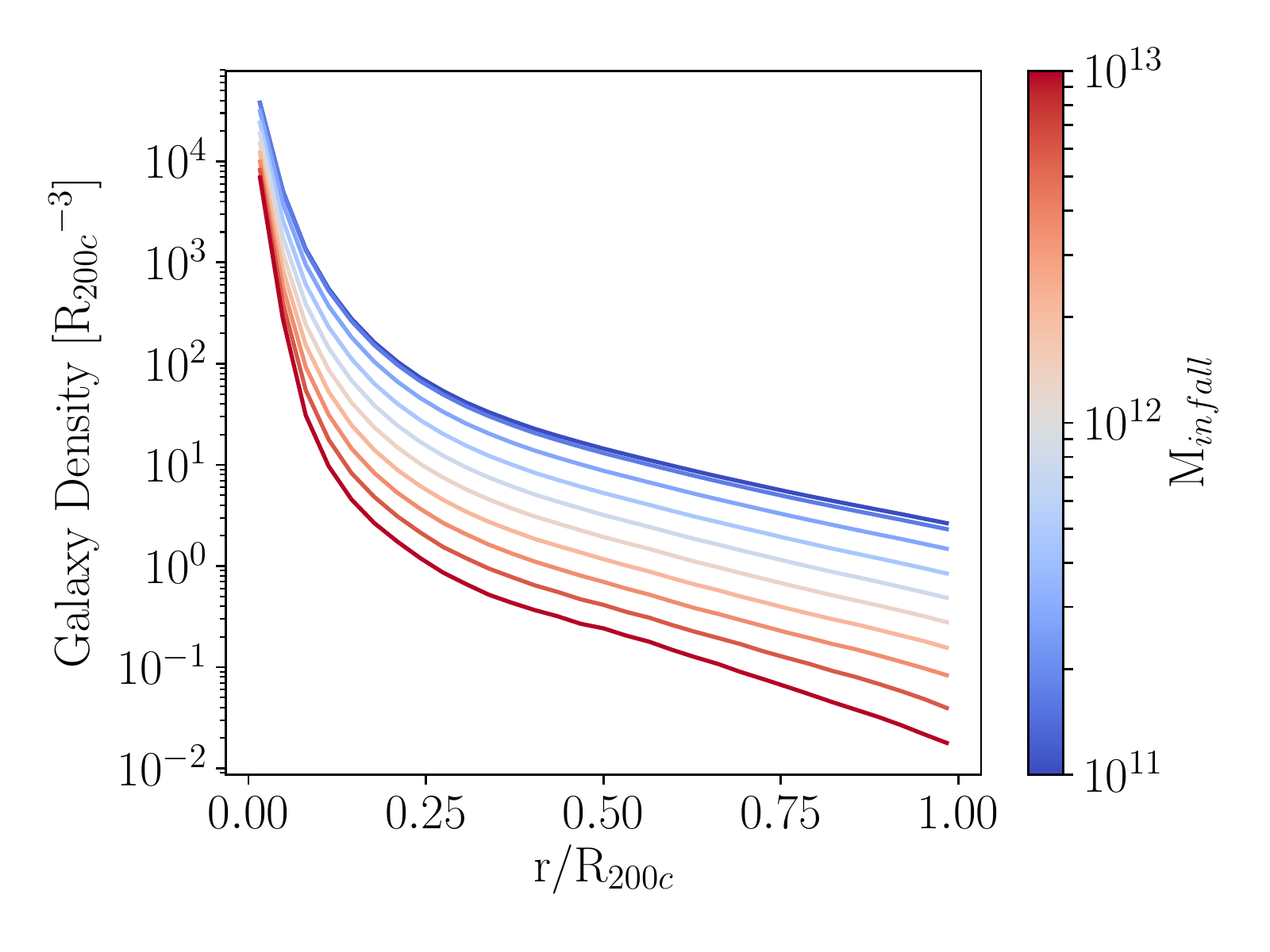}
  \caption{Surface density profiles of cores for the Mi-model (varying \minfall) for halos $ 14.0 < \log_{10}(M_{200c}) < 14.25$ in the Outer Rim simulation at $z=0.24$. Varying
    \minfall\ significantly modifies the overall normalization of the
    profile and also creates a more peaked profile shape for higher values of \minfall.       }
  \label{fig:vary_m_infall}
\end{figure}

The qualitative effect of varying the \minfall ~threshold on cluster core profiles is
shown in Figure~\ref{fig:vary_m_infall}. The total number of modeled
galaxies in clusters depends strongly on the threshold and the galaxy profile
shape is affected as well. For higher \minfall\ thresholds, the profile
is more peaked at the center. The cores from more massive halos tend
to reside closer to the potential center of the halo and thus
increasing the threshold does not remove as many galaxies from the
center as it does from the outskirts.
\\

\subsubsection{Core Disruption}
\label{sec:core_disruption}
Next, we study the surface density profile for the MiRd-model. In this case, the number of cores is controlled by the infall mass threshold \minfall\ and possible core disruption, modeled via \rdisrupt. Strong gravitational tidal fields within a cluster environment can cause a galaxy to be stripped of its stellar component (e.g.~\citealt{Mayer01, Kravtsov04, Bahe2019}). (Ram-pressure stripping of halo gas is a relevant physical effect but is missing in gravity-only simulations.) Because stars are subject primarily to gravitational forces, the simulation particles that represent the core experience similar gravitational forces as would the stellar component of galaxies. An important caveat is that galaxies can be more compact than the cores that we use to represent them, so mapping between galaxy dynamics and core dynamics is not expected to be exact. The MiRd and MiRdRm core models of galaxies both use the disruption of cores as a mechanism for galaxy removal. 

We apply a simple model of galaxy disruption using the cores as proxies. If the core particles remain tightly clustered, we assume that the galaxy the core represents has not lost enough stellar mass to fall below the luminosity threshold or be completely disrupted. If the particles are widely spread apart, the modeled galaxy is assumed to be disrupted or fallen below the luminosity threshold. The spread of the particles is measured using the core effective radius as defined in Eq.~\ref{eq:core_radius}. The threshold value distinguishing compact galaxies and disrupted galaxies is a model parameter that we label as \rdisrupt. The parameter is in comoving distance and has no assumed redshift or infall mass dependence.

Investigating the distribution of core radii at fixed redshift, there is no clear distinguishing feature in the satellite core distribution that would naturally set a value for \rdisrupt\ (see Figure~\ref{fig:core_radii_pop}, top panel) This is to be expected since the final set of cores has a complex history with individual core members having gone through a large number of interactions as they traversed their branches in the associated merger tree. Cores that have recently fallen in have smaller radii than those that have spent more time as satellites in their host halo (see Figure~\ref{fig:core_radii_pop}, bottom panel).

\begin{figure}[t]
  \centering
  \includegraphics[width=0.45\textwidth]{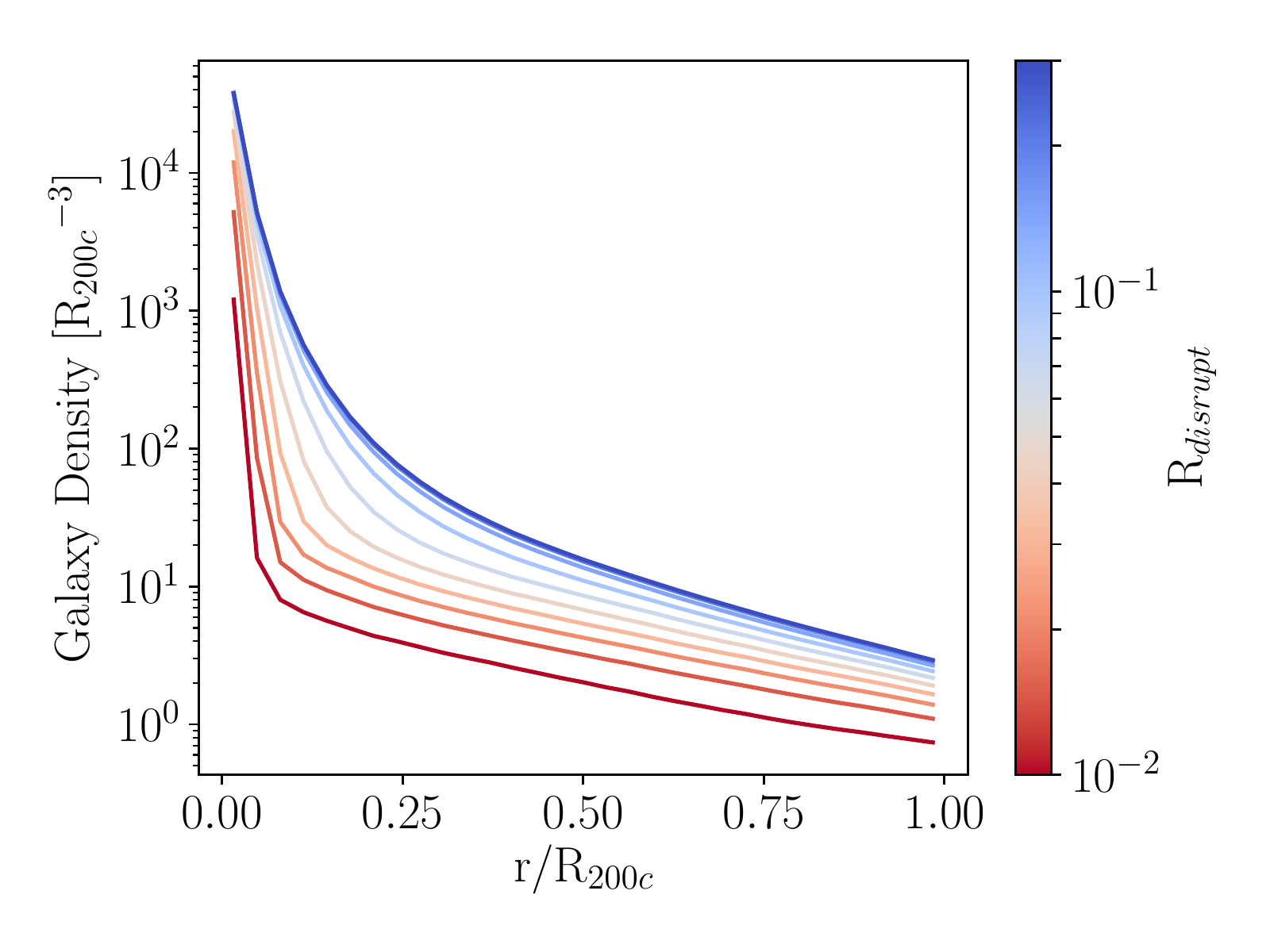}
  \caption{Surface density profiles of cores for the MiRd-model
  (varying \rdisrupt) for halos $ 14.0 < \log_{10}(M_{200c}) <  14.25$. Changing \rdisrupt\ modifies the shape of the profile. }
  \label{fig:vary_disruption}
\end{figure}

The effect of varying \rdisrupt\ (with no \minfall\ cut) on the surface galaxy profiles in halos is shown in Figure~\ref{fig:vary_disruption}. The \rdisrupt\ parameter affects the total number of galaxies, but it also affects the shape of the profile more strongly than varying \minfall. The change in the profile is greatest near the halo center where the cores have spent more time interacting with the strong tidal field. Cores in
the halo outskirts are more likely to have recently merged and have not yet experienced strong disruptive forces. In addition, a larger fraction of the surface density at the outskirts arises from cores in interloping halos that happen to project into the target halo's profile. 

\subsubsection{Core Mergers}
\label{sec:core_merger}

Finally, we investigate the MiRm-model to study the effect of core mergers on the surface density profile of cores. We use core mergers as an approximation for galaxy mergers. If two candidate cores that pass the \minfall\ threshold of $10^{12}$\hmdot, are closer than a threshold set for the 3D comoving distance, they are merged into one modeled galaxy. The threshold, labeled \rmerge, is a free parameter in the model. Multiple cores can merge to form one object: an FoF algorithm  on core centers with a linking length of \rmerge\ is used to identify two or more merged cores. The position of the new object is taken as the mean position of all the core centers identified as having being merged by the FoF algorithm. We do not allow cores with infall mass below \minfall\ to merge into more massive objects. Additionally, \rmerge\ has no redshift or core mass dependence. In the spirit of simplicity, we do not include any temporal or relative velocity information for determining the merger threshold. In principle, core merger trees carry detailed information about core mergers and are a significant step beyond the simple model considered here (for more discussion, see Section~\ref{sec:conclusion}).

\begin{figure}[t]
  \centering
  \includegraphics[width=0.5\textwidth]{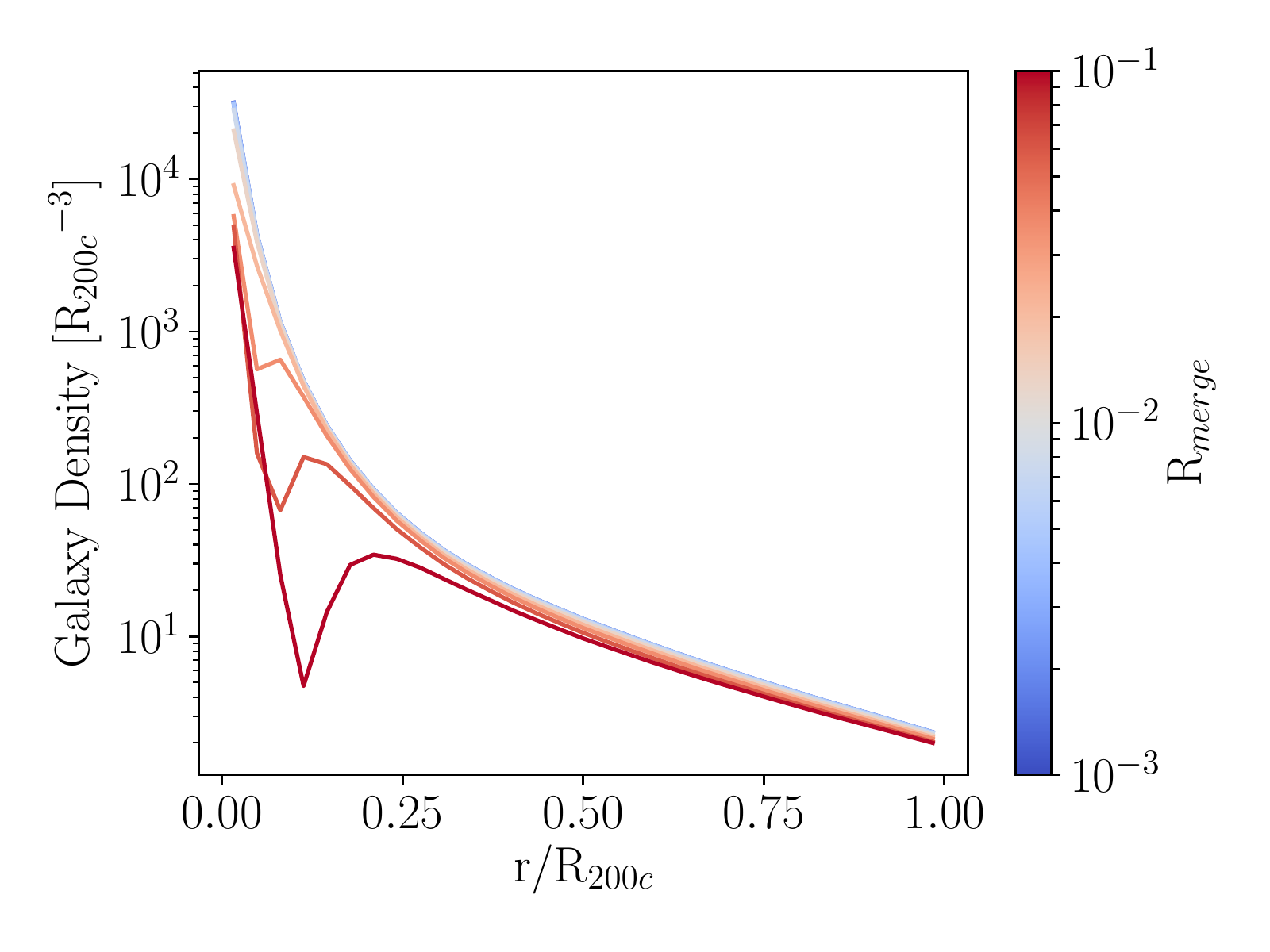}
  \caption{
    Surface density profiles of cores for the MiRm-model (varying \rmerge) for halos with masses $ 14.0 < \log_{10}(M_{200c}) < 14.25$. Because of the percolation nature of the FoF merging criterion, the cores have been truncated to have an abundance similar to $>$\lstar~galaxies: \minfall\ threshold of $10^{12}$\hmdot~and \rdisrupt\ threshold of 20~\hkpc. At large \rmerge, unphysical spikes form at small radii, clearly signaling overmerging. 
    }
  \label{fig:vary_merger}
\end{figure}

The effect of varying \rmerge\ on the galaxy profile in clusters is shown in Figure~\ref{fig:vary_merger}. At very high and unphysical \rmerge, the FoF linking of cores can percolate to almost the full core population and remove a significant fraction of the cores. At smaller \rmerge, most of the merging occurs at the center of the halo where the density of cores is the highest.

After having investigated the behavior of the surface density profiles of cores for three different mechanisms that allow us to change the number of cores via an infall mass threshold, core merging and core disruption, we now compare the four models of Section~\ref{sec:core-galaxies-models} to observational data, finding the optimal fitting parameters for each model to match the observations as closely as possible. For this study, we employ cluster data taken from the Sloan Digital Sky Survey. 

\section{Observations: SDSS Data Reduction}
\label{sec:sdss_data}
Empirical methods, by construction, rely on observational data to determine modeling parameters. In this paper, we use the number of cluster galaxies above a luminosity threshold, as well as the spatial distribution of these galaxies, as a function of cluster halo mass, as the input information to tune our core-galaxy models. Using this information, we investigate how well a modeling approach based on cores might work, which model assumptions are most important to successfully reconstruct the spatial distribution, and how robustly the model parameters can be determined. 

The source of the optical galaxy information here is derived from the SDSS, which restricts us to relatively low redshifts. We use optically identified redMaPPer clusters~\citep{Rykoff2014} as our observational cluster catalog. For each cluster within $0.1<z<0.35$, we obtain all galaxies within $R_{200}$ from the SDSS DR15 main galaxy sample~\citep{SDSS_DR15} to construct radial galaxy
surface density profiles. 

We use the redMaPPer catalog v6.3, which finds clusters of red galaxies and calculates a color- and position-weighted galaxy richness. The probability of each cluster galaxy member to be the central galaxy is weighted by the luminosity of the galaxy, color, photo-$z$ and local environment. The cluster richness can be converted to an estimate of the cluster mass. From the cluster richness and redshift, we calculate the cluster M$_{200m}$ mass using the weak lensing calibration determined in~\cite{2019MNRAS.482.1352M}:

\begin{eqnarray}
  M_{200m} &=& 3.081\times 10^{14}\frac{1.08\lambda}{40}^{1.356}\nonumber\\
  &&\times\left[(1+z)/1.35\right]^{-0.30}\rm{M}_\odot,
\end{eqnarray}
where $\lambda$ is the cluster richness as determined by redMaPPer. The factor of $1.08$ accounts for the average difference between
SDSS richness used in this paper and the DES Y1 richness used in
\cite{2019MNRAS.482.1352M}. As a last step, the estimated \mass{200m} is converted to \mass{200c} assuming an NFW profile and obeys the concentration-mass relationship specified in \cite{Child2018}. Several richness-mass calibrations are
explored in Section~\ref{sec:remapper_mass_richess} and most give similar results. The number of clusters in our \redmapper sample is given in Table~\ref{tab:halo_counts} in Section~\ref{sec:methods}.

\subsection{Obtaining SDSS Galaxies}
In order to construct individual galaxy surface density profiles, we
first must determine the angular size of \radius{200c} on the sky for
each cluster. For SOD halos the physical
mass and physical radius are directly linked via:
\begin{equation}
  R_{200c} = \left(\frac{3M_{200c}}{4 \pi~200\rho_{c}}\right)^{1/3} ,
\end{equation}
where $\rho_{c}$ is the critical density at the cluster's
redshift.  Given the physical radius of the cluster, the angular
radius on the sky (knowing the cluster's redshift) is calculated using
the Astropy cosmology package~\citep{Astropy2013}. 

\begin{figure}[t]
  \includegraphics[width=0.47\textwidth]{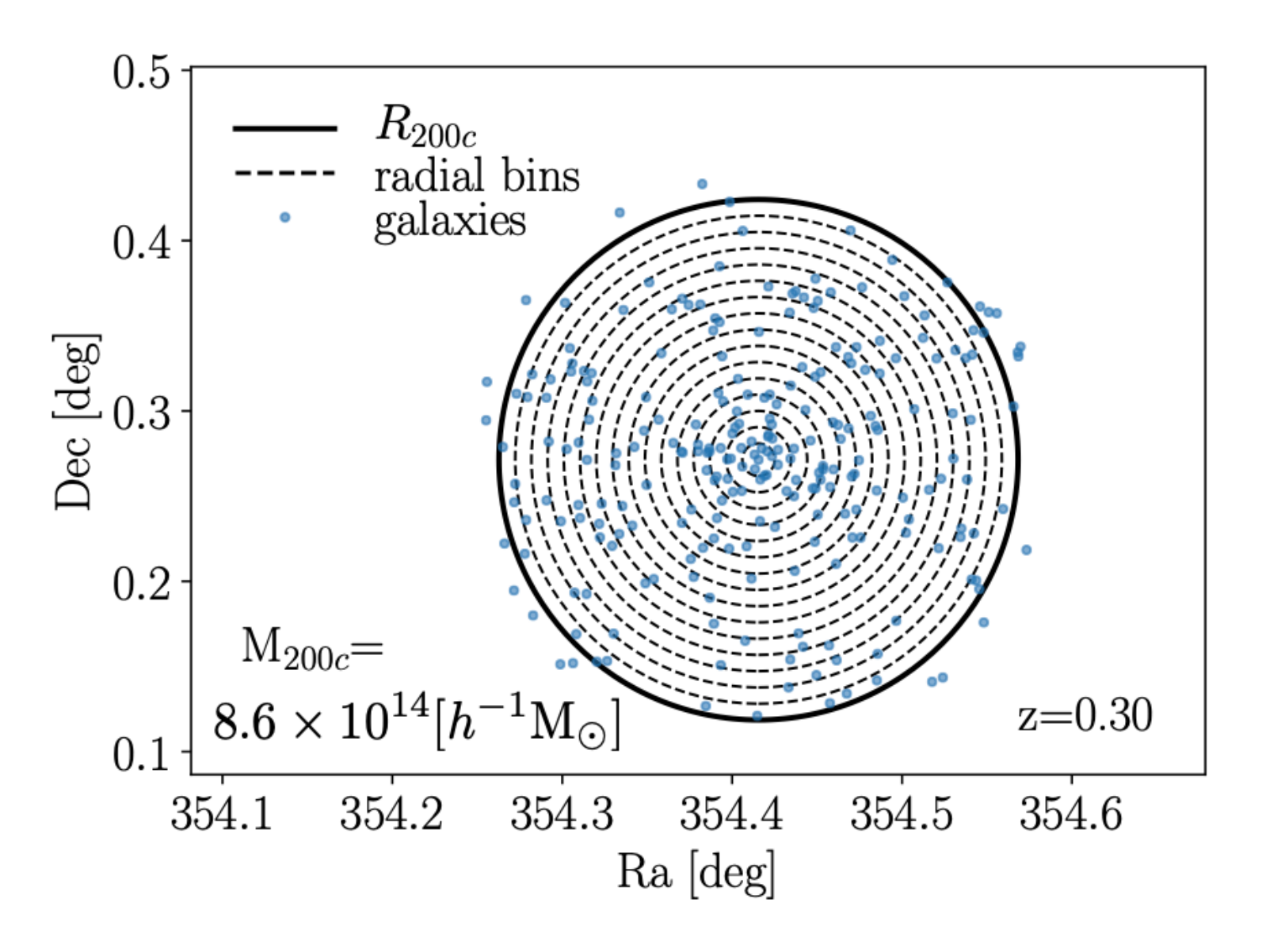}
  \caption{SDSS galaxies queried around a \redmapper galaxy
    cluster. The galaxies below the $>0.4$\lstar~threshold and with bad
    flags (see text) have been removed. $10\times10^{14}$}
  \label{fig:sdss_example}
\end{figure}

\begin{table}[]
    \centering
    \caption{Galaxy brightness thresholds modeled}
    \begin{tabular}{cc}
Luminosity & Magnitude \\
    \hline\hline
    $>2.50$\lstar &  $<$\mstar$- 1.0$ \\
    $\gtrapprox 1.58$\lstar & $<$\mstar$- 0.5$ \\
    $>1.00$\lstar & $<$\mstar$- 0.0$ \\
    $\gtrapprox 0.63$\lstar & $<$\mstar$+ 0.5$ \\
    $>0.40$\lstar & $<$\mstar$+1.0$ \\
    \end{tabular}
    \label{tab:my_label}
\end{table}

For each cluster, all galaxies in the SDSS DR15 main galaxy sample that are within the angular size set by R$_{200c}$ from the cluster center are gathered (see Figure~\ref{fig:sdss_example}); galaxies with flags BRIGHT, SATURATED, SATUR-CENTER, NOPETRO, DEBLENDED\_AS\_MOVING set in the $g$, $r$ or $i$-bands are removed. Any cluster with a BRIGHT\_STAR or BLEED mask that overlaps with the cluster's R$_{200}$ is excluded from the analysis. A total of 18.5\% of clusters are removed due to an interloping mask. While there is correlation between cluster sky area and clusters being masked, there is no significant bias in cluster mass. Those masked are often a result of bright stars in the Milky Way and these are uncorrelated with external galaxies.

We construct profiles for several rest frame SDSS $i$-band luminosity
thresholds. The five galaxy luminosity thresholds used in this paper are given in Table~\ref{tab:my_label}. 

\lstar~and
\mstar\ are the bend in the galaxy luminosity and magnitude function,
assumed to be $2.25 \times 10^{10}$L$_\odot$ and $-22.18$,
respectively. Galaxies brighter than the threshold are selected by a
k-corrected $i$-band magnitude cut. We use the same polynomial
approximation for $M_\star(z)$ as in \cite{Rykoff2014}:
\begin{equation}
  M_\star^{i}(z) = 12.27 + 62.36z - 289.79z^2 + 729.69z^3 - 709.42z^4.
    \label{eq:mstar}
\end{equation}
In \cite{Rykoff2014}, the rest frame magnitude of \mstar~is $-21.22$.

\begin{figure}[b]
  \includegraphics[width=0.45\textwidth]{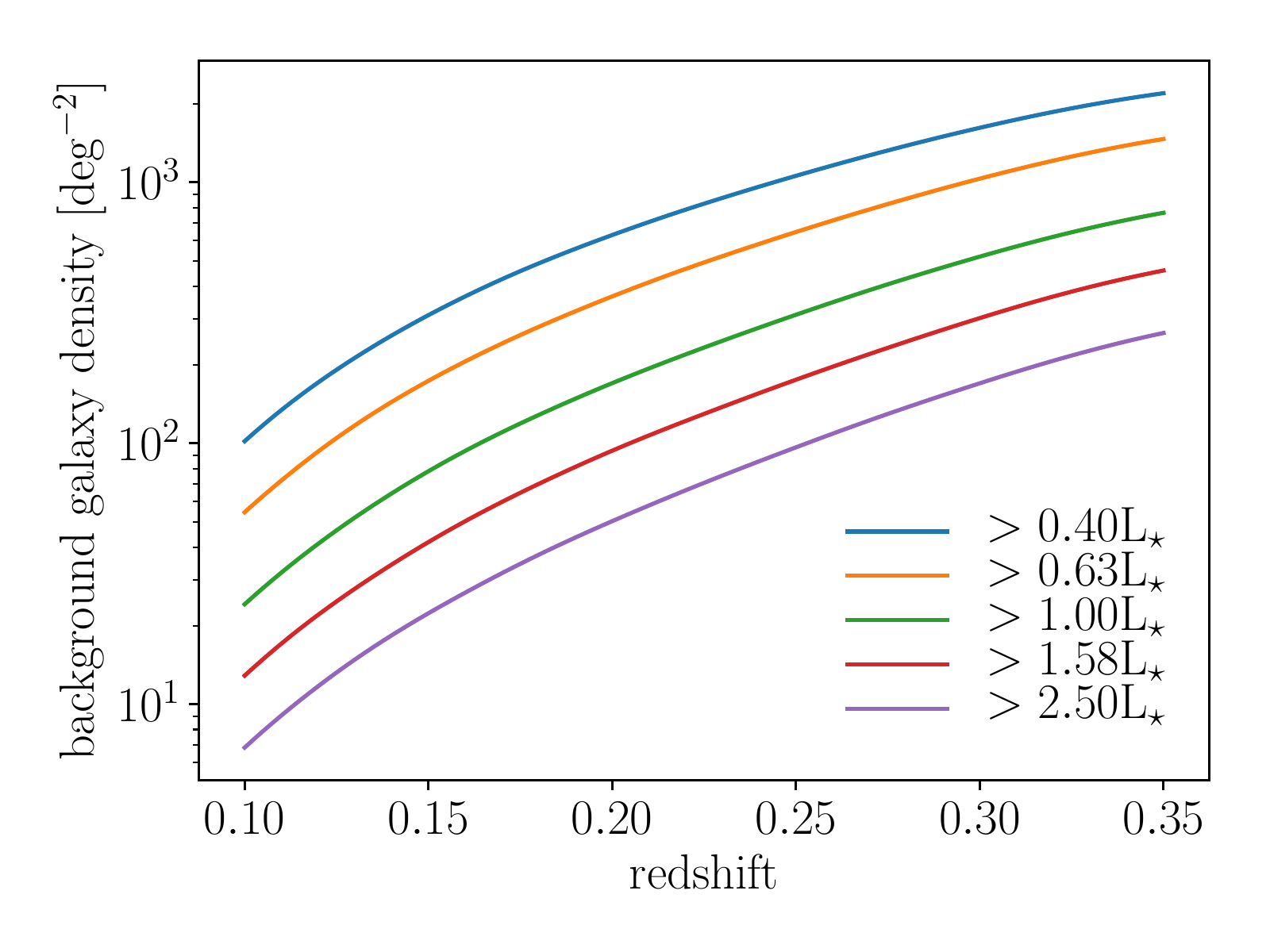}
  \caption{Surface density of background galaxies for different
    intrinsic luminosity thresholds as a function of redshift. Clusters at
    higher redshift have more interloper galaxies as the apparent
    brightness threshold is lower at fixed intrinsic luminosity.}
  \label{fig:background_estimate}
\end{figure} 

\subsection{Background Estimation and Subtraction}
\label{sec:background}

\begin{figure*}[t]
  \centering \includegraphics[width=0.5\textwidth]{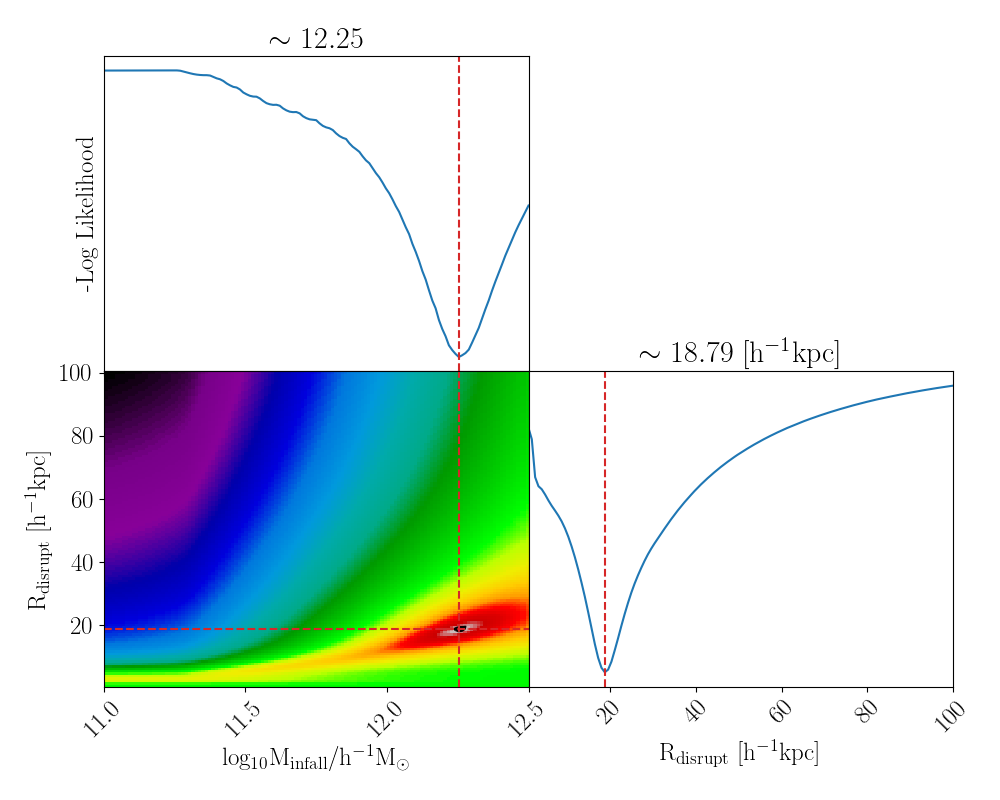}\includegraphics[width=0.5\textwidth]{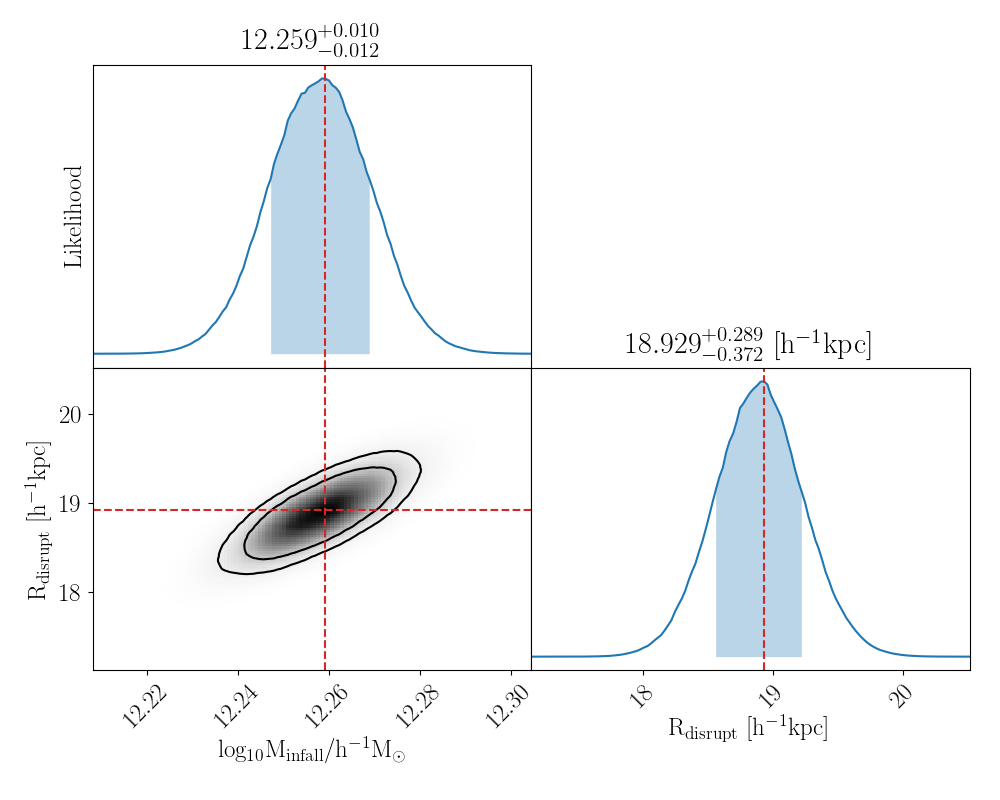}

  \caption{Parameter optimization results from the grid likelihood evaluation for one (MiRd model and $>$1.00~\lstar) of the 20 different scenarios investigated. For the luminosity threshold we choose the values given in Table~\ref{tab:my_label} and we vary the infall mass threshold \minfall\ and the disruption radius \rdisrupt, summarized as the MiRd $>$\lstar~galaxy model. The left triangle plot shows the
    negative log likelihood calculated on a wide parameter grid
    scan. The right triangle plot shows the likelihood calculated on a
    zoomed-in region identified in the left panel. Results are similar for the other 19 scenarios.}
  \label{fig:example_likelihood}
\end{figure*}

Not all galaxies projected within the galaxy cluster R$_{200}$
belong to the galaxy cluster. Galaxies uncorrelated to the halo can
appear either in front of or behind the cluster at different
redshifts. To remove the interloper galaxy contribution to the cluster
galaxy profile, we assume that the uncorrelated galaxies have a
constant and flat spatial distribution that can be estimated and
subtracted out. Any excess galaxy counts above the background level
should correspond to the contribution from the cluster. We calculate the expected background galaxy density per square degree (Figure~\ref{fig:background_estimate}), as a function of the magnitude
threshold and subtract it from individual cluster profiles, similar to
the process described in \cite{Gladders2005} and  \cite{Hansen2005}.

While the background galaxies have a constant observed magnitude
distribution, the magnitude cut applied to clusters depends both on
the cluster redshift and the rest-frame luminosity threshold that we are using
to construct the galaxy profiles. To this end, we require a background
galaxy density estimate for each cluster to take both
the luminosity threshold and the redshift into account. To calculate the background galaxy surface density, we use redMaPPer's random catalog, which points to
fairly sampled random locations in the SDSS footprint. For each
luminosity threshold, we calculated the average counts of galaxies per
steradian using Eq.~\ref{eq:mstar} at 12 linearly spaced redshifts
spanning redshifts from $z=0$ to $z=0.4$. The background density at a specific
redshift is interpolated from these 12 redshifts using a spline fit.
 
Once the raw galaxy counts are obtained for each cluster in our
sample, the count of galaxies in each radial bin is reduced by the
product of background surface density at the cluster redshift and the
sky area of the radial bin.
 
\subsection{Galaxy Radial Profile Stacks}

The observed clusters are binned into five logarithmic mass bins
from $M_{200c}=10^{14.00}~h^{-1}\text{M}_\odot$ to
$M_{200c}=10^{15.25}~h^{-1}\text{M}_\odot$. The number of halos in
each simulation and catalog are given in Table~\ref{tab:halo_counts}.
We took a conservative approach by excluding low richness clusters ($\lambda<40$)
from the analysis. These clusters are not well suited for our approach since they exhibit a large scatter between mass and richness, are plagued by projection effects, and only provide low galaxy counting statistics \citep{Myles2021, Grandis2021,Wu2022}. We confirmed that including them does not qualitatively change the modeling results. For each cluster, we bin the galaxies above the luminosity threshold into 16 linearly spaced radial bins (8 bins for $>$2.5\lstar~profiles) from $r=0$ to $r=R_{200c}$, calculate the surface density for each bin and subtract the expected background surface density. The radial galaxy density profile of the clusters are stacked within each mass bin. The error in each radial bin is taken as the Poisson fluctuation of galaxy counts in the bin.

\section{Core-based Modeling of SDSS Cluster Galaxy Profiles}
\label{sec:results}

In this section we compare our different core-based galaxy models to observation of SDSS clusters. This will enable us to evaluate the roles of core disruption and merging, both separately and in combination, as part of our core-galaxy modeling approach. 

\subsection{Parameter Tuning of Core Models}
We derive best-fit parameters for the four models described in Section~\ref{sec:core-galaxies-models} for the five different galaxy luminosity thresholds given in Table~\ref{tab:my_label}, leading to 20 sets of tuned parameters. We exclude the lowest cluster mass bin, 10$^{14}~h^{-1}$M$_\odot$ to 10$^{14.25}~h^{-1}$M$_\odot$, due to the high mass-richness scatter in the observational data. Our optimization procedure is carried out across all four cluster mass bins at once. However, each of the models with the corresponding luminosity thresholds is tuned independently. The different luminosity thresholds were investigated to ensure that our approach is flexible and works across a range of different assumptions with regard to the modeled galaxy population. The luminosity thresholds also allow us to investigate that our modeling approach captures the physics of the galaxy-core connection in a sensible way. For example, if we allow for a lower luminosity threshold and the clusters therefore host more galaxies, we expect \minfall\ to be smaller and to have smaller merger and disruption rates. 

\begin{table*}[t]
\begin{center}
\caption{Best-fit core model parameters and the reduced chi-squared statistic, $\tilde\chi^2$, for the four models and five luminosity thresholds each fit across all four mass bins simultaneously}
\begin{tabular}{cccccccc}
 & & \multicolumn{5}{c}{Galaxy Luminosity Threshold} &  \\
\cline{3-7} 
Model & Parameter & $>$0.40~\lstar & $>$0.63~\lstar & $>$1.00~\lstar & $>$1.58~\lstar & $>$2.50~\lstar & Units\\
\hline
\hline
\multirow{2}{*}{Mi} &$M_{\rm{infall}}$ & $12.709^{+0.007}_{-0.008}$ & $12.892^{+0.014}_{-0.009}$ & $13.090^{+0.011}_{-0.013}$ & $13.412^{+0.014}_{-0.017}$ & $13.787^{+0.034}_{-0.029}$ & log$_{10}$(\hmdot{}) \\
 & $\tilde\chi^2$ & $77$ & $403$ & $18$ & $6.7$ & $1.6$ \\
\hline
\multirow{3}{*}{MiRd} &$M_{\rm{infall}}$ & $11.631^{+0.005}_{-0.006}$ & $11.906^{+0.008}_{-0.007}$ & $12.258^{+0.010}_{-0.011}$ & $12.777^{+0.026}_{-0.011}$ & $13.444^{+0.031}_{-0.026}$ & $~h^{-1}$M$_\odot$ \\
 & $R_{\rm{disrupt}}$ & $16.7^{+0.2}_{-0.2}$ & $17.7^{+0.3}_{-0.2}$ & $18.9^{+0.3}_{-0.4}$ & $21.6^{+0.6}_{-0.6}$ & $25.8^{+0.9}_{-1.3}$ & $~h^{-1}$kpc\\
 & $\tilde\chi^2$ & $1.2$ & $0.53$ & $0.62$ & $1.2$ & $0.33$ \\
\hline
\multirow{3}{*}{MiRm} &$M_{\rm{infall}}$ & $11.846^{+0.005}_{-0.006}$ & $12.112^{+0.006}_{-0.010}$ & $12.449^{+0.011}_{-0.011}$ & $12.993^{+0.016}_{-0.016}$ & $13.497^{+0.027}_{-0.023}$  & log$_{10}$(\hmdot{}) \\
 & $R_{\rm{merge}}$ & $0.160^{+0.005}_{-0.004}$ & $0.200^{+0.004}_{-0.010}$ & $0.245^{+0.011}_{-0.008}$ & $0.225^{+0.010}_{-0.007}$ & $0.423^{+0.023}_{-0.028}$ & $~h^{-1}$Mpc\\
 & $\tilde\chi^2$ & $1.4$ & $1.3$ & $1.3$ & $1.6$ & $1.1$ \\
\hline
\multirow{4}{*}{MiRdRm} &$M_{\rm{infall}}$ & $11.786^{+0.009}_{-0.007}$ & $11.952^{+0.011}_{-0.012}$ & $12.260^{+0.012}_{-0.011}$ & $12.821^{+0.019}_{-0.018}$ & $13.390^{+0.029}_{-0.029}$  & log$_{10}$(\hmdot{})\\
 & $R_{\rm{disrupt}}$ & $53.5^{+2.3}_{-2.4}$ & $23.3^{+1.4}_{-1.2}$ & $19.0^{+0.4}_{-0.4}$ & $27.5^{+1.4}_{-1.3}$ & $27.6^{+1.3}_{-1.4}$ & $~h^{-1}$kpc\\
 & $R_{\rm{merge}}$ & $0.104^{+0.011}_{-0.011}$ & $0.060^{+0.022}_{-0.012}$ & $0.000^{+0.011}_{-0.000}$ & $0.201^{+0.020}_{-0.020}$ & $0.321^{+0.036}_{-0.035}$ & $~h^{-1}$Mpc\\
 & $\tilde\chi^2$ & $0.66$ & $0.49$ & $0.62$ & $0.90$ & $0.20$ \\  
 \end{tabular}
\label{tab:best_fit_parameteres}
\end{center}
\end{table*}

The tuning was carried out using a grid scan over the parameter space. This approach worked well due to the low dimensionality of the optimization problem, which includes only between one and three parameters for the four different models. We confirmed that the results are consistent with an MCMC approach for a subset of our models but opted for the grid scan due to speed advantages. For each choice of the parameters in the grid, we calculated a log-likelihood and from that the bounding 68\% limits. The grid scans were carried out iteratively with a fixed resolution in each dimension of the considered parameters depending on the model (see \ref{tab:resolutions}). The two sets of initial parameter limits were used depending on the model under investigation and the luminosity threshold (see \ref{tab:limits}). The ``higher'' limits were used for galaxy luminosity threshold above $>$0.63~\lstar and the Mi model for any threshold. 

\begin{table}[t]
\begin{center}
    \label{tab:resolutions}
    \caption{Grid scan resolutions used in parameter fitting}
    \begin{tabular}{cccc}
        Model & $M_{\rm{infall}}$ & $R_{\rm{disrupt}}$ & $R_{\rm{merge}}$ \\
        \hline
        \hline
        Mi & 1024 & & \\
        MiRd & 128 & 128 & \\
        MiRm & 64 & & 16\\
        MiRdRm & 24 & 24 & 8\\
    \end{tabular}
\end{center}
\end{table}

\begin{table}[t]
\begin{center}
    \label{tab:limits}
    \caption{Initial grid scan parameter limits}
    \begin{tabular}{cccc}
         Parameter & Lower & Upper & Units \\
         \hline
         \hline
         $M_{\rm{infall}}$ (low) & 11.0 & 12.5 & log$_{10}$(\hmdot{}) \\
        $M_{\rm{infall}}$ (high) & 12.5 & 14 & log$_{10}$(\hmdot{}) \\
       $R_{\rm{disrupt}}$ & 1 & 300 & \hkpc{} \\
       $R_{\rm{merge}}$ & 0.001 & 0.4 & \hmpc{} \\
    \end{tabular}
    \end{center}
    \end{table}
Figure~\ref{fig:example_likelihood} shows an example of the optimization for one model, MiRd, at one luminosity threshold, $>$1.00~\lstar. The first iteration is shown in the right panel of the figure and covers a wide range in parameter space. For each following iteration, the grid size is kept the same but the parameter upper and lower limits are adjusted such that they correspond to three times the 68\% likelihood bounds in each direction. The left panel in Figure~\ref{fig:example_likelihood} shows the final result.

\subsection{Results for the Different Core-based Galaxy Models}
\label{sec:model_flavors}

We now discuss the results from fitting the four core-based galaxy models of Section~\ref{sec:fitting} to the redMaPPer cluster data for five luminosity thresholds. In order to obtain a well-matched galaxy surface density profile from the core modeling approach, the number of galaxies in the clusters above the chosen luminosity threshold has to match the number of cores that are used as galaxy markers and the radial distribution has to be correct. Our four models allow for the reduction of the numbers of cores being considered via the three mechanisms discussed above: the infall mass threshold (the higher the mass threshold, the fewer cores are considered galaxy markers), the merging of cores (reducing the number of cores after infall), and the disruption of cores (also reducing the number of cores after infall). The infall mass threshold is clearly the most impactful parameter. We note that our models do not take into account the infall time of halos and therefore a possible mass loss of cores. This additional modeling component was considered in \cite{LJ2} and will be integrated with our current approach in future work. As we will show below, the rather simple models we have developed in the current paper already lead to excellent agreement with regard to describing the redMaPPer data.

\begin{figure}[t]
  \centering
  \includegraphics[width=0.37\textwidth]{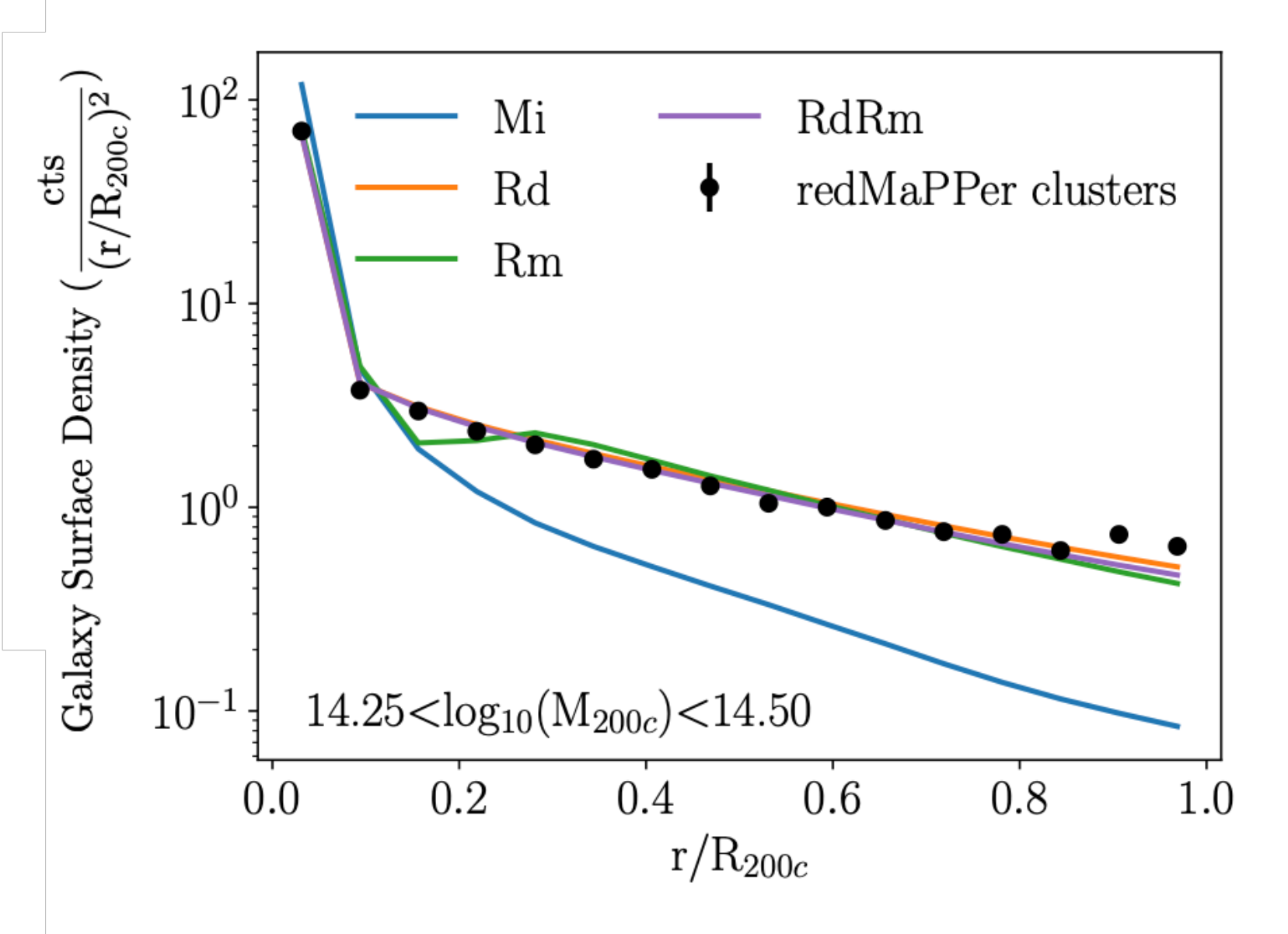}
  \includegraphics[width=0.37\textwidth]{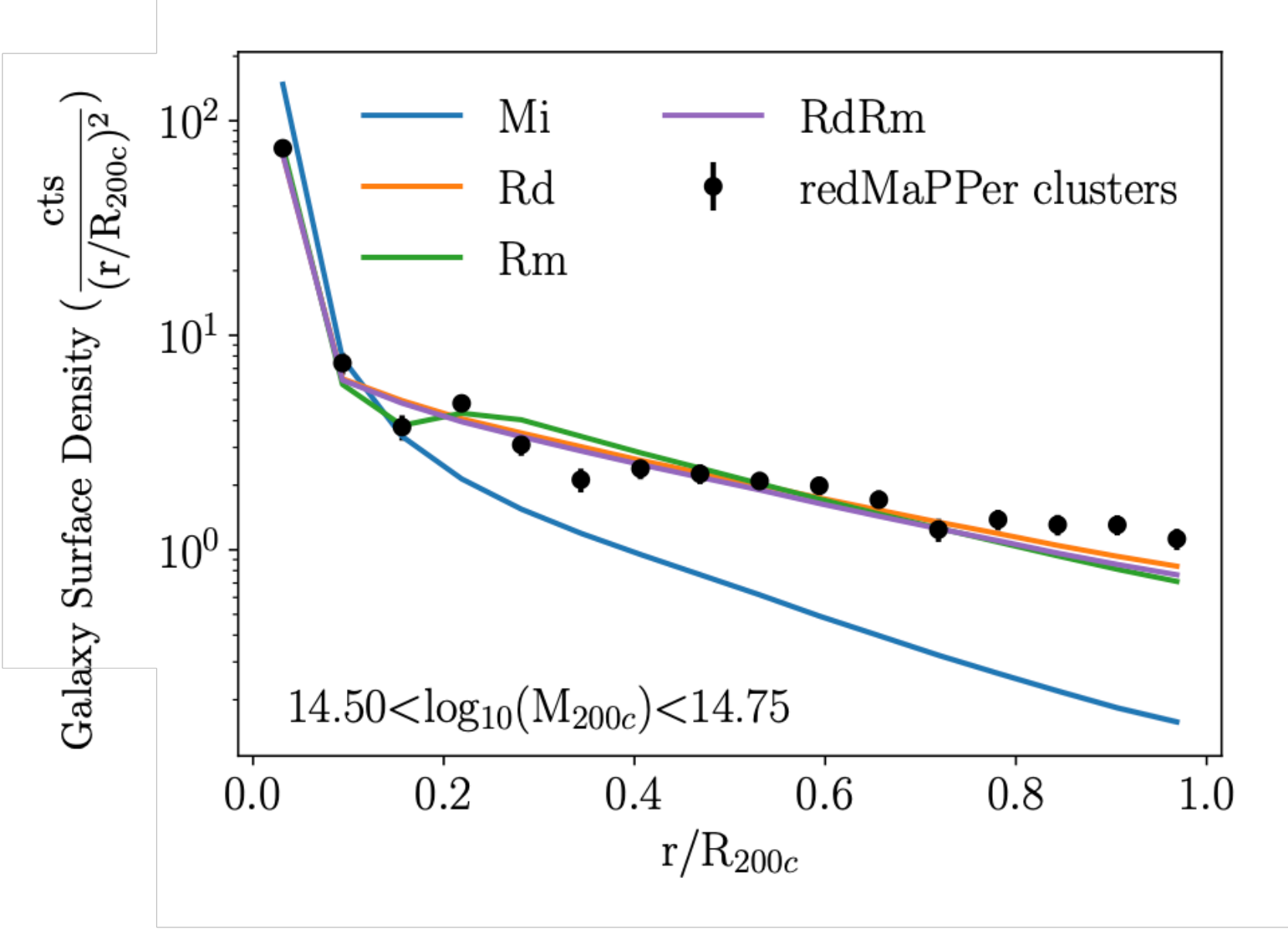}
\includegraphics[width=0.37\textwidth]{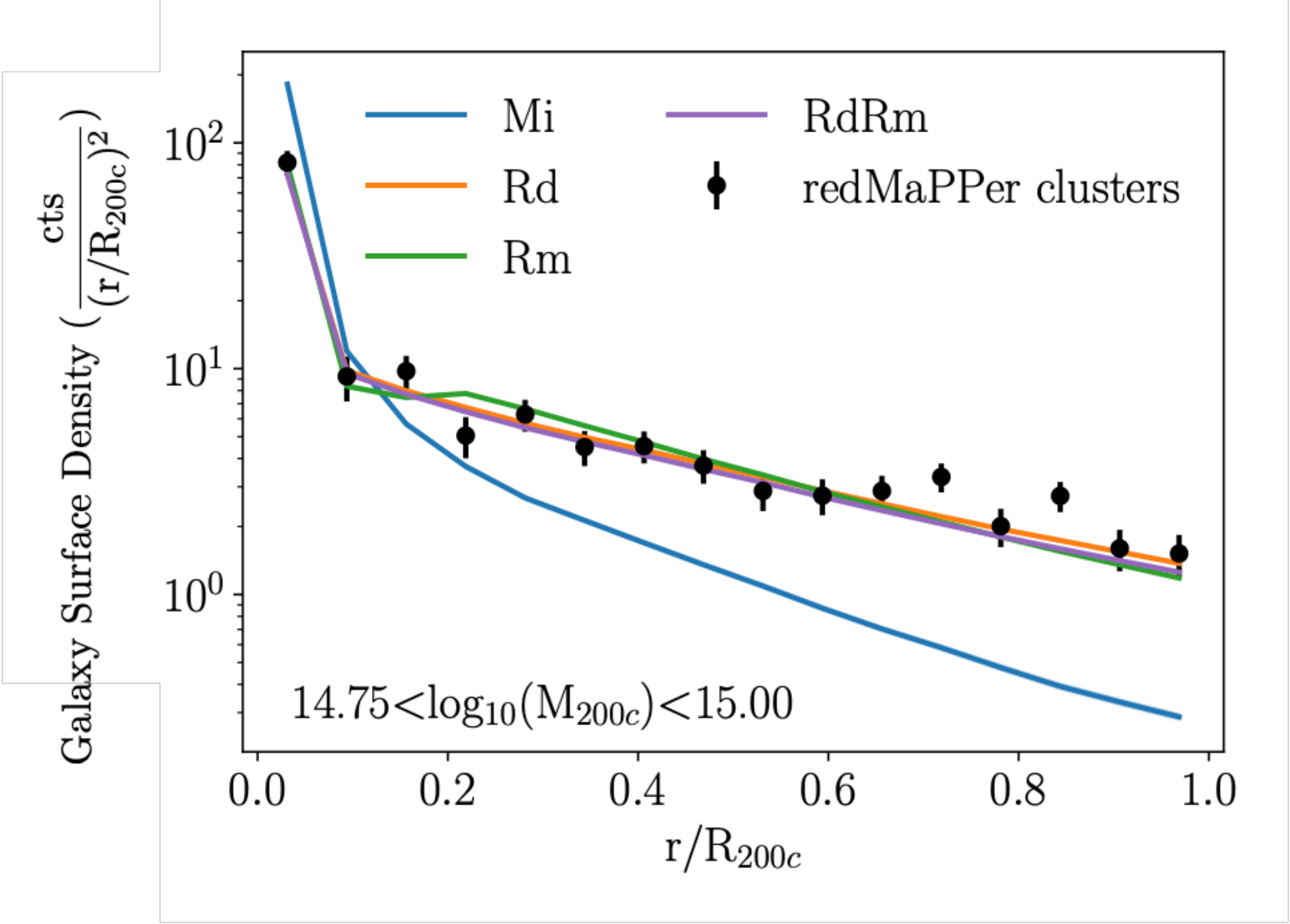}
\includegraphics[width=0.37\textwidth]{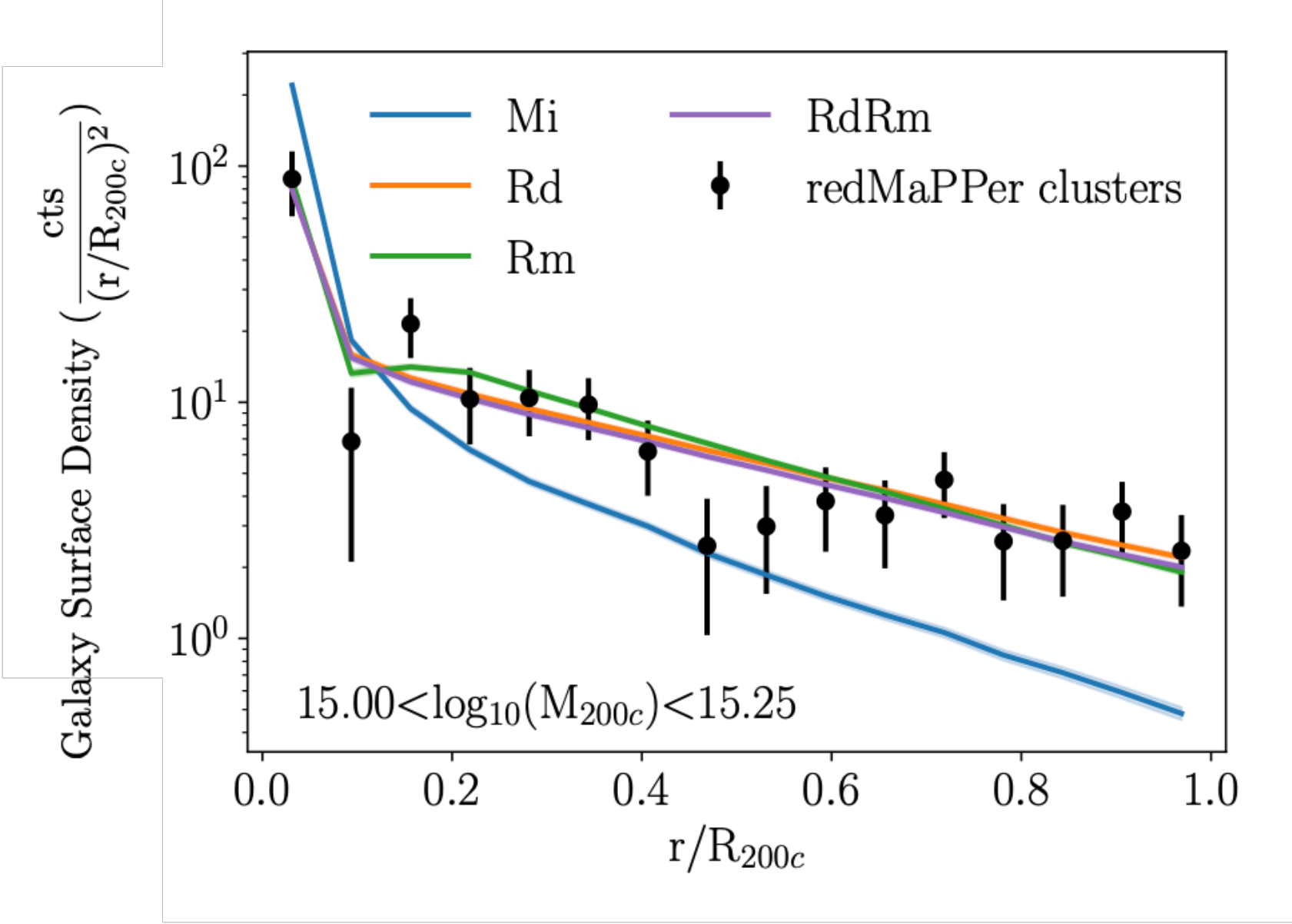}
  \caption{\label{fig:model_profiles}Radial $>$1.00~\lstar~galaxy surface density profiles from
    observations and core galaxy models for several halo mass
    bins. The markers and errors bars are the mean and error on the
    mean, respectively, for SDSS galaxy surface density profiles of
    observed clusters. The lines and shaded regions (too small to see) are the average
    and the error on the mean, respectively, of the profiles of
    modeled galaxies in the Outer Rim simulation using the best-fit parameters to
    the observed redMaPPer clusters. The density profiles for the other luminosity thresholds are shown in Appendix~\ref{sec:profiles}.}
  \label{fig:or_mstar0_profiles}  
\end{figure}

The results for the best-fit parameters for the different models and luminosity thresholds are summarized in Table~\ref{tab:best_fit_parameteres}. We also list the $\tilde\chi^{2}$ value we find for each fit. In addition, in Figure~\ref{fig:or_mstar0_profiles} we show a comparison of the redMaPPer data for the four models for three cluster mass bins for one luminosity threshold, $>$1.00~\lstar. The results for the other luminosity thresholds are very similar and the complete set of profiles is shown in Appendix~\ref{sec:profiles}. 

The simple SHAM-analog model (Mi), fails to reproduce the galaxy surface density profile reliably. In particular, the lower luminosity thresholds are difficult to accommodate, and the $\tilde\chi^{2}$ values reported in Table~\ref{tab:best_fit_parameteres} for the two lower luminosity thresholds are by far the worst from all our experiments, at 77 and 403. But even for the higher luminosity thresholds, the fits are not very good, demonstrating that a pure thresholding technique applied to the core distribution does not reproduce cluster galaxy profiles. All the
other models were able to reconstruct the galaxy profiles of the
\redmapper cluster samples. The $\tilde\chi^{2}$ for non-Mi models ranged from $\sim$0.2 to 1.3. In particular, the fourth model, MiRdRm, which combines all three parameters, performs extremely well, with $\tilde\chi^{2}$ values between 0.2 and 0.9. This good performance of MiRdRm is not surprising, given that it has the most parameters.

\begin{figure}[t]
  \centering
  \includegraphics[width=0.48\textwidth]{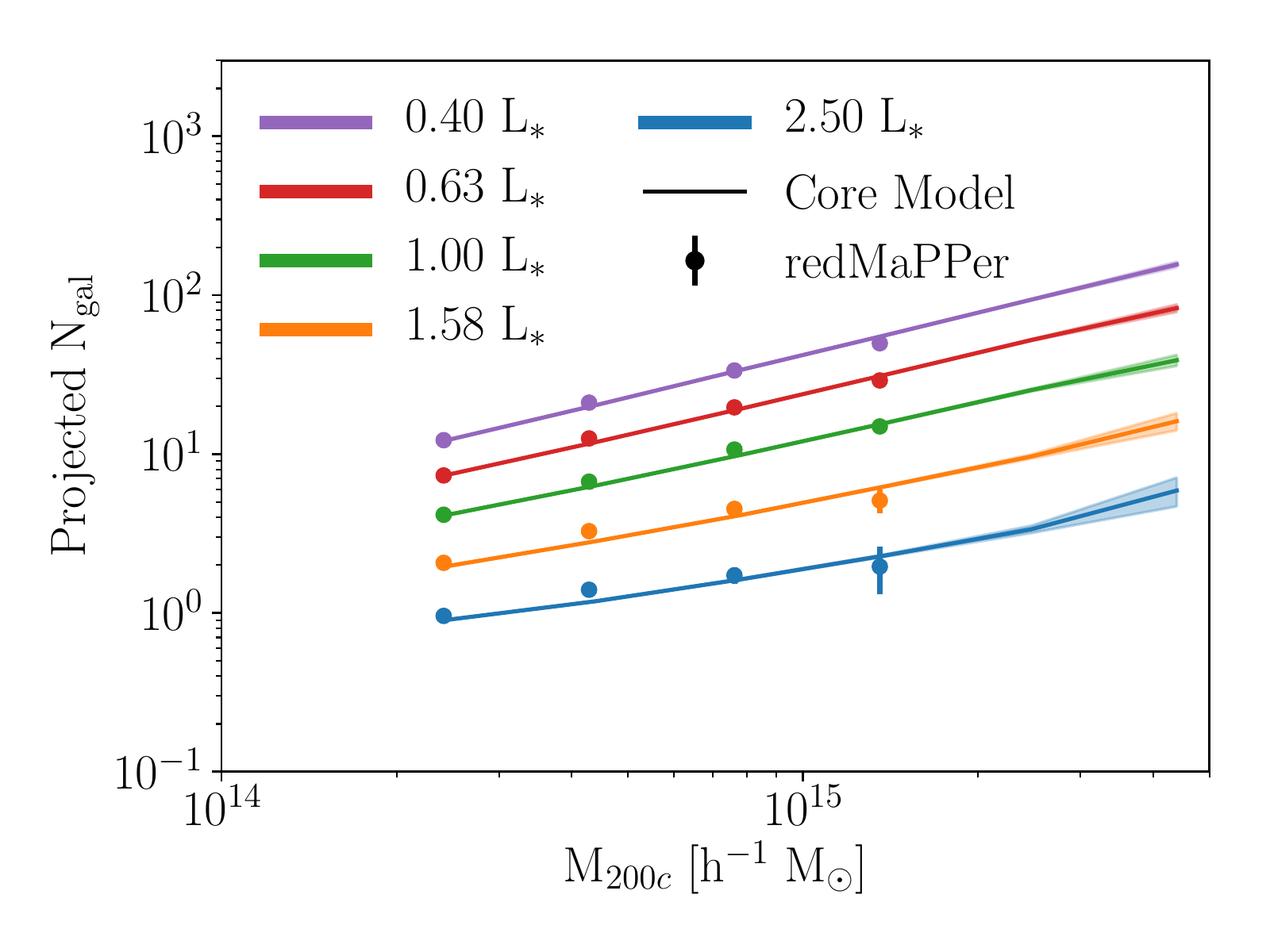}
  \caption{N$_{gal}$ as a function of cluster mass for the \redmapper
    cluster catalogs and for the MiRd galaxy model of cores. The cluster
    N$_{gal}$ is calculated by integrating the background subtracted
    galaxy counts within \radius{200c} of the cluster center. The MiRd
    model uses the best-fit model parameters to the redMaPPer cluster
    profiles.}
  \label{fig:or_ngal}
\end{figure}

Comparing the disruption and merger models, MiRd and MiRm, the disruption approach performs slightly better. The MiRm model does well in reproducing the outskirts of
halos and the central bin density but fails to reproduce the profilenear the center. At around $0.2r/R_{200}$, the MiRm model profile has an unphysical dip which does not exist in the observational data. The dip is most visible in the top panel of Figure~\ref{fig:or_mstar0_profiles}. The Mi and MiRm models perform similarly for $>$2.5\lstar~galaxy profiles and the larger mass bin, indicating that merging of $>$2.5\lstar~galaxies in massive clusters does not play a significant role in the profile shape. However, the quality of the observations in this case is not as good as for the other cases (see Appendix~\ref{sec:profiles} for more details).

The MiRd (modeled galaxy only removed through disruption) and MiRdRm (removal via both disruption and merging) models perform similarly well -- the MiRd model is a
subset of the MiRdRm model where the merging radius is reduced to
zero. The change in $\tilde\chi^2$ in including the merging mechanism
is small. This may point to the fact that satellite core (or subhalo) mergers in the infall mass range we are investigating are relatively rare events. For
the $>$\lstar~and the $>$0.4\lstar~profiles, the best-fit MiRdRm model has a
small merging radius, not different from zero at a statistically significant level. 
A merging length that is statistically consistent with zero 
would indicate no merging and the MiRdRm model would effectively be the same as the MiRd model. Finally, it is interesting to note that the best-fit values for \minfall\ are very similar for all models once we allow for one additional mechanism to reduce the number of cores, either by merging or disruption. 

To summarize, three of our four models perform well in capturing the radial galaxy surface density profiles as measured from SDSS \redmapper clusters across cluster masses and galaxy luminosity thresholds. The addition of one mechanism to reduce the number of cores beyond a cut on the infall mass provides enough flexibility to capture the number and distribution of galaxies in \redmapper clusters.

Next, we investigate in more detail how the different models perform given our luminosity thresholds that range from $>$0.4\lstar~to $>$2.5\lstar. As expected, the
\minfall\ parameter for all models increases with luminosity: brighter
galaxies tend to be more rare and originate from more massive
halos. The Mi model performs significantly worse at all luminosity ranges
as compared to the other models. The MiRm model does not perform as well as
the MiRd and the MiRdRm models. As the luminosity threshold is increased, however,
the Mi model improves. When compared to lower luminosity thresholds, the highest threshold of $>$2.5\lstar,
the Mi and MiRm models have the have only slightly differing \minfall\ and $\tilde\chi^2 \sim
1$. This may indicate that loss mechanisms for these bright galaxies
play a less important role.

The \rdisrupt\ parameter in the MiRd model also increases with galaxy
luminosity. A higher \rdisrupt\ parameter value can be physically
interpreted as the statement that these more massive galaxies are harder to disrupt. The
MiRdRm model does not have a significant change in the
\rdisrupt\ parameter across the full luminosity range and has a much
wider uncertainty bound as compared to the MiRd model. Most of the change
in the loss mechanisms arising in the MiRdRm model is restricted to merging. It
appears that the merging and disruption mechanisms are somewhat degenerate
in their effect on the galaxy profiles.

In Figure~\ref{fig:or_ngal} we provide a different view for the comparison of the data with the best-fit model. Here we show the projected $N_{\rm{gal}}$, the count of modeled galaxies above a luminosity threshold and within a projected \radius{200m} averaged from all angles (we note that our definition of of $N_{\rm{gal}}$ differs from \citealt{Koester2007}), as a function of cluster mass for the five different luminosity thresholds we have chosen. We focus on the results for one model, MiRd. The model provides an excellent description for all five luminosity thresholds and across the cluster mass range examined. 

\section{Robustness and Convergence Tests}
\label{sec:robustness_convergence_tests}
In order to investigate the robustness of our
results, we performed several checks. These include investigations of choices made in our analyses, such as the definition of core mass and effective cluster volume, and possible observational systematics, such as cluster miscentering and the mass-richness relation that was used to calibrate cluster masses. The core mass definition concerns our simulations, while the other investigations are focused on the SDSS data analysis. Throughout this section we use the $>$\lstar~MiRd model for our investigations. We confirmed that our conclusions hold for other models and luminosity thresholds as well.

\subsection{Comparison of Peak Mass and Infall Mass}
\label{sec:peak_infall_mass}
We first investigate the effect of using the peak mass of the halo along its past history
just before the merger, instead of using the halo mass at infall. We
find that the difference is minimal in the cluster halo mass regime of interest. On
applying the peak mass instead of infall mass in the AlphaQ
simulation, we find that the average mass associated with cores above
an infall mass of $10^{12}$~\hmdot\ increases by only 1.5\%, with a
standard deviation of 5\%. Such a small change is not unexpected. The
difference between infall mass and peak mass becomes significant when
one considers subhalos. A subhalo can have multiple merger events --
with each merger, it can lose a large fraction of its mass, thus
leading to a final infall mass that may be much smaller than the peak
mass. As described in Section~\ref{sec:core_tracking}, the core model
uses merger trees for halos only, and has at most a single infall
event for each halo. For the peak and infall mass to differ
significantly, the halo itself should have lost mass by some other
mechanism, such as, for example,  a glancing fly-by with a larger halo.

\begin{figure}[t]
  \centering
  \includegraphics[width=0.48\textwidth]{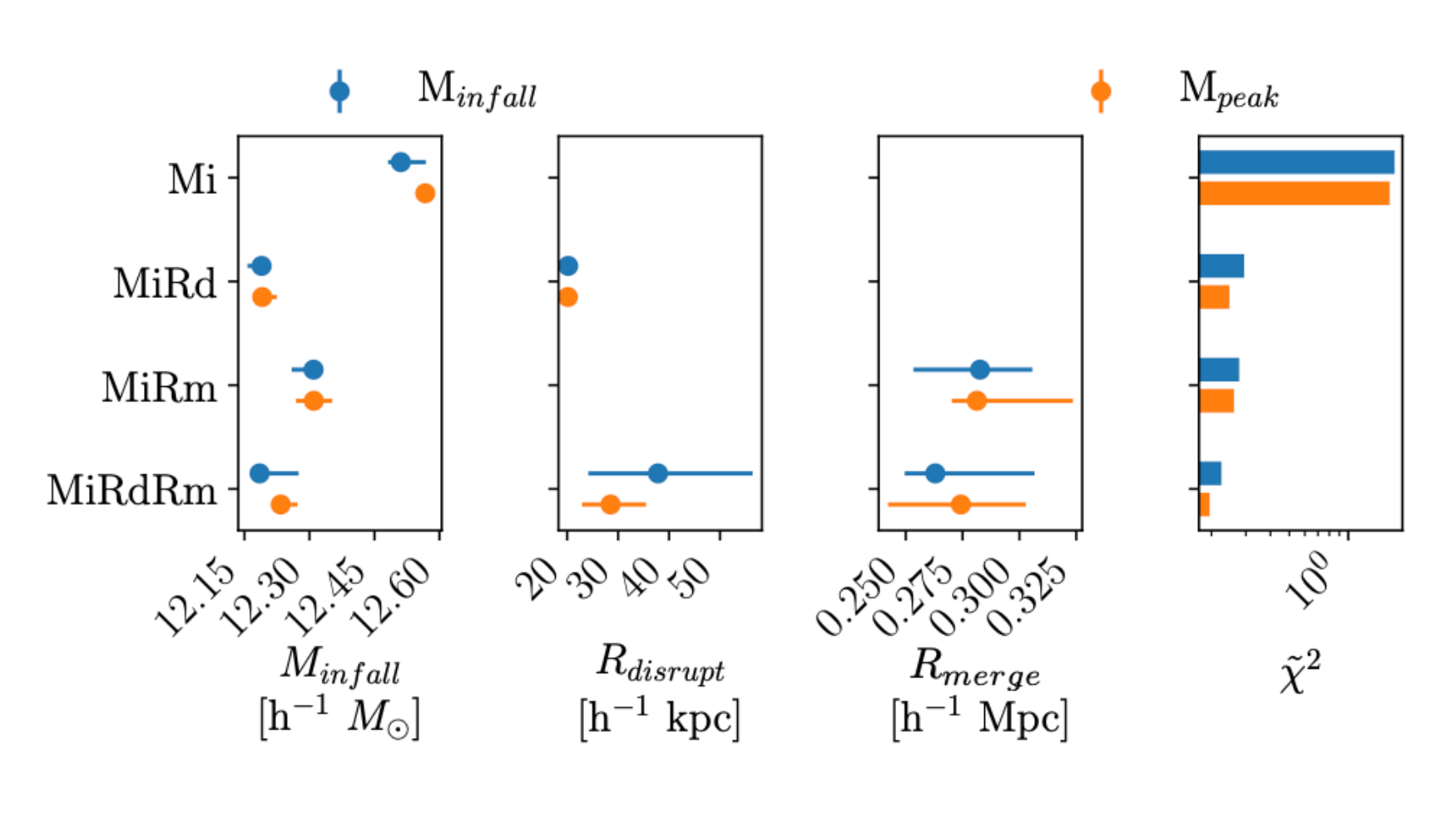}
  \caption{Best-fit parameters comparing \minfall\ and
    M$_{peak}$ as mass definitions for the cores. Neither definition
    significantly affects how well the model is able to reproduce
    observational data and the best-fit parameters remain within
    statistical confidence bounds.}
  \label{fig:best_fit_m_peak}
\end{figure}

Fitting the core models to $>$\lstar~galaxies, we find no
significant difference in the profile shapes, with only slight
improvement in $\tilde\chi^{2}$. The best-fit parameters remain within the fiducial
1$\sigma$ confidence bounds and are shown in
Figure~\ref{fig:best_fit_m_peak}. The best-fit \rdisrupt\ does not
change significantly, remaining well within the statistical error
bounds. We note that the systematic uncertainty for \minfall\ is much
greater for different choices of the projection volume (Section~\ref{sec:cluster
  volume}).

\begin{figure}[b]
  \centering
  \includegraphics[width=0.48\textwidth]{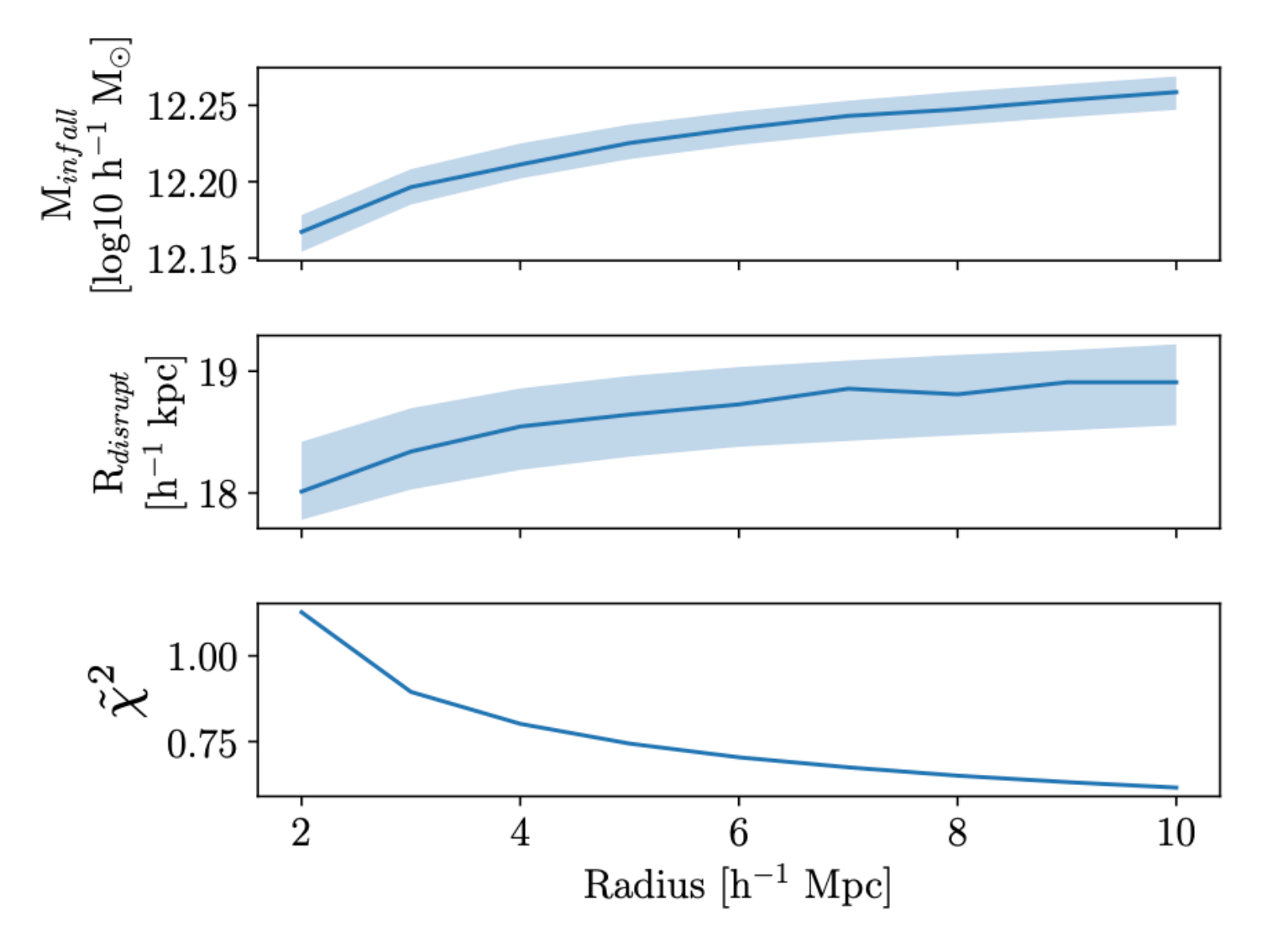}
    \caption{Best-fit model parameters and $\tilde\chi^2$ for the MiRd $>$\lstar~galaxy model as a function of projection radius. All cores within the projection radius are used to compute the projected modeled galaxy surface density.}
  \label{fig:cluster_volume}
\end{figure}

\subsection{Cluster Volume and Projection Effects}
\label{sec:cluster volume}
The SDSS galaxy profiles from the \redmapper cluster catalogs are constructed using projected galaxies. While we do carry out an average background subtraction (Section~\ref{sec:background}) to calculate galaxy profiles of clusters, the environment around clusters is more dense with both matter and galaxies than the global average as large-scale structure filaments of the cosmic web guide inflows into the cluster nodes. As a consequence, the background subtraction does not fully remove the contribution of galaxies near the cluster. To include these projection effects into the core model as well, we include all cores
within a fixed 10 \hmpc~comoving distance of the cluster center. By increasing the
volume used for projection, a larger fraction of galaxies within the
projected \radius{200c} will be interloper galaxies and not lie within the
physical \radius{200c} of the galaxy cluster. If we increase the
projected radius too much, we would include modeled galaxies that
would be removed by the average background subtraction.

We measure the effect of varying the cluster projection volume for one specific example, the MiRd model for $>$\lstar~galaxies; the best-fit parameters as a function of projection volume are shown in Figure~\ref{fig:cluster_volume}. Increasing the projection radius will always increase the number of interloper galaxies, so the parameters will shift to decrease the fraction of cores that pass both the mass and radius cuts. The best-fit \rdisrupt\ ~model parameter does not change significantly past 4~\hmpc. Most of the compensation for the increased volume occurs in the \minfall\ ~parameter. From 4~\hmpc\ ~to 10~\hmpc, \minfall~increases by 0.06 dex, which is significantly greater
than the typical statistical uncertainty of $\sim$0.01 dex. Increasing \minfall, as shown in Section~\ref{sec:core_m_infall}, strongly affects
the normalization of the profile, but does not change the shape
as strongly. The choice of a fixed cluster volume (10 \hmpc~in this work) introduces a systematic uncertainty greater than the statistical uncertainty on the precise value \minfall\ in the core galaxy model.  

\subsection{Cluster Miscentering}
\label{sec:cluster_miscentering}

The incorrect identification of the cluster center can cause a change in profile \citep{2008JCAP...08..006M,2017MNRAS.466.3103S,2017MNRAS.469.4899M,2019MNRAS.482.1352M}. The strongest effect is near the halo center where the sharp central peak may be flattened out. The \redmapper algorithm identifies central galaxy candidates by using luminosity and intracluster positions. The candidates are then assigned a probability of being a central -- such center definitions do not always line up with X-ray detected centers \citep{kart08, 2012ApJ...757....2G,2019MNRAS.484.1946G}. (We note that for unrelaxed clusters, the two centers may well be different.) Estimates on the SDSS \redmapper miscentering place the miscentering fraction at 14-32\% \citep{Rozo14, 2019MNRAS.487.2578Z} depending on the criterion used.

\begin{figure}[t]
  \centering
  \includegraphics[width=0.48\textwidth]{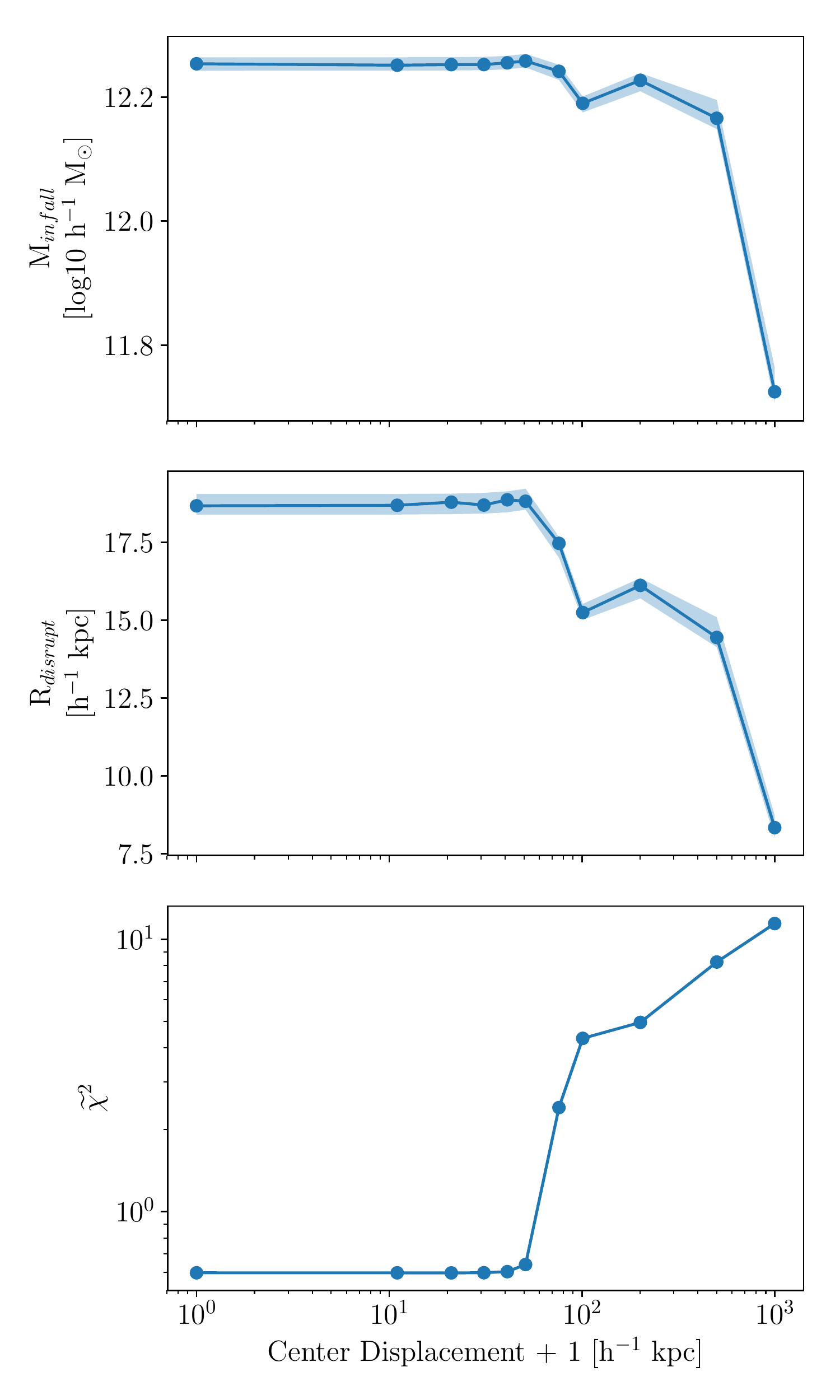}
  \caption{Effect of randomly displacing halo centers measured in the simulation on best-fit parameters and $\tilde\chi^2$ for the MiRd model for \redmapper $>$\lstar clusters. The points are individual fits, while the shaded regions are interpolated 1$\sigma$ error bands. The best-fit parameters are not significantly affected until a displacement of 100~$h^{-1}$kpc or greater, where the goodness of fit is significantly worse.}
  \label{fig:miscentering_effect}
\end{figure}

While we expect miscentering in our observational cluster samples, the core model of galaxies by its very construction does not include miscentering. The central galaxy is always placed at the halo center (with the exception of fragment halos, see Section~\ref{sec:core_tracking}). As a result, the core model of galaxies may display a systematic bias relative to the observations. The effect of miscentering on the best-fit $>$\lstar~MiRd model is shown in Figure~\ref{fig:miscentering_effect}, where we
randomly shift all halo centers and their central core in a random direction by a fixed distance. A significant effect in the best-fit parameters does not appear until the miscentering distance is about $>$50~\hkpc, where the best-fit parameters significantly shift and the goodness of fit decreases.  

\begin{figure}[t]
  \centering
  \includegraphics[width=0.48\textwidth]{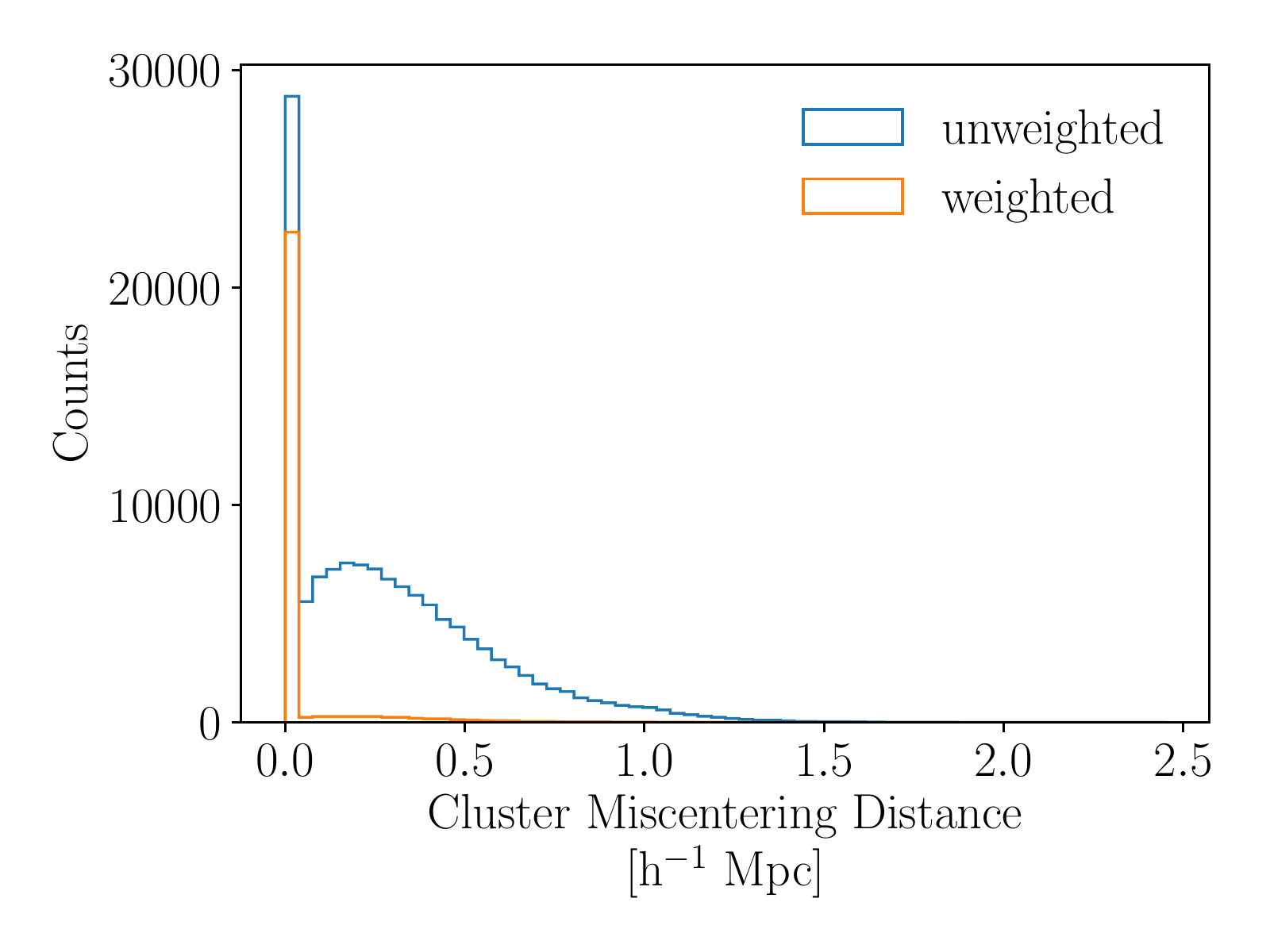}
  \caption{Probability distribution for the miscentering distance of
    \redmapper clusters, as calculated from the v6.3 catalog. Most of the non-primary
    candidate central galaxies have an estimated low probability of being the central which causes a
    large difference between the probability weighted and unweighted
    histograms.}
  \label{fig:redmapper_miscentering_distribution}
\end{figure}

For a more accurate characterization of miscentering within the core model, we introduce miscentering at a similar degree
as estimated in \redmapper and examine the shift in the best-fit model parameters. The \redmapper catalog provides the probability of the top five candidate galaxies to be the central galaxy in a cluster. From all \redmapper clusters, we use these candidates to generate the probability density function of the miscentering distance. First, we take the highest probability galaxy as the center and calculate the projected distance of each candidate from the center. A fixed distance, three-dimensional displacement in a random direction will have a smaller average measured displacement once it is projected onto two dimensions. We increase the distance by a factor of 1.273 to account for the that difference. Finally,
we weigh each distance measurement by the probability of being a
central. The unweighted and weighted distributions are shown
Figure~\ref{fig:redmapper_miscentering_distribution}. Using this method we estimate the miscentering fraction to be about 14\%, with an average of 317~\hkpc, if miscentered, and a total average of 48~\hkpc.

\begin{figure}[t]
  \centering
  \includegraphics[width=0.48\textwidth]{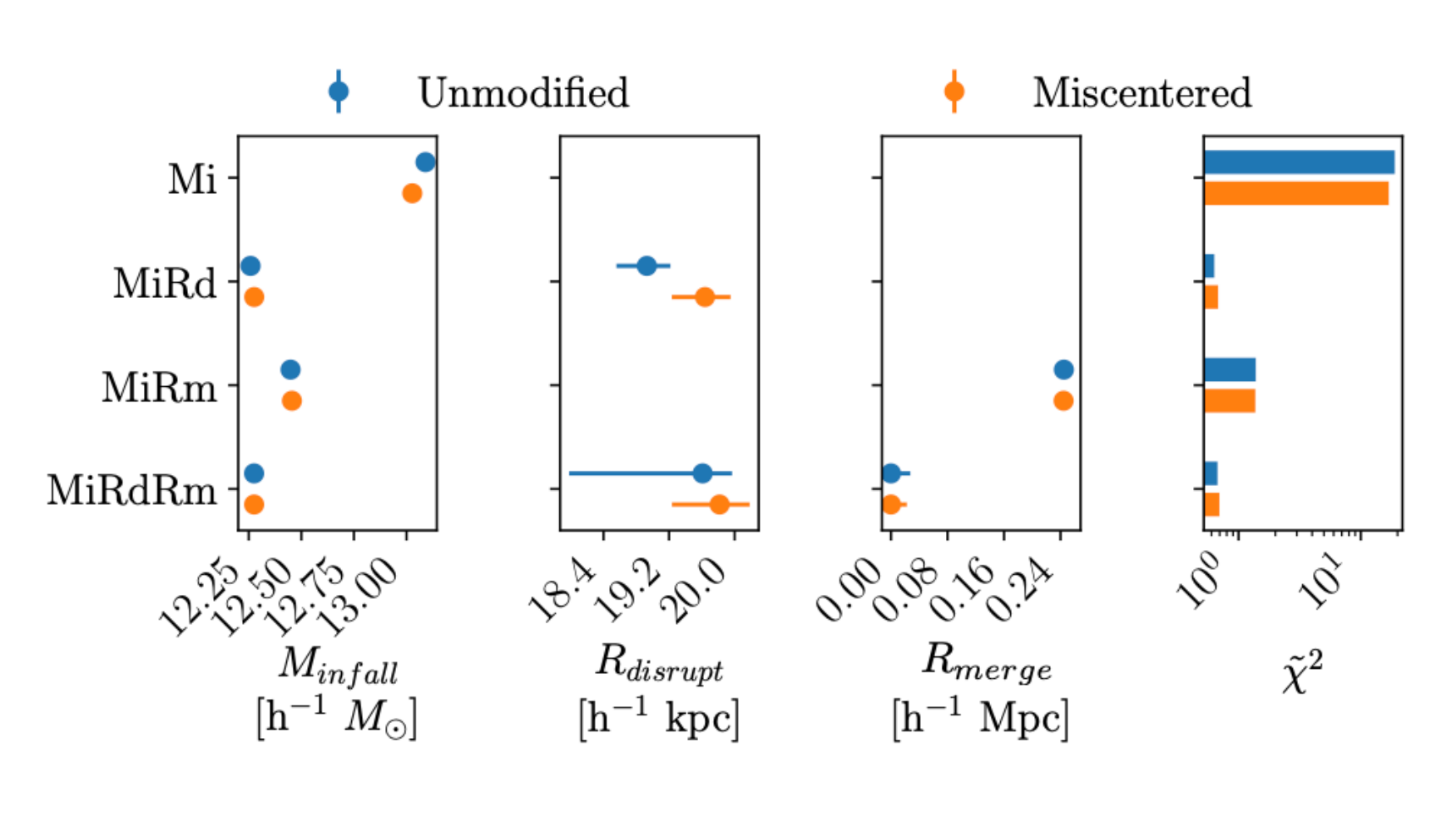}
  \caption{Best-fit model parameters and $\tilde\chi^2$ of $>$\lstar~galaxy profiles for unmodified (blue) and miscentered (orange). Error bars show the 1 sigma statistical confidence bounds. The centers are miscentered into a random direction with a distance distribution following what was found in the \redmapper catalog.}
  \label{fig:miscentering_fits}
\end{figure}

We apply the miscentering distribution to the core model by randomly
shifting each halo center and its central core by a distance sampled
from the miscentering distribution. Only $\sim 14\%$ of the centers and
central cores move as a result. All galaxy model profiles are then built around the
new halo center. The best-fit parameters are shown for
$>$\lstar~models of galaxies for unmodified and miscentered
simulation halos in Figure~\ref{fig:miscentering_fits}. For the models that perform well, miscentering did not significantly affect \rdisrupt, \rmerge, or \minfall~best-fit parameters. Therefore, miscentering in observations
does not lead to a large systemic uncertainty or bias in the core
model.

\subsection{RedMaPPer Mass-Richness Scaling}
\label{sec:remapper_mass_richess}

RedMaPPer does not provide galaxy cluster masses directly. The
mass must be inferred from calibrated \redmapper richness and halo
mass scaling. The choice of calibration modifies the observed cluster
mass and represents an uncertainty in the observations.  

The mass-richness scaling relationships have been calibrated by a diverse set
of techniques ranging from weak gravitational lensing \citep{2019MNRAS.482.1352M,2019PASJ...71..107M,2020MNRAS.491.1643P,2020MNRAS.495..428C,2021MNRAS.502.1494K}, X-ray observations \citep{2019MNRAS.486.1594C}, two-point clustering
\citep{2020MNRAS.498.2030C}, and velocity dispersion~\citep{andreon2010scaling, saro2013toward, bocquet2015mass}. The
calibration either converts cluster mass into an expected richness
or richness into an expected cluster halo mass. Due to scatter in the
relationship and the steepness of the halo mass function, the
conversions are not symmetric.

\begin{figure}[t]
  \centering
  \includegraphics[width=0.48\textwidth]{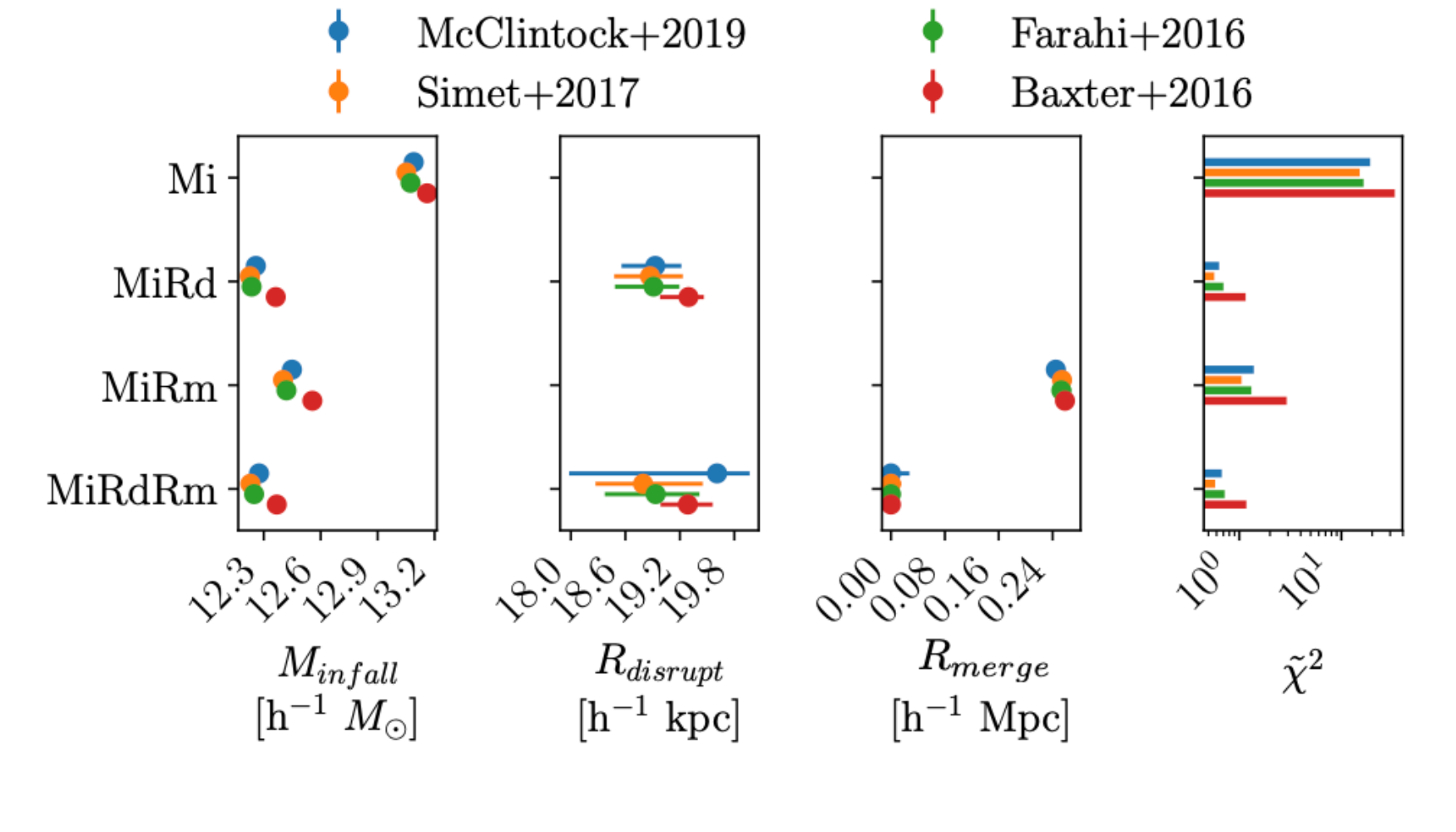}
  \caption{Best-fit parameters for $>$\lstar~galaxies for different mass-richness scaling relations. Overall, the results for the different scaling relations agree well. The goodness-of-fit using results from~\cite{Baxter2016} is slightly worse.}
  \label{fig:best_fit_mass_def}
\end{figure}

We examine the effect of four different mass calibrations on the best-fit parameters of the $>$\lstar~galaxy model (see Figure~\ref{fig:best_fit_mass_def}). The calibrations from \cite{2017MNRAS.466.3103S} and \cite{2019MNRAS.482.1352M} are based on weak lensing measurements, the one from \cite{Baxter2016} is based on angular clustering, and \cite{Farahi2016} use stacked velocity dispersion.

With the exception of the calibration from~\cite{Baxter2016} (which is an outlier from the other relations in terms of amplitude and slope constraints and notes that its large (18\%) uncertainty in mass calibration is driven by the theoretical uncertainty in the halo mass-bias relation), the
remaining three calibration approaches perform similarly in terms of the 
goodness-of-fit and produce similar best-fit model parameters. The
choice of richness-mass calibration does not qualitatively change the
findings, but leads to a shift of the best-fit parameters by small
amounts. 

\begin{table*}{
\begin{center}
\caption{Best-fit parameters for $>$\lstar~galaxies for different mass-richness scaling relations }
\begin{tabular}{cccccc}
 & & \multicolumn{4}{c}{Mass Definition} \\
\cline{3-6}
model & parameter & Baxter+2016 & Simet+2017 & Farahi+2016 & McClintock+2019 \\
\hline 
\hline 
\multirow{1}{*}{Mi} &$M_{\rm{infall}}$ &$13.161^{+0.004}_{-0.013}$ & $13.074^{+0.005}_{-0.019}$ & $13.051^{+0.015}_{-0.007}$ & $13.090^{+0.011}_{-0.013}$ \\
 & $\chi^2$ & $32$ & $16$ & $15$ & $18$ \\ 
\hline 
\multirow{2}{*}{Rd} &$M_{\rm{infall}}$ & $12.363^{+0.007}_{-0.008}$ & $12.236^{+0.013}_{-0.010}$ & $12.227^{+0.011}_{-0.014}$ & $12.258^{+0.010}_{-0.011}$ \\
 & $R_{\rm{disrupt}}$ & $19.3^{+0.2}_{-0.3}$ & $18.9^{+0.3}_{-0.4}$ & $18.9^{+0.4}_{-0.4}$ & $18.9^{+0.3}_{-0.4}$ \\ 
 & $\chi^2$ & $1.1$ & $0.68$ & $0.55$ & $0.62$ \\
\hline 
\multirow{2}{*}{Rm} &$M_{\rm{infall}}$ & $12.556^{+0.009}_{-0.007}$ & $12.420^{+0.010}_{-0.015}$ & $12.401^{+0.013}_{-0.012}$ & $12.449^{+0.011}_{-0.011}$ \\
 & $R_{\rm{merge}}$ & $0.258^{+0.003}_{-0.012}$ & $0.253^{+0.015}_{-0.006}$ & $0.254^{+0.014}_{-0.007}$ & $0.245^{+0.011}_{-0.008}$ \\
 & $\chi^2$ &  $2.8$ & $1.3$ & $1.0$ & $1.3$ \\
\hline 
\multirow{3}{*}{RdRm} &$M_{\rm{infall}}$ & $12.369^{+0.009}_{-0.010}$ & $12.249^{+0.006}_{-0.024}$ & $12.230^{+0.014}_{-0.017}$ & $12.260^{+0.012}_{-0.011}$ \\
 & $R_{\rm{disrupt}}$ & $19.3^{+0.3}_{-0.3}$ & $18.9^{+0.5}_{-0.6}$ & $18.8^{+0.7}_{-0.5}$ & $19.0^{+0.4}_{-0.4}$ \\ 
 & $R_{\rm{merge}}$ & $0.000^{+0.010}_{-0.000}$ & $0.000^{+0.014}_{-0.000}$ & $0.000^{+0.015}_{-0.000}$ & $0.000^{+0.011}_{-0.000}$ \\
 & $\chi^2$ & $1.1$ & $0.70$ & $0.56$ & $0.62$ \\
\end{tabular} 
\end{center}
  \begin{tablenotes}
          \item 
          \centering\small Overall, the results for the different scaling relations agree well. The goodness-of-fit using the results from~\cite{Baxter2016} is slightly worse than for the other relations.
        \end{tablenotes}}
\end{table*}

\section{Summary and Future Directions}
\label{sec:conclusion}

The development of large, realistic,  synthetic survey catalogs is a key element in the cosmological analyses of survey observations. For these purposes, gravity-only simulations are still the main ingredient, as opposed to far more computationally expensive hydrodynamical calculations. Additionally, within this approach, post-simulation modeling of the galaxy-halo connection allows for a great deal of modeling flexibility. This methodology for constructing galaxy catalogs comes with its own challenges, however, and these need to be addressed carefully, given the unprecedented datasets now becoming available from cosmological surveys. In this paper we focus on a new physically motivated approach to describe the galaxy distribution above different luminosity thresholds within galaxy cluster-sized halos. Simple SHAM approaches have been shown to fail for this problem; our approach adds further ingredients to allow an extension of SHAM-like ideas with encouraging initial results. 

Our method is based on the identification and tracking of substructures within halos in cosmological simulations. The method is based on following halo cores after infall into another halo and then applying core selection via infall mass thresholds, and merger and disruption criteria. We track the evolution of the core-member particle set throughout the remainder of the simulation. This information is included in detailed halo merger trees where the cores are tracked as substructure components after they have fallen into another halo. At the same time, a range of core properties is evaluated, including measures of the distribution of the core particles over time.  

We use this approach to develop a set of models for the galaxy-halo connection to address the galaxy distribution in clusters of galaxies. One advantage of explicitly tracking halo cores is that it is computationally more efficient than identifying and tracking subhalos. In addition, the approach avoids some of the complexities and difficulties associated with robustly identifying, characterizing and constructing merger trees for subhalos. In particular, the method more easily accounts for so-called `orphan' galaxies -- galaxies that have been largely stripped of their dark halo.

We use halo core tracking to motivate four models for galaxy identification within galaxy cluster-sized halos, applying them to describe the galaxy distribution using different choices for luminosity thresholds. The models are based on simple physical assumptions to reproduce both number and spatial distributions of the target galaxies. The mechanisms in the models are implemented via three parameters: 1) a galaxy luminosity proxy, 2) a single variable to account for the effect of tidal 
disruptions, and 3) a proxy for galaxy mergers. We explored each mechanism separately, combining all of them in a final version. In order to determine the free parameters in the models, we use the galaxy surface density profiles of \redmapper~clusters observed in the Sloan Digital Sky Survey.  We fit the model parameters to observed stacked radial distribution of galaxies in SDSS \redmapper~clusters with a mass range of
$14.25<\log_{10}(M_{200c})<15.25$ and explored five galaxy luminosity thresholds.

Excellent
agreement was found for the classes of models that allow for core disruption or merging (or both) along with an infall (or peak) mass threshold. However, the simplest model that neglects merging and disruption, and is therefore closest to a traditional SHAM approach, does not provide a good description of the SDSS cluster data.

We considered several sources of systematic errors. The role of the `core mass' definition was examined by studying two choices, either by using the mass of the associated halo at the (first) infall event or the peak mass of the halo along its past (pre-infall) history. The infall mass and the peak mass for cores differed only slightly and therefore 
did not significantly affect the best-fit model parameters, nor the associated radial profiles. 

The possible effect of cluster miscentering was studied by reproducing the
\redmapper cluster miscentering distribution in the simulation model; there was no significant effect on the resulting galaxy model, both
in best-fit parameters and goodness of fit. A much larger effect on the
\minfall~parameter is connected to the choice made for the included volume around the cluster when constructing the projected profiles as discussed in Section~\ref{sec:cluster volume}.

We also examined the effect of different \redmapper richness-mass
scaling relations as determined in different calibrations and found little difference
with the exception of the result in \cite{Baxter2016}, which had slightly worse performance and offset best-fit parameters as
compared to the other calibrations.

There are several directions to explore in future work. An obvious application is the construction of models for field galaxies with the goal of
reproducing the two-point correlation function and the abundance for the targeted galaxy sample. Although the
simple model of associating galaxies with cores is able to reproduce the galaxy
distribution inside galaxy cluster sized halos, such a model cannot be directly applied to field galaxies. Work on this topic will be detailed elsewhere. 

The approach described in this paper does not take into account the post-infall time spent by the cores within larger host halos. The longer a core resides inside a halo as substructure, the more likely it will be to lose mass; a simple mass loss model is reported in~\cite{LJ2}. In that paper, however, we did not consider core merging or disruption separately. Combining the considerations in this paper with the approach  of~\cite{LJ2} is another step for future work, as it forms an important ingredient of a new framework for developing semi-analytic models for galaxy formation.

A more detailed analysis of the core-galaxy connection based on the use of core merger trees, the analog of halo (and subhalo) merger trees, is in progress. In particular, this construction provides a direct path to obtaining core merger statistics, and provides a backbone on which galaxy modeling can be grafted using a variety of empirical approaches. Finally, the assumptions and results of the core model and its core tree extension can also be investigated by comparing to results from hydrodynamic cosmological simulations, which model galaxy formation directly. We have begun such a program by running paired gravity-only and hydrodynamic simulations with the hydro-extension of the HACC code -- CRK-HACC~\citep{frontiere_crkhacc}. Initial results are very encouraging and the key assumptions underlying the core model appear to be very robust.

\begin{acknowledgments}
We are grateful to the referee for a detailed and valuable report. We are indebted to Mike Gladders, Andrew Hearin, Eve Kovacs, Yao-Yuan Mao, and Malin Renneby for helpful discussions. We thank Michael Buehlmann, Isabele Vitorio, and Azton Wells for their work on early tests of the core model with hydrodynamic simulations using CRK-HACC. Work at Argonne National Laboratory was supported under the U.S. Department of Energy contract DE-AC02-06CH11357. This research used resources of the Argonne Leadership Computing Facility, which is supported by DOE/SC under
contract DE-AC02-06CH11357.
\end{acknowledgments}

\bibliographystyle{yahapj}
\bibliography{ref}

\begin{thebibliography}{}
\providecommand\natexlab[1]{#1}
\providecommand\JournalTitle[1]{#1}

\bibitem[{Aguado {et~al.}(2019)}]{SDSS_DR15}
Aguado, D.~S., {et~al.} 2019,
  \href{http://dx.doi.org/10.3847/1538-4365/aaf651}{\JournalTitle{\apjs}, 240,
  23}

\bibitem[{Andreon \& Hurn(2010)}]{andreon2010scaling}
Andreon, S., \& Hurn, M.~A. 2010, \JournalTitle{\mnras}, 404, 1922

\bibitem[{{Astropy Collaboration} {et~al.}(2013){Astropy Collaboration},
  {Robitaille}, {Tollerud}, {Greenfield}, {Droettboom}, {Bray}, {Aldcroft},
  {Davis}, {Ginsburg}, {Price-Whelan}, {Kerzendorf}, {Conley}, {Crighton},
  {Barbary}, {Muna}, {Ferguson}, {Grollier}, {Parikh}, {Nair}, {Unther},
  {Deil}, {Woillez}, {Conseil}, {Kramer}, {Turner}, {Singer}, {Fox}, {Weaver},
  {Zabalza}, {Edwards}, {Azalee Bostroem}, {Burke}, {Casey}, {Crawford},
  {Dencheva}, {Ely}, {Jenness}, {Labrie}, {Lim}, {Pierfederici}, {Pontzen},
  {Ptak}, {Refsdal}, {Servillat}, \& {Streicher}}]{Astropy2013}
{Astropy Collaboration}, {Robitaille}, T.~P., {Tollerud}, E.~J., {et~al.} 2013,
  \href{http://dx.doi.org/10.1051/0004-6361/201322068}{\JournalTitle{\aap},
  558, A33}

\bibitem[{Bah{\'{e}} {et~al.}(2019)Bah{\'{e}}, Schaye, Barnes,
  {Dalla Vecchia}, Kay, Bower, Hoekstra, McGee, \& Theuns}]{Bahe2019}
Bah{\'{e}}, Y.~M., Schaye, J., Barnes, D.~J., {et~al.} 2019,
  \href{http://dx.doi.org/10.1093/mnras/stz361}{\JournalTitle{\mnras}, 485,
  2287}

\bibitem[{Baxter {et~al.}(2016)Baxter, Rozo, Jain, Rykoff, \&
  Wechsler}]{Baxter2016}
Baxter, E.~J., Rozo, E., Jain, B., Rykoff, E., \& Wechsler, R.~H. 2016,
  \href{http://dx.doi.org/10.1093/mnras/stw1939}{\JournalTitle{\mnras}, 463,
  205}

\bibitem[{Behroozi {et~al.}(2010)Behroozi, Conroy, \& Wechsler}]{Behroozi2010}
Behroozi, P.~S., Conroy, C., \& Wechsler, R.~H. 2010,
  \href{http://dx.doi.org/10.1088/0004-637X/717/1/379}{\JournalTitle{\apj},
  717, 379}

\bibitem[{Berlind \& Weinberg(2002)}]{Berlind2002}
Berlind, A.~A., \& Weinberg, D.~H. 2002,
  \href{http://dx.doi.org/10.1086/341469}{\JournalTitle{\apj}, 575, 587}

\bibitem[{Blumenthal {et~al.}(1986)Blumenthal, Faber, Flores, \&
  Primack}]{Blumenthal1986}
Blumenthal, G.~R., Faber, S.~M., Flores, R., \& Primack, J.~R. 1986,
  \href{http://dx.doi.org/10.1086/163867}{\JournalTitle{\apj}, 301, 27}

\bibitem[{Bocquet {et~al.}(2015)Bocquet, Saro, Mohr, Aird, Ashby, Bautz,
  Bayliss, Bazin, Benson, Bleem, {et~al.}}]{bocquet2015mass}
Bocquet, S., Saro, A., Mohr, J., {et~al.} 2015, \JournalTitle{\apj}, 799, 214

\bibitem[{Budzynski {et~al.}(2012)Budzynski, Koposov, Mccarthy, Mcgee, \&
  Belokurov}]{Budzynski2012}
Budzynski, J.~M., Koposov, S.~E., Mccarthy, I.~G., Mcgee, S.~L., \& Belokurov,
  V. 2012,
  \href{http://dx.doi.org/10.1111/j.1365-2966.2012.20663.x}{\JournalTitle{\mnras},
  423, 104}

\bibitem[{{Campbell} {et~al.}(2018){Campbell}, {van den Bosch}, {Padmanabhan},
  {Mao}, {Zentner}, {Lange}, {Jiang}, \& {Villarreal}}]{2018MNRAS.477..359C}
{Campbell}, D., {van den Bosch}, F.~C., {Padmanabhan}, N., {et~al.} 2018,
  \href{http://dx.doi.org/10.1093/mnras/sty495}{\JournalTitle{\mnras}, 477,
  359}

\bibitem[{{Capasso} {et~al.}(2019){Capasso}, {Mohr}, {Saro}, {Biviano},
  {Clerc}, {Finoguenov}, {Grandis}, {Collins}, {Erfanianfar}, {Damsted},
  {Kirkpatrick}, \& {Kukkola}}]{2019MNRAS.486.1594C}
{Capasso}, R., {Mohr}, J.~J., {Saro}, A., {et~al.} 2019,
  \href{http://dx.doi.org/10.1093/mnras/stz931}{\JournalTitle{\mnras}, 486,
  1594}

\bibitem[{Child {et~al.}(2018)Child, Habib, Heitmann, Frontiere, Finkel, Pope,
  \& Morozov}]{Child2018}
Child, H.~L., Habib, S., Heitmann, K., {et~al.} 2018,
  \href{http://dx.doi.org/10.3847/1538-4357/aabf95}{\JournalTitle{\apj}, 859,
  55}

\bibitem[{{Chiu} {et~al.}(2020{\natexlab{a}}){Chiu}, {Okumura}, {Oguri},
  {Agrawal}, {Umetsu}, \& {Lin}}]{2020MNRAS.498.2030C}
{Chiu}, I.~N., {Okumura}, T., {Oguri}, M., {et~al.} 2020{\natexlab{a}},
  \href{http://dx.doi.org/10.1093/mnras/staa2440}{\JournalTitle{\mnras}, 498,
  2030}

\bibitem[{{Chiu} {et~al.}(2020{\natexlab{b}}){Chiu}, {Umetsu}, {Murata},
  {Medezinski}, \& {Oguri}}]{2020MNRAS.495..428C}
{Chiu}, I.~N., {Umetsu}, K., {Murata}, R., {Medezinski}, E., \& {Oguri}, M.
  2020{\natexlab{b}},
  \href{http://dx.doi.org/10.1093/mnras/staa1158}{\JournalTitle{\mnras}, 495,
  428}

\bibitem[{Conroy {et~al.}(2006)Conroy, Wechsler, \& Kravtsov}]{Conroy2006}
Conroy, C., Wechsler, R.~H., \& Kravtsov, A.~V. 2006,
  \href{http://dx.doi.org/10.1086/503602}{\JournalTitle{\apj}, 647, 201}

\bibitem[{Davis {et~al.}(1985)Davis, Efstathiou, Frenk, \& White}]{Davis1985}
Davis, M., Efstathiou, G., Frenk, C.~S., \& White, S. D.~M. 1985,
  \href{http://dx.doi.org/10.1086/163168}{\JournalTitle{\apj}, 292, 371}

\bibitem[{De~Lucia \& Blaizot(2007)}]{DeLucia2007}
De~Lucia, G., \& Blaizot, J. 2007,
  \href{http://dx.doi.org/10.1111/j.1365-2966.2006.11287.x}{\JournalTitle{\mnras},
  375, 2}

\bibitem[{Dubinski(1998)}]{Dubinski1998}
Dubinski, J. 1998, \href{http://dx.doi.org/10.1086/305901}{\JournalTitle{\apj},
  502, 141}

\bibitem[{Fall \& Efstathiou(1980)}]{Fall1980}
Fall, S.~M., \& Efstathiou, G. 1980,
  \href{http://dx.doi.org/10.1093/mnras/193.2.189}{\JournalTitle{\mnras}, 193,
  189}

\bibitem[{Farahi {et~al.}(2016)Farahi, Evrard, Rozo, Rykoff, \&
  Wechsler}]{Farahi2016}
Farahi, A., Evrard, A.~E., Rozo, E., Rykoff, E.~S., \& Wechsler, R.~H. 2016,
  \href{http://dx.doi.org/10.1093/mnras/stw1143}{\JournalTitle{\mnras}, 460,
  3900}

\bibitem[{{Frontiere} {et~al.}(2022){Frontiere}, {Emberson}, {Buehlmann},
  {Adamo}, {Habib}, {Heitmann}, \& {Faucher-Gigu{\`e}re}}]{frontiere_crkhacc}
{Frontiere}, N., {Emberson}, J.~D., {Buehlmann}, M., {et~al.} 2022,
  \JournalTitle{arXiv e-prints}, arXiv:2202.02840

\bibitem[{Gao {et~al.}(2004{\natexlab{a}})Gao, {De Lucia}, White, \&
  Jenkins}]{Gao2004}
Gao, L., {De Lucia}, G., White, S.~D., \& Jenkins, A. 2004{\natexlab{a}},
  \href{http://dx.doi.org/10.1111/j.1365-2966.2004.08098.x}{\JournalTitle{\mnras},
  352, L1}

\bibitem[{Gao {et~al.}(2004{\natexlab{b}})Gao, Loeb, Peebles, White, \&
  Jenkins}]{Gao2004b}
Gao, L., Loeb, A., Peebles, P. J.~E., White, S. D.~M., \& Jenkins, A.
  2004{\natexlab{b}},
  \href{http://dx.doi.org/10.1086/423444}{\JournalTitle{\apj}, 614, 17}

\bibitem[{{Ge} {et~al.}(2019){Ge}, {Sun}, {Rozo}, {Sehgal}, {Vikhlinin},
  {Forman}, {Jones}, \& {Nagai}}]{2019MNRAS.484.1946G}
{Ge}, C., {Sun}, M., {Rozo}, E., {et~al.} 2019,
  \href{http://dx.doi.org/10.1093/mnras/stz088}{\JournalTitle{\mnras}, 484,
  1946}

\bibitem[{{George} {et~al.}(2012){George}, {Leauthaud}, {Bundy}, {Finoguenov},
  {Ma}, {Rykoff}, {Tinker}, {Wechsler}, {Massey}, \&
  {Mei}}]{2012ApJ...757....2G}
{George}, M.~R., {Leauthaud}, A., {Bundy}, K., {et~al.} 2012,
  \href{http://dx.doi.org/10.1088/0004-637X/757/1/2}{\JournalTitle{\apj}, 757,
  2}

\bibitem[{Gill {et~al.}(2005)Gill, Knebe, \& Gibson}]{Stuart2005}
Gill, S. P.~D., Knebe, A., \& Gibson, B.~K. 2005,
  \href{http://dx.doi.org/10.1111/j.1365-2966.2004.08562.x}{\JournalTitle{\mnras},
  356, 1327}

\bibitem[{Gladders \& Yee(2000)}]{Gladders_2000}
Gladders, M.~D., \& Yee, H. K.~C. 2000,
  \href{http://dx.doi.org/10.1086/301557}{\JournalTitle{The Astronomical
  Journal}, 120, 2148}

\bibitem[{Gladders \& Yee(2005)}]{Gladders2005}
---. 2005, \href{http://dx.doi.org/10.1086/427327}{\JournalTitle{\apjs}, 157,
  1}

\bibitem[{{Grandis} {et~al.}(2021){Grandis}, {Mohr}, {Costanzi}, {Saro},
  {Bocquet}, {Klein}, {Aguena}, {Allam}, {Annis}, {Ansarinejad}, {Bacon},
  {Bertin}, {Bleem}, {Brooks}, {Burke}, {Carnero Rosel}, {Carrasco Kind},
  {Carretero}, {Castander}, {Choi}, {da Costa}, {De Vincente}, {Desai},
  {Diehl}, {Dietrich}, {Doel}, {Eifler}, {Everett}, {Ferrero}, {Floyd},
  {Fosalba}, {Frieman}, {Garc{\'\i}a-Bellido}, {Gaztanaga}, {Gruen}, {Gruendl},
  {Gschwend}, {Gupta}, {Gutierrez}, {Hinton}, {Hollowood}, {Honscheid},
  {James}, {Jeltema}, {Kuehn}, {Lahav}, {Lidman}, {Lima}, {Maia}, {March},
  {Marshall}, {Melchior}, {Menanteau}, {Miquel}, {Morgan}, {Myles}, {Ogando},
  {Palmese}, {Paz-Chinch{\'o}n}, {Plazas}, {Reichardt}, {Romer}, {Sanchez},
  {Scarpine}, {Serrano}, {Sevilla-Noarbe}, {Singh}, {Smith}, {Suchyta},
  {Swanson}, {Tarle}, {Thomas}, {To}, {Weller}, {Wilkinson}, \&
  {Wu}}]{Grandis2021}
{Grandis}, S., {Mohr}, J.~J., {Costanzi}, M., {et~al.} 2021,
  \href{http://dx.doi.org/10.1093/mnras/stab869}{\JournalTitle{\mnras}, 504,
  1253}

\bibitem[{{Guo} \& {White}(2014)}]{2014MNRAS.437.3228G}
{Guo}, Q., \& {White}, S. 2014,
  \href{http://dx.doi.org/10.1093/mnras/stt2116}{\JournalTitle{\mnras}, 437,
  3228}

\bibitem[{Guo {et~al.}(2010)Guo, White, Li, \& Boylan-Kolchin}]{Guo2010}
Guo, Q., White, S., Li, C., \& Boylan-Kolchin, M. 2010,
  \href{http://dx.doi.org/10.1111/j.1365-2966.2010.16341.x}{\JournalTitle{\mnras},
  404, 1111}

\bibitem[{Habib {et~al.}(2013)Habib, Morozov, Frontiere, Finkel, Pope, \&
  Heitmann}]{HACC}
Habib, S., Morozov, V., Frontiere, N., {et~al.} 2013,
  \href{http://dx.doi.org/10.1145/2503210.2504566}{in Proceedings of the
  International Conference on High Performance Computing, Networking, Storage
  and Analysis, SC '13} (New York, NY, USA: ACM), 6:1

\bibitem[{Habib {et~al.}(2016)Habib, Pope, Finkel, Frontiere, Heitmann, Daniel,
  Fasel, Morozov, Zagaris, Peterka, Vishwanath, Luki\'c, Sehrish, \&
  Liao}]{Habib:2014uxa}
Habib, S., Pope, A., Finkel, H., {et~al.} 2016,
  \href{http://dx.doi.org/10.1016/j.newast.2015.06.003}{\JournalTitle{New
  Astronomy}, 42, 49}

\bibitem[{{Han} {et~al.}(2018){Han}, {Cole}, {Frenk}, {Benitez-Llambay}, \&
  {Helly}}]{2018MNRAS.474..604H}
{Han}, J., {Cole}, S., {Frenk}, C.~S., {Benitez-Llambay}, A., \& {Helly}, J.
  2018, \href{http://dx.doi.org/10.1093/mnras/stx2792}{\JournalTitle{\mnras},
  474, 604}

\bibitem[{Hansen {et~al.}(2005)Hansen, McKay, Wechsler, Annis, Sheldon, \&
  Kimball}]{Hansen2005}
Hansen, S.~M., McKay, T.~A., Wechsler, R.~H., {et~al.} 2005,
  \href{http://dx.doi.org/10.1086/444554}{\JournalTitle{\apj}, 633, 122}

\bibitem[{{Heitmann} {et~al.}(2019){Heitmann}, {Finkel}, {Pope}, {Morozov},
  {Frontiere}, {Habib}, {Rangel}, {Uram}, {Korytov}, {Child}, {Flender},
  {Insley}, \& {Rizzi}}]{OuterRim}
{Heitmann}, K., {Finkel}, H., {Pope}, A., {et~al.} 2019,
  \href{http://dx.doi.org/10.3847/1538-4365/ab4da1}{\JournalTitle{\apjs}, 245,
  16}

\bibitem[{{Heitmann} {et~al.}(2021){Heitmann}, {Frontiere}, {Rangel}, {Larsen},
  {Pope}, {Sultan}, {Uram}, {Habib}, {Finkel}, {Korytov}, {Kovacs}, {Rizzi},
  {Insley}, \& {Knowles}}]{LJ1}
{Heitmann}, K., {Frontiere}, N., {Rangel}, E., {et~al.} 2021,
  \href{http://dx.doi.org/10.3847/1538-4365/abcc67}{\JournalTitle{\apjs}, 252,
  19}

\bibitem[{Ho {et~al.}(2009)Ho, Lin, Spergel, \& Hirata}]{Ho2009}
Ho, S., Lin, Y.-T., Spergel, D., \& Hirata, C.~M. 2009,
  \href{http://dx.doi.org/10.1088/0004-637x/697/2/1358}{\JournalTitle{\apj},
  697, 1358}

\bibitem[{Jones \& Forman(1984)}]{Jones1984}
Jones, C., \& Forman, W. 1984,
  \href{http://dx.doi.org/10.1086/161591}{\JournalTitle{\apj}, 276, 38}

\bibitem[{Kaiser(1984)}]{kaiser1984}
Kaiser, N. 1984, \JournalTitle{\apj}, 284, L9

\bibitem[{Kartaltepe {et~al.}(2008)Kartaltepe, Ebeling, Ma, \&
  Donovan}]{kart08}
Kartaltepe, J.~S., Ebeling, H., Ma, C., \& Donovan, D. 2008,
  \JournalTitle{\mnras}, 389, 1240

\bibitem[{{Kiiveri} {et~al.}(2021){Kiiveri}, {Gruen}, {Finoguenov}, {Erben},
  {van Waerbeke}, {Rykoff}, {Miller}, {Hagstotz}, {Dupke}, {Patrick Henry},
  {Kneib}, {Gozaliasl}, {Kirkpatrick}, {Cibirka}, {Clerc}, {Costanzi},
  {Cypriano}, {Rozo}, {Shan}, {Spinelli}, {Valiviita}, \&
  {Weller}}]{2021MNRAS.502.1494K}
{Kiiveri}, K., {Gruen}, D., {Finoguenov}, A., {et~al.} 2021,
  \href{http://dx.doi.org/10.1093/mnras/staa3936}{\JournalTitle{\mnras}, 502,
  1494}

\bibitem[{Knebe {et~al.}(2013)Knebe, Pearce, Lux, Ascasibar, Behroozi, Casado,
  Moran, Diemand, Dolag, Dominguez-Tenreiro, Elahi, Falck, Gottl{\"{o}}ber,
  Han, Klypin, Luki{\'{c}}, Maciejewski, McBride, Merch{\'{a}}n, Muldrew,
  Neyrinck, Onions, Planelles, Potter, Quilis, Rasera, Ricker, Roy, Ruiz,
  Sgr{\'{o}}, Springel, Stadel, Sutter, Tweed, \& Zemp}]{Knebe2013}
Knebe, A., Pearce, F.~R., Lux, H., {et~al.} 2013,
  \href{http://dx.doi.org/10.1093/mnras/stt1403}{\JournalTitle{\mnras}, 435,
  1618}

\bibitem[{{Koester} {et~al.}(2007){Koester}, {McKay}, {Annis}, {Wechsler},
  {Evrard}, {Bleem}, {Becker}, {Johnston}, {Sheldon}, {Nichol}, {Miller},
  {Scranton}, {Bahcall}, {Barentine}, {Brewington}, {Brinkmann}, {Harvanek},
  {Kleinman}, {Krzesinski}, {Long}, {Nitta}, {Schneider}, {Sneddin}, {Voges},
  \& {York}}]{Koester2007}
{Koester}, B.~P., {McKay}, T.~A., {Annis}, J., {et~al.} 2007,
  \href{http://dx.doi.org/10.1086/509599}{\JournalTitle{\apj}, 660, 239}

\bibitem[{{Komatsu} {et~al.}(2011){Komatsu}, {Smith}, {Dunkley}, {Bennett},
  {Gold}, {Hinshaw}, {Jarosik}, {Larson}, {Nolta}, {Page}, {Spergel},
  {Halpern}, {Hill}, {Kogut}, {Limon}, {Meyer}, {Odegard}, {Tucker}, {Weiland},
  {Wollack}, \& {Wright}}]{2011ApJS..192...18K}
{Komatsu}, E., {Smith}, K.~M., {Dunkley}, J., {et~al.} 2011,
  \href{http://dx.doi.org/10.1088/0067-0049/192/2/18}{\JournalTitle{\apjs},
  192, 18}

\bibitem[{Kravtsov {et~al.}(2004{\natexlab{a}})Kravtsov, Berlind, Wechsler,
  Klypin, Gottlober, Allgood, \& Primack}]{Kravtsov2004}
Kravtsov, A.~V., Berlind, A.~A., Wechsler, R.~H., {et~al.} 2004{\natexlab{a}},
  \href{http://dx.doi.org/10.1086/420959}{\JournalTitle{\apj}, 609, 35}

\bibitem[{Kravtsov {et~al.}(2004{\natexlab{b}})Kravtsov, Gnedin, \&
  Klypin}]{Kravtsov04}
Kravtsov, A.~V., Gnedin, O.~Y., \& Klypin, A.~A. 2004{\natexlab{b}},
  \href{http://dx.doi.org/10.1086/421322}{\JournalTitle{The Astrophysical
  Journal}, 609, 482}

\bibitem[{{Lacey} \& {Cole}(1994)}]{1994MNRAS.271..676L}
{Lacey}, C., \& {Cole}, S. 1994,
  \href{http://dx.doi.org/10.1093/mnras/271.3.676}{\JournalTitle{\mnras}, 271,
  676}

\bibitem[{Lin {et~al.}(2004)Lin, Mohr, \& Stanford}]{Lin2004}
Lin, Y.-T., Mohr, J.~J., \& Stanford, S.~A. 2004,
  \href{http://dx.doi.org/10.1086/421714}{\JournalTitle{\apj}, 610, 745}

\bibitem[{{LSST Dark Energy Science Collaboration: LSST DESC}
  {et~al.}(2021){LSST Dark Energy Science Collaboration: LSST DESC},
  {Abolfathi}, {Alonso}, {Armstrong}, {Aubourg}, {Awan}, {Babuji}, {Bauer},
  {Bean}, {Beckett}, {Biswas}, {Bogart}, {Boutigny}, {Chard}, {Chiang},
  {Claver}, {Cohen-Tanugi}, {Combet}, {Connolly}, {Daniel}, {Digel},
  {Drlica-Wagner}, {Dubois}, {Gangler}, {Gawiser}, {Glanzman}, {Gris}, {Habib},
  {Hearin}, {Heitmann}, {Hernandez}, {Hlo{\v{z}}ek}, {Hollowed}, {Ishak},
  {Ivezi{\'c}}, {Jarvis}, {Jha}, {Kahn}, {Kalmbach}, {Kelly}, {Kovacs},
  {Korytov}, {Krughoff}, {Lage}, {Lanusse}, {Larsen}, {Le Guillou}, {Li},
  {Longley}, {Lupton}, {Mandelbaum}, {Mao}, {Marshall}, {Meyers}, {Moniez},
  {Morrison}, {Nomerotski}, {O'Connor}, {Park}, {Park}, {Peloton}, {Perrefort},
  {Perry}, {Plaszczynski}, {Pope}, {Rasmussen}, {Reil}, {Roodman}, {Rykoff},
  {S{\'a}nchez}, {Schmidt}, {Scolnic}, {Stubbs}, {Tyson}, {Uram}, {Villarreal},
  {Walter}, {Wiesner}, {Wood-Vasey}, \& {Zuntz}}]{dc2}
{LSST Dark Energy Science Collaboration: LSST DESC}, {Abolfathi}, B., {Alonso},
  D., {et~al.} 2021,
  \href{http://dx.doi.org/10.3847/1538-4365/abd62c}{\JournalTitle{\apjs}, 253,
  31}

\bibitem[{Ludlow {et~al.}(2009)Ludlow, Navarro, Springel, Jenkins, Frenk, \&
  Helmi}]{Ludlow2009}
Ludlow, A.~D., Navarro, J.~F., Springel, V., {et~al.} 2009,
  \href{http://dx.doi.org/10.1088/0004-637x/692/1/931}{\JournalTitle{\apj},
  692, 931}

\bibitem[{{Mandelbaum} {et~al.}(2008){Mandelbaum}, {Seljak}, \&
  {Hirata}}]{2008JCAP...08..006M}
{Mandelbaum}, R., {Seljak}, U., \& {Hirata}, C.~M. 2008,
  \href{http://dx.doi.org/10.1088/1475-7516/2008/08/006}{\JournalTitle{JCAP},
  2008, 006}

\bibitem[{Mayer {et~al.}(2001)Mayer, Governato, Colpi, Moore, Quinn, Wadsley,
  Stadel, \& Lake}]{Mayer01}
Mayer, L., Governato, F., Colpi, M., {et~al.} 2001,
  \href{http://dx.doi.org/10.1086/318898}{\JournalTitle{The Astrophysical
  Journal}, 547, L123}

\bibitem[{{McClintock} {et~al.}(2019){McClintock}, {Varga}, {Gruen}, {Rozo},
  {Rykoff}, {Shin}, {Melchior}, {DeRose}, {Seitz}, {Dietrich}, {Sheldon},
  {Zhang}, {von der Linden}, {Jeltema}, {Mantz}, {Romer}, {Allen}, {Becker},
  {Bermeo}, {Bhargava}, {Costanzi}, {Everett}, {Farahi}, {Hamaus}, {Hartley},
  {Hollowood}, {Hoyle}, {Israel}, {Li}, {MacCrann}, {Morris}, {Palmese},
  {Plazas}, {Pollina}, {Rau}, {Simet}, {Soares-Santos}, {Troxel}, {Vergara
  Cervantes}, {Wechsler}, {Zuntz}, {Abbott}, {Abdalla}, {Allam}, {Annis},
  {Avila}, {Bridle}, {Brooks}, {Burke}, {Carnero Rosell}, {Carrasco Kind},
  {Carretero}, {Castander}, {Crocce}, {Cunha}, {D'Andrea}, {da Costa}, {Davis},
  {De Vicente}, {Diehl}, {Doel}, {Drlica-Wagner}, {Evrard}, {Flaugher},
  {Fosalba}, {Frieman}, {Garc{\'\i}a-Bellido}, {Gaztanaga}, {Gerdes},
  {Giannantonio}, {Gruendl}, {Gutierrez}, {Honscheid}, {James}, {Kirk},
  {Krause}, {Kuehn}, {Lahav}, {Li}, {Lima}, {March}, {Marshall}, {Menanteau},
  {Miquel}, {Mohr}, {Nord}, {Ogando}, {Roodman}, {Sanchez}, {Scarpine},
  {Schindler}, {Sevilla-Noarbe}, {Smith}, {Smith}, {Sobreira}, {Suchyta},
  {Swanson}, {Tarle}, {Tucker}, {Vikram}, {Walker}, {Weller}, \& {DES
  Collaboration}}]{2019MNRAS.482.1352M}
{McClintock}, T., {Varga}, T.~N., {Gruen}, D., {et~al.} 2019,
  \href{http://dx.doi.org/10.1093/mnras/sty2711}{\JournalTitle{\mnras}, 482,
  1352}

\bibitem[{Mehrtens {et~al.}(2012)Mehrtens, Romer, Hilton, Lloyd-Davies, Miller,
  Stanford, Hosmer, Hoyle, Collins, Liddle, Viana, Nichol, Stott, Dubois, Kay,
  Sahlén, Young, Short, Christodoulou, Watson, Davidson, Harrison, Baruah,
  Smith, Burke, Mayers, Deadman, Rooney, Edmondson, West, Campbell, Edge, Mann,
  Sabirli, Wake, Benoist, da~Costa, Maia, \& Ogando}]{Mehrtens2012}
Mehrtens, N., Romer, A.~K., Hilton, M., {et~al.} 2012,
  \href{http://dx.doi.org/10.1111/j.1365-2966.2012.20931.x}{\JournalTitle{\mnras},
  423, 1024}

\bibitem[{{Melchior} {et~al.}(2017){Melchior}, {Gruen}, {McClintock}, {Varga},
  {Sheldon}, {Rozo}, {Amara}, {Becker}, {Benson}, {Bermeo}, {Bridle},
  {Clampitt}, {Dietrich}, {Hartley}, {Hollowood}, {Jain}, {Jarvis}, {Jeltema},
  {Kacprzak}, {MacCrann}, {Rykoff}, {Saro}, {Suchyta}, {Troxel}, {Zuntz},
  {Bonnett}, {Plazas}, {Abbott}, {Abdalla}, {Annis}, {Benoit-L{\'e}vy},
  {Bernstein}, {Bertin}, {Brooks}, {Buckley-Geer}, {Carnero Rosell}, {Carrasco
  Kind}, {Carretero}, {Cunha}, {D'Andrea}, {da Costa}, {Desai}, {Eifler},
  {Flaugher}, {Fosalba}, {Garc{\'\i}a-Bellido}, {Gaztanaga}, {Gerdes},
  {Gruendl}, {Gschwend}, {Gutierrez}, {Honscheid}, {James}, {Kirk}, {Krause},
  {Kuehn}, {Kuropatkin}, {Lahav}, {Lima}, {Maia}, {March}, {Martini},
  {Menanteau}, {Miller}, {Miquel}, {Mohr}, {Nichol}, {Ogando}, {Romer},
  {Sanchez}, {Scarpine}, {Sevilla-Noarbe}, {Smith}, {Soares-Santos},
  {Sobreira}, {Swanson}, {Tarle}, {Thomas}, {Walker}, {Weller}, \&
  {Zhang}}]{2017MNRAS.469.4899M}
{Melchior}, P., {Gruen}, D., {McClintock}, T., {et~al.} 2017,
  \href{http://dx.doi.org/10.1093/mnras/stx1053}{\JournalTitle{\mnras}, 469,
  4899}

\bibitem[{Mo {et~al.}(2011)Mo, van~den Bosch, \& White}]{Mo_2011}
Mo, H., van~den Bosch, F., \& White, S. D.~M. 2011, Galaxy Formation and
  Evolution (Cambridge: Cambridge University Press)

\bibitem[{Moster {et~al.}(2010)Moster, Somerville, Maulbetsch, {Van Den Bosch},
  MacCi{\`{o}}, Naab, \& Oser}]{Moster2010}
Moster, B.~P., Somerville, R.~S., Maulbetsch, C., {et~al.} 2010,
  \href{http://dx.doi.org/10.1088/0004-637X/710/2/903}{\JournalTitle{\apj},
  710, 903}

\bibitem[{{Muldrew} {et~al.}(2011){Muldrew}, {Pearce}, \&
  {Power}}]{2011MNRAS.410.2617M}
{Muldrew}, S.~I., {Pearce}, F.~R., \& {Power}, C. 2011,
  \href{http://dx.doi.org/10.1111/j.1365-2966.2010.17636.x}{\JournalTitle{\mnras},
  410, 2617}

\bibitem[{{Murata} {et~al.}(2019){Murata}, {Oguri}, {Nishimichi}, {Takada},
  {Mandelbaum}, {More}, {Shirasaki}, {Nishizawa}, \&
  {Osato}}]{2019PASJ...71..107M}
{Murata}, R., {Oguri}, M., {Nishimichi}, T., {et~al.} 2019,
  \href{http://dx.doi.org/10.1093/pasj/psz092}{\JournalTitle{\pasj}, 71, 107}

\bibitem[{{Myles} {et~al.}(2021){Myles}, {Gruen}, {Mantz}, {Allen}, {Morris},
  {Rykoff}, {Costanzi}, {To}, {DeRose}, {Wechsler}, {Rozo}, {Jeltema},
  {Carrasco}, {Kremin}, \& {Kron}}]{Myles2021}
{Myles}, J., {Gruen}, D., {Mantz}, A.~B., {et~al.} 2021,
  \href{http://dx.doi.org/10.1093/mnras/stab1243}{\JournalTitle{\mnras}, 505,
  33}

\bibitem[{Navarro {et~al.}(1996)Navarro, Frenk, \& White}]{Navarro1996}
Navarro, J.~F., Frenk, C.~S., \& White, S. D.~M. 1996,
  \href{http://dx.doi.org/10.1086/177173}{\JournalTitle{\apj}, 462, 563}

\bibitem[{Onions {et~al.}(2012)Onions, Knebe, Pearce, Muldrew, Lux, Knollmann,
  Ascasibar, Behroozi, Elahi, Han, Maciejewski, Merch{\'{a}}n, Neyrinck, Ruiz,
  Sgr{\'{o}}, Springel, \& Tweed}]{Onions2012}
Onions, J., Knebe, A., Pearce, F.~R., {et~al.} 2012,
  \href{http://dx.doi.org/10.1111/j.1365-2966.2012.20947.x}{\JournalTitle{\mnras},
  423, 1200}

\bibitem[{Peacock \& Smith(2000)}]{Peacock2000}
Peacock, J.~A., \& Smith, R.~E. 2000,
  \href{http://dx.doi.org/10.1046/j.1365-8711.2000.03779.x}{\JournalTitle{\mnras},
  318, 1144}

\bibitem[{{Phriksee} {et~al.}(2020){Phriksee}, {Jullo}, {Limousin}, {Shan},
  {Finoguenov}, {Komonjinda}, {Wannawichian}, \&
  {Sawangwit}}]{2020MNRAS.491.1643P}
{Phriksee}, A., {Jullo}, E., {Limousin}, M., {et~al.} 2020,
  \href{http://dx.doi.org/10.1093/mnras/stz3049}{\JournalTitle{\mnras}, 491,
  1643}

\bibitem[{Rangel {et~al.}(2018)Rangel, Frontiere, Habib, Heitmann, Liao,
  Agrawal, \& Choudhary}]{Rangel2018}
Rangel, E., Frontiere, N., Habib, S., {et~al.} 2018,
  \href{http://dx.doi.org/10.1109/HiPC.2017.00052}{in Proceedings - 24th IEEE
  International Conference on High Performance Computing, HiPC 2017, Vol.
  2017-Decem} (IEEE), 398

\bibitem[{Rees \& Ostriker(1977)}]{Rees1977}
Rees, M.~J., \& Ostriker, J.~P. 1977,
  \href{http://dx.doi.org/10.1093/mnras/179.4.541}{\JournalTitle{\mnras}, 179,
  541}

\bibitem[{{Rossi} {et~al.}(2021){Rossi}, {Choi}, {Moon}, {Bautista},
  {Gil-Mar{\'\i}n}, {Paviot}, {Vargas-Maga{\~n}a}, {de la Torre}, {Fromenteau},
  {Ross}, {{\'A}vila}, {Burtin}, {Dawson}, {Escoffier}, {Habib}, {Heitmann},
  {Hou}, {Mueller}, {Percival}, {Smith}, {Zhao}, \& {Zhao}}]{rossi}
{Rossi}, G., {Choi}, P.~D., {Moon}, J., {et~al.} 2021,
  \href{http://dx.doi.org/10.1093/mnras/staa3955}{\JournalTitle{\mnras}, 505,
  377}

\bibitem[{{Rozo} \& {Rykoff}(2014)}]{Rozo14}
{Rozo}, E., \& {Rykoff}, E.~S. 2014,
  \href{http://dx.doi.org/10.1088/0004-637X/783/2/80}{\JournalTitle{\apj}, 783,
  80}

\bibitem[{Rykoff {et~al.}(2014)Rykoff, Rozo, Busha, Cunha, Finoguenov, Evrard,
  Hao, Koester, Leauthaud, Nord, Pierre, Reddick, Sadibekova, Sheldon, \&
  Wechsler}]{Rykoff2014}
Rykoff, E.~S., Rozo, E., Busha, M.~T., {et~al.} 2014,
  \href{http://dx.doi.org/10.1088/0004-637x/785/2/104}{\JournalTitle{\apj},
  785, 104}

\bibitem[{Sales {et~al.}(2007)Sales, Navarro, Abadi, \& Steinmetz}]{Sales2007}
Sales, L.~V., Navarro, J.~F., Abadi, M.~G., \& Steinmetz, M. 2007,
  \href{http://dx.doi.org/10.1111/j.1365-2966.2007.12026.x}{\JournalTitle{\mnras},
  379, 1475}

\bibitem[{Saro {et~al.}(2013)Saro, Mohr, Bazin, \& Dolag}]{saro2013toward}
Saro, A., Mohr, J.~J., Bazin, G., \& Dolag, K. 2013, \JournalTitle{\apj}, 772,
  47

\bibitem[{{Shin} {et~al.}(2021){Shin}, {Jain}, {Adhikari}, {Baxter}, {Chang},
  {Pandey}, {Salcedo}, {Weinberg}, {Amsellem}, {Battaglia}, {Belyakov},
  {Dacunha}, {Goldstein}, {Kravtsov}, {Varga}, {Abbott}, {Aguena}, {Alarcon},
  {Allam}, {Amon}, {Andrade-Oliveira}, {Annis}, {Bacon}, {Bechtol}, {Becker},
  {Bernstein}, {Bertin}, {Bocquet}, {Bond}, {Brooks}, {Buckley-Geer}, {Burke},
  {Campos}, {Rosell}, {Kind}, {Carretero}, {Chen}, {Choi}, {Costanzi}, {da
  Costa}, {DeRose}, {Desai}, {De Vicente}, {Devlin}, {Diehl}, {Dietrich},
  {Dodelson}, {Doel}, {Doux}, {Drlica-Wagner}, {Eckert}, {Elvin-Poole},
  {Everett}, {Ferraro}, {Ferrero}, {Fert{\'e}}, {Flaugher}, {Frieman},
  {Gallardo}, {Gatti}, {Gaztanaga}, {Gerdes}, {Gruen}, {Gruendl}, {Gutierrez},
  {Harrison}, {Hartley}, {Hill}, {Hilton}, {Hinton}, {Hollowood}, {Hughes},
  {James}, {Jarvis}, {Jeltema}, {Koopman}, {Krause}, {Kuehn}, {Kuropatkin},
  {Lahav}, {Lima}, {Lokken}, {MacCrann}, {Madhavacheril}, {Maia}, {McCullough},
  {McMahon}, {Melchior}, {Menanteau}, {Miquel}, {Mohr}, {Moodley}, {Morgan},
  {Myles}, {Nati}, {Navarro-Alsina}, {Niemack}, {Ogando}, {Page}, {Palmese},
  {Partridge}, {Paz-Chinch{\'o}n}, {Pereira}, {Pieres}, {Malag{\'o}n}, {Prat},
  {Raveri}, {Rodriguez-Monroy}, {Rollins}, {Romer}, {Rykoff}, {Salatino},
  {S{\'a}nchez}, {Sanchez}, {Santiago}, {Scarpine}, {Schillaci}, {Secco},
  {Serrano}, {Sevilla-Noarbe}, {Sheldon}, {Sherwin}, {Sif{\'o}n}, {Smith},
  {Soares-Santos}, {Staggs}, {Suchyta}, {Swanson}, {Tarle}, {Thomas}, {To},
  {Troxel}, {Tutusaus}, {Vavagiakis}, {Weller}, {Wollack}, {Yanny}, {Yin}, \&
  {Zhang}}]{Shin2021}
{Shin}, T., {Jain}, B., {Adhikari}, S., {et~al.} 2021,
  \href{http://dx.doi.org/10.1093/mnras/stab2505}{\JournalTitle{\mnras}, 507,
  5758}

\bibitem[{{Simet} {et~al.}(2017){Simet}, {McClintock}, {Mandelbaum}, {Rozo},
  {Rykoff}, {Sheldon}, \& {Wechsler}}]{2017MNRAS.466.3103S}
{Simet}, M., {McClintock}, T., {Mandelbaum}, R., {et~al.} 2017,
  \href{http://dx.doi.org/10.1093/mnras/stw3250}{\JournalTitle{\mnras}, 466,
  3103}

\bibitem[{Somerville \& Dav\'e(2015)}]{Somerville_2015}
Somerville, R.~S., \& Dav\'e, R. 2015,
  \href{http://dx.doi.org/10.1146/annurev-astro-082812-140951}{\JournalTitle{Annual
  Review of Astronomy and Astrophysics}, 53, 51}

\bibitem[{{Sultan} {et~al.}(2021){Sultan}, {Frontiere}, {Habib}, {Heitmann},
  {Kovacs}, {Larsen}, \& {Rangel}}]{LJ2}
{Sultan}, I., {Frontiere}, N., {Habib}, S., {et~al.} 2021,
  \href{http://dx.doi.org/10.3847/1538-4357/abf4fe}{\JournalTitle{\apj}, 913,
  109}

\bibitem[{Taylor(1997)}]{Taylor_1997}
Taylor, J. 1997, An Introduction to Error Analysis (Sausalito, California:
  University Science Books)

\bibitem[{Vale \& Ostriker(2004)}]{Val2004}
Vale, A., \& Ostriker, J.~P. 2004,
  \href{http://dx.doi.org/10.1111/j.1365-2966.2004.08059.x}{\JournalTitle{\mnras},
  353, 189}

\bibitem[{{van den Bosch}(2017)}]{2017MNRAS.468..885V}
{van den Bosch}, F.~C. 2017,
  \href{http://dx.doi.org/10.1093/mnras/stx520}{\JournalTitle{\mnras}, 468,
  885}

\bibitem[{{van den Bosch} \& {Ogiya}(2018)}]{2018MNRAS.475.4066V}
{van den Bosch}, F.~C., \& {Ogiya}, G. 2018,
  \href{http://dx.doi.org/10.1093/mnras/sty084}{\JournalTitle{\mnras}, 475,
  4066}

\bibitem[{Vogelsberger {et~al.}(2020)Vogelsberger, Marinacci, Torrey, \&
  Puchwein}]{vogelsberger2019cosmological}
Vogelsberger, M., Marinacci, F., Torrey, P., \& Puchwein, E. 2020, Cosmological
  Simulations of Galaxy Formation

\bibitem[{Wechsler \& Tinker(2018)}]{Wechsler_2018}
Wechsler, R.~H., \& Tinker, J.~L. 2018,
  \href{http://dx.doi.org/10.1146/annurev-astro-081817-051756}{\JournalTitle{Annual
  Review of Astronomy and Astrophysics}, 56, 435}

\bibitem[{Wetzel {et~al.}(2009)Wetzel, Cohn, \& White}]{Wetzel2009}
Wetzel, A.~R., Cohn, J.~D., \& White, M. 2009,
  \href{http://dx.doi.org/10.1111/j.1365-2966.2009.14424.x}{\JournalTitle{\mnras},
  395, 1376}

\bibitem[{Wetzel \& White(2010)}]{Wetzel2010}
Wetzel, A.~R., \& White, M. 2010,
  \href{http://dx.doi.org/10.1111/j.1365-2966.2009.16191.x}{\JournalTitle{\mnras},
  403, 1072}

\bibitem[{White {et~al.}(1987)White, Davis, Efstathioui, \& Frenk}]{white1987}
White, S.~D., Davis, M., Efstathioui, G., \& Frenk, C.~S. 1987,
  \JournalTitle{Nature}, 330, 451

\bibitem[{White \& Rees(1978)}]{White1978}
White, S. D.~M., \& Rees, M.~J. 1978,
  \href{http://dx.doi.org/10.1093/mnras/183.3.341}{\JournalTitle{\mnras}, 183,
  341}

\bibitem[{{Wu} {et~al.}(2022){Wu}, {Costanzi}, {To}, {Salcedo}, {Weinberg},
  {Annis}, {Bocquet}, {da Silva Pereira}, {DeRose}, {Esteves}, {Farahi},
  {Grandis}, {Rozo}, {Rykoff}, {Varga}, {Wechsler}, {Zeng}, {Zhang}, {Zhang},
  \& {DES Collaboration}}]{Wu2022}
{Wu}, H.-Y., {Costanzi}, M., {To}, C.-H., {et~al.} 2022,
  \href{http://dx.doi.org/10.1093/mnras/stac2048}{\JournalTitle{\mnras}, 515,
  4471}

\bibitem[{{Zehavi} {et~al.}(2019){Zehavi}, {Kerby}, {Contreras}, {Jim{\'e}nez},
  {Padilla}, \& {Baugh}}]{Zehavi19}
{Zehavi}, I., {Kerby}, S.~E., {Contreras}, S., {et~al.} 2019,
  \href{http://dx.doi.org/10.3847/1538-4357/ab4d4d}{\JournalTitle{\apj}, 887,
  17}

\bibitem[{{Zhang} {et~al.}(2019){Zhang}, {Jeltema}, {Hollowood}, {Everett},
  {Rozo}, {Farahi}, {Bermeo}, {Bhargava}, {Giles}, {Romer}, {Wilkinson},
  {Rykoff}, {Mantz}, {Diehl}, {Evrard}, {Stern}, {Gruen}, {von der Linden},
  {Splettstoesser}, {Chen}, {Costanzi}, {Allen}, {Collins}, {Hilton}, {Klein},
  {Mann}, {Manolopoulou}, {Morris}, {Mayers}, {Sahlen}, {Stott}, {Vergara
  Cervantes}, {Viana}, {Wechsler}, {Allam}, {Avila}, {Bechtol}, {Bertin},
  {Brooks}, {Burke}, {Carnero Rosell}, {Carrasco Kind}, {Carretero},
  {Castander}, {da Costa}, {De Vicente}, {Desai}, {Dietrich}, {Doel},
  {Flaugher}, {Fosalba}, {Frieman}, {Garc{\'\i}a-Bellido}, {Gaztanaga},
  {Gruendl}, {Gschwend}, {Gutierrez}, {Hartley}, {Honscheid}, {Hoyle},
  {Krause}, {Kuehn}, {Kuropatkin}, {Lima}, {Maia}, {Marshall}, {Melchior},
  {Menanteau}, {Miller}, {Miquel}, {Ogando}, {Plazas}, {Sanchez}, {Scarpine},
  {Schindler}, {Serrano}, {Sevilla-Noarbe}, {Smith}, {Soares-Santos},
  {Suchyta}, {Swanson}, {Tarle}, {Thomas}, {Tucker}, {Vikram}, {Wester}, \&
  {DES Collaboration}}]{2019MNRAS.487.2578Z}
{Zhang}, Y., {Jeltema}, T., {Hollowood}, D.~L., {et~al.} 2019,
  \href{http://dx.doi.org/10.1093/mnras/stz1361}{\JournalTitle{\mnras}, 487,
  2578}

\bibitem[{Zheng {et~al.}(2005)Zheng, Berlind, Weinberg, Benson, Baugh, Cole,
  Dave, Frenk, Katz, \& Lacey}]{Zheng2005}
Zheng, Z., Berlind, A.~A., Weinberg, D.~H., {et~al.} 2005,
  \href{http://dx.doi.org/10.1086/466510}{\JournalTitle{\apj}, 633, 791}

\end{thebibliography}

\appendix 

\section{Best-Fit Profiles}
\label{sec:profiles}

\begin{figure*}
\centering\includegraphics[width=1.0\textwidth]{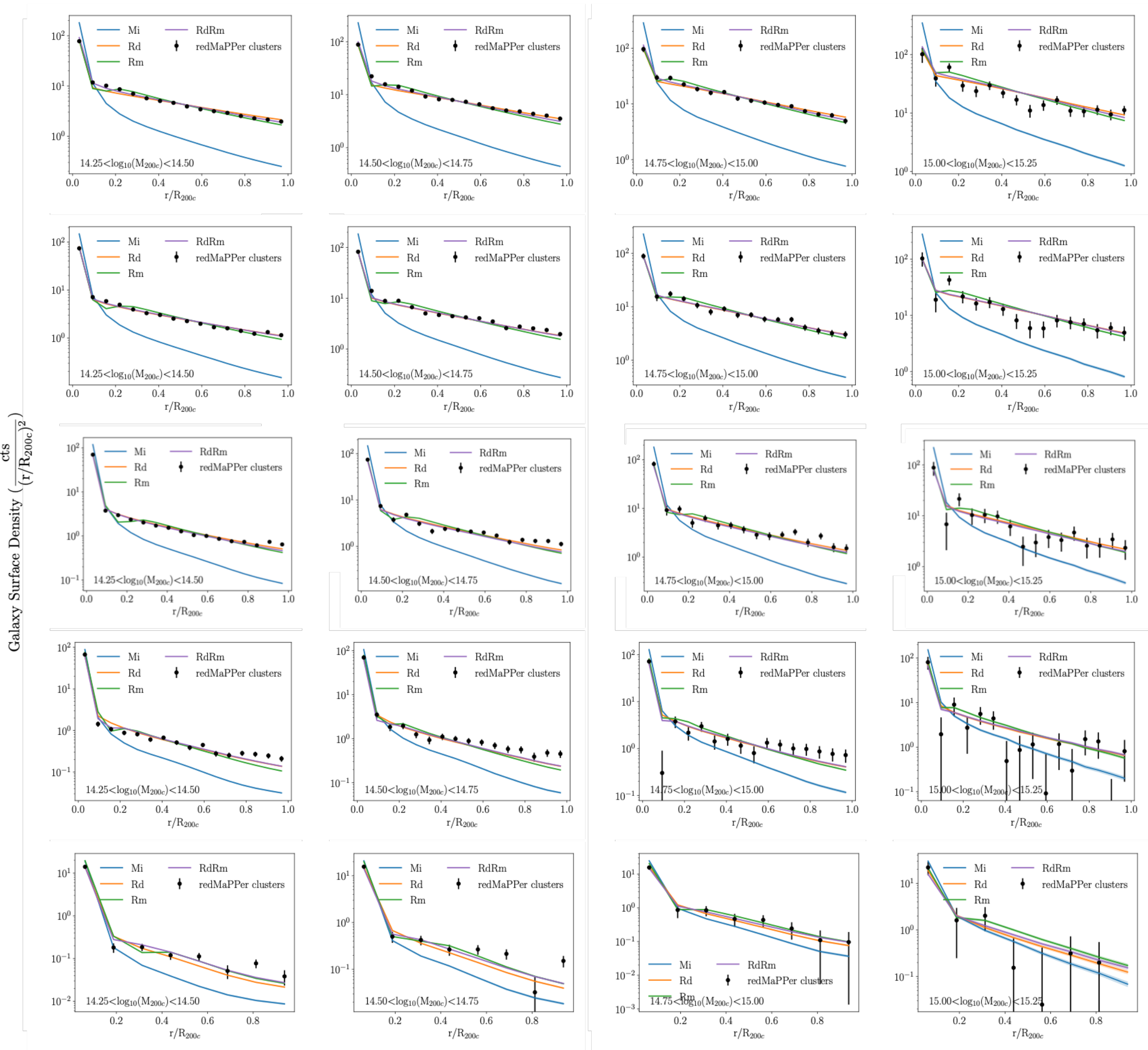}
\caption{\label{fig:all_profiles}Surface galaxy density profiles from SDSS \redmapper clusters and best-fit core models. From top to bottom the luminosity threshold is varied following Table~\ref{tab:my_label}: $>$0.40~\lstar, $>$0.63~\lstar, $>$1.00~\lstar, $>$1.58~\lstar and $>$2.50~\lstar. From the left to right we vary the cluster mass bin considered. The mass bin edges are $\log_{10}(M_{200\mathrm{c}})$=14.25, 14.50, 14.75, 15.0 and 15.25. Each panel shows the results for all of the four models used. The three models that take into account an additional parameter beyond just \minfall\ all provide good fits.}
\end{figure*}

In this Appendix, we present results for all the best-fit profiles for the four different cluster mass bins and the five different luminosity thresholds in Figure~\ref{fig:all_profiles}. The images are a visual representation of the information shown in Table~\ref{tab:best_fit_parameteres} in Section~\ref{sec:model_flavors}. The second row includes the three panels that are shown in the main paper in Figure~\ref{fig:model_profiles}. 

The panels in Figure~\ref{fig:all_profiles} are organized as follows. Each row shows the results for four different cluster mass bins, the lighter masses are on the left with mass increasing towards the right. Each column represents the five luminosity thresholds listed in Table~\ref{tab:my_label}. The top row shows the lowest luminosity threshold and the bottom row the highest (and therefore containing the bright end of cluster galaxies). Ordered in this way, the number of galaxies available to measure the surface galaxy density profile in each panel decreases rapidly when approaching the panel in the lower right corner. Each of the twenty panels shows the four different models that we investigate throughout the paper, a simple model that only varies the infall mass threshold (Mi, blue line), the model that allows in addition for core disruption (MiRd, orange line), the model that includes core mergers (MiRm, green line) and the model that combines all three parameters (MiRdRm, purple line).  

The compilation of all cases shows very clearly that the Mi model always leads to an underestimation of the profile beyond $\sim 0.2r/R_{200}$. Only for the clusters where we restrict the cluster galaxies to the brightest sample (last row), does it provide a reasonable match. The figure also clearly shows that the other three models provide very satisfactory results. 

\end{document}